\documentclass[a4paper,12pt]{article}

\usepackage{listings}
\usepackage{anysize}
\marginsize{2cm}{2cm}{2cm}{2cm}
\usepackage{amsmath, amsthm, amssymb, amsfonts} %%%for math environment
\usepackage{mathtools}
\usepackage{graphicx}
\usepackage{lscape}
\usepackage{float}
\usepackage{setspace}
\restylefloat{table}
\usepackage{epstopdf}
\usepackage{xcolor}
\usepackage{url}

\newcommand{\mc}[3]{\multicolumn{#1}{#2}{#3}}

\makeatother

\title{Sovereign Debt Default and Climate Risk}

\author{Emilio Barucci\thanks{Department of Mathematics, Politecnico di Milano, Italy - emilio.barucci@polimi.it},
\ Daniele Marazzina\thanks{Department of Mathematics, Politecnico di Milano, Italy - daniele.marazzina@polimi.it},
\ Aldo Nassigh\thanks{Department of Mathematics, Politecnico di Milano, Italy - aldo.nassigh@polimi.it}
}

\begin{document}
%\nocite{*}
\global\long\def\sgn{\mathop{\mathrm{sgn}}}
\linespread{1.5}

\maketitle

\abstract{We explore the interplay between sovereign debt default/renegotiation and environmental factors (e.g., pollution from land use, natural resource exploitation). Pollution contributes to the likelihood of natural disasters and influences economic growth rates. The country can default on its debt at any time while also deciding whether to invest in pollution abatement. The framework provides insights into the credit spreads of sovereign bonds and explains the observed relationship between bond spread and a country's climate vulnerability. Through calibration for developing and low-income countries, we demonstrate that there is limited incentive for these countries to address climate risk, and the sensitivity of bond spreads to climate vulnerability remains modest. Climate risk does not play a relevant role on the decision to default on sovereign debt. 
Financial support for climate abatement expenditures can effectively foster climate adaptation actions, instead renegotiation conditional upon pollution abatement does not produce any effect.}

\vspace{1cm}

%\begin{flushleft}
 %     {\bf Keywords:} \\
%\end{flushleft}

%\newpage
\section{Introduction}

The dangerous {\em liaison} between climate risk and public debt mostly concerns developing and emerging countries.
These countries are responsible in a limited way for greenhouse gas emissions, but are more exposed than developed countries to climate risk according to vulnerability indexes such as 
the Notre Dame Global Adaptation Initiative or the Climate Vulnerability Monitor, see  \cite{BOLTO,BURKE,CHAN,CHEN,DARA}. 

These countries do not take actions to address climate risk mostly because the effort would jeopardize economic growth and private green financing is lacking, see \cite{ALIG,PRASA}. 
Private financing is limited because developing and emerging countries are not fully integrated in financial markets and 
``risk adaptation'' investments (e.g., drought-resistant seeds, resilient buildings, sea walls) are less likely to generate financial returns compared to ``risk mitigation'' investments (reduction of greenhouse gas emissions) and, therefore, 
attract less funds, see \cite{CLIM}.

Risk mitigation refers to actions to decrease gas emissions and temperature, risk adaptation to actions to abate the effects of global warming and of natural disasters. The latter mostly refers to the conservation of 
natural resources, e.g. reforestation, land regeneration, maintenance of biodiversity. 
While developed countries are suffering in limited way from global warming and some of them 
(mostly in the northern hemisphere) may even get a benefit, climate change is significantly affecting 
developing and emerging countries. In these countries, risk adaptation investment is an urgent issue.   
According to \cite{UNEP}, developing countries need \$70 billion per year to cover adaptation costs,
but the figure will double by 2030 and will be around 280-500 billion by 2050. 
Adaptation costs are estimated to amount to $1/4$ percent of world GDP, 1-2\% on average 
of GDP in emerging and low income countries, 1 billion every year for Pacific island countries (7-9\% of their GDP), see 
\cite{ALIG,IMF0}. In small island countries, a significant portion of the population faces heightened risks due to rising sea levels.
It is worthwhile to observe that risk adaptation in emerging and developing countries is also linked to risk mitigation 
as it concerns land use which plays a key role in greenhouse emissions and in their containment.

As developing and emerging countries suffer of little private financing to tackle climate risk, they
mostly rely on public funds and international transfers. 
In this environment, the surge of public debt in the 
aftermath of the pandemic and of the Ukrainian war represents a significant obstacle, 
see \cite{CHAB}. According to \cite{IMF,UNNAT}, 29 out 69 countries eligible to access concessional funds under the IMF's Poverty Reduction and Growth Trust at the same time show a high debt and are climate vulnerable. Emerging and developing economies have 
limited fiscal space and are often at high risk of debt distress, in many cases adaptation cost 
estimates exceed the available fiscal space, see \cite{ALIG,CHAM}.
This motivated the debate on how to promote climate action in developing and emerging countries 
with the support of developed countries and new financial devices, e.g., debt for climate swaps and climate conditional grants,
see statements and decisions of COP29 and  \cite{CHAM,IMF,IMF0,DEBTR}. 

This paper aims to contribute to this debate. We deal with climate risk and sovereign debt renegotiation/sustainability.
Our goal is to investigate the optimal management of debt when the economy is exposed to climate risk 
and the country may take actions to mitigate its effects. 
We are interested in addressing three main research questions:
How strong is the trade-off between climate risk adaptation actions and growth in developing and emerging countries?
What is the relationship between climate risk and debt service?
Can international financial support foster climate actions in developing countries?

The paper is at the interception between two strands of literature: debt default/renegotiation and climate risk.
We build on the models of \cite{AGU,ARE} for debt default/renegotiation of a small economy.
In the spirit of \cite{NUNO,REBE}, we consider a pure endowment economy, its GDP follows a jump-diffusion process that allows
for rare disasters.
At each instant of time the country may decide to default,
which implies a reduction of GDP and, for a random time, the inhibition to participate in financial markets issuing public debt.
When the country is admitted again to financial markets its debt is set to a fraction of the level at the default time
thanks to a renegotiation process. 
We assume that the economic activity impacts physical pollution,
i.e., depletion of natural resources of the country, mostly because of agricultural production and mining,
see \cite{LAMB}. The depletion of natural resources negatively impacts the output through two channels:
a reduction of the growth rate of the economy and a higher probability of a natural disaster.
As shown in \cite{AGU4}, permanent shocks drive fluctuations in emerging countries.
Rare disasters do not occur at an exogenous rate as in \cite{BAR,REBE}, their arrival rate is linked to the stock of physical pollution. 
The connection between the probability of a disaster and pollution is similar to the assumption in \cite{HAMB22},
where it is connected to global warming. In our setting climate risk is mostly environmental risk, the words climate and environmental risk are used as synonymous.

The papers closest to ours are \cite{HONG,HONG2}. In these papers, disasters are modeled through a jump-diffusion process.
\cite{HONG2} deals with mitigation of disasters,
i.e., adaptation risk according to our taxonomy: firms invest in
disaster exposure mitigation which allows them to reduce the capital loss
associated with a natural disaster. They show that mitigation is under-supplied in competitive markets relative to the first best 
planner's solution, an outcome that can be achieved via capital taxation and subsidy schemes. 
In \cite{HONG}, 
emission removals are proportional to the de-carbonization capital stock, firms can invest to increase it.
The probability of a natural disaster increases in emissions, that are proportional to the capital stock, and decreases in emission removals.
Our model provides a parsimonious setting based on the stock of physical pollution (depletion of natural resources) which affects both the
growth rate of the economy and the probability of a disaster.
Notice that in our model pollution is located in the territory, a phenomenon quite common in developing countries, and is not strictly related to temperature as in 
\cite{BARN,VAN_VAN,HAMB21,HAMB22}.
The country is allowed to invest in reducing the stock of pollution replenishing natural resources (adaptation expenditure).

The connection between climate risk and public debt has been investigated in several papers from an empirical point of view, see \cite{AGA,BEIRN,BEIRNA,BOLTO,CEVIK,KLING}.
Considering 98 advanced and developing countries over the period 1995-2017,
\cite{CEVIK} looks at sovereign bond yields and spreads with respect to the U.S. benchmark. As explanatory variables they consider
vulnerability (country's exposure, sensitivity, and capacity to adapt to the impacts of climate change) and resilience (country's capacity to apply economic investments and convert them to adaptation actions). 
They observe that a 1\% increase in climate vulnerability increases long-term bond spread by 3\%.
\cite{KLING} covers 38 countries (7 advanced and 31 emerging and developing economies) over the period 1998–2018; 
according to their analysis exposure to climate risk leads to a significant increase in borrowing costs. \cite{BEIRNA} concentrates  
on southeast Asia showing a strong positive correlation between climate vulnerability and sovereign spreads. 
\cite{BEIRN} shows that climate vulnerability is associated with higher spreads in emerging economies, 
but there is no connection for advanced economies. \cite{BOLTO} confirm the relationship between spreads and 
climate vulnerability in developing and emerging countries, the magnitudo increases over time.

These papers empirically show the connection between climate vulnerability and public finance in emerging countries 
providing the route for a vicious circle: high public debt leads to less action on climate risk 
which leads to higher costs for natural catastrophes, higher cost of debt and so forth less fiscal space and 
less investment to face climate risk.

Our model provides a theoretical framework to analyze the climate risk-public debt nexus. The main results can be summarized as follows.
Debt default occurs for a very high debt/GDP ratio, the decision to default on debt does not represent the optimal choice for the countries in scope of our analysis. 
Climate action can occur both in the default and in the no-default regime and is not necessarily monotonic in GDP, however the incentive to take adaptation risk actions is very limited for the countries considered in the analysis. 
A positive relation is observed between pollution and spread of sovereign debt, but the sensitivity is limited and
there is no strong conflict between climate risk and debt sustainability.
International support, under the form of contribution to investment in risk  adaptation, is effective in inducing countries to take action against adaptation risk, instead renegotiation on debt conditional upon a pollution goal is not effective to foster climate action. Although pollution negatively affects consumption, it doesn't provide an incentive to default on debt. 

The paper is organized as follows.
In Sections \ref{MOD} and \ref{OPT} we present and solve the model.
Section \ref{BOND} deals with the pricing of government bonds.
In Section \ref{CALIBR} we address the estimation and calibration of the model, while Section \ref{NUM} provides the main results.
In Section \ref{SENS} we deal with a sensitivity analysis and we evaluate policy options to induce countries to invest in adaptation risk expenditure.
Finally, in Section \ref{CONC} we provide conclusive remarks. Appendix \ref{CALIB} contains a detailed analysis of the estimation and calibration of the model, in Appendix \ref{SCHEME} we present the numerical procedure, while Appendix \ref{Sensitivity} deals with the sensitivity analysis of our model with respect to some key parameters.

\section{The model}
\label{MOD}
We consider a pure endowment economy with two state variables: output ($y$) and the stock of physical pollution ($f$).
We build our analysis on the models proposed in \cite{NUNO,REBE},
that are continuous time extensions of \cite{AGU,ARE}, including climate risk.

We deal with a small emerging or developing country, whose economy is mostly based on agriculture or the exploitation of natural resources. 
In this environment we may restrict our attention to a pure endowment economy without considering
capital accumulation which is more relevant in a developed economy, 
e.g., see \cite{BARN,CAI,HAMB21,HAMB22,HONG,VAN_VAN}. 
A pure endowment economy allows us to concentrate on four key decisions: consumption, pollution abatement investment, debt issuance, and default.

Climate risk affects the endowment of the economy through the following channel: consumption
of natural resources, which is directly linked to output,
impacts the stock of physical pollution and depletes the amount of natural resources, which in turn
affects both the growth rate of the economy and the probability of a natural disaster.
Therefore, we model two types of climate risk effects: long term effects concerning the rate
of growth of the economy and the probability of rare disasters. The first one
is modeled assuming that the stock of pollution $f$
reduces the drift of the economy through a damage function, the second one is modeled assuming that the probability of
a natural disaster, yielding a negative jump in the endowment process, depends on the stock of pollution.

The output of the economy $y$ evolves as a jump-diffusion process:
\begin{equation}
\label{OUTPUT}
\frac{dy(t)}{y(t)}=\mu (1-D(f(t))dt+\sigma_y  dW_1(t) - (1-Z) dN(t).
\end{equation}
The drift is affected by a damage function which is a function of the stock of pollution as in \cite{HAMB21,VAN_VAN} for temperature:
\begin{equation}
\label{OUTPUT2}
D(f) = {\phi} f^{1+{\theta}},  \ \ \theta \ge -1.
\end{equation}
The function describes how the long-term growth rate of the output is impacted by the stock of pollution,
$\theta$ characterizes the convexity of the damage ratio as a function of  the stock of pollution.

Dealing with the sources of randomness, $W_1$ is a Wiener process, independent of the disasters' counting process $N(t)$. In several papers, disasters are modeled through a jump process that depends upon pollution/temperature,
e.g., see \cite{HAMB22}. In our setting the frequency of a disaster depends upon  the stock of pollution.
$Z$ represents the fraction of output recovered in case of a disaster, its cumulative distribution function $F(Z)$ being governed by a power law which is
independent of climate risk
\begin{equation}\label{Z}
F(Z) = Z^\beta, \; 0 \le Z \le 1,
\end{equation}
and $N(t)$ is a non homogeneous Poisson process with intensity $\nu$, which nonlinearly depends on the stock of pollution:
\begin{equation}
\label{NU_T}
\nu = \nu_0 + \nu_1 f^{\psi_f},
\end{equation}
where $\psi_f$ characterizes the convexity of the probability of a disaster with respect to the stock of pollution. 
As in \cite{REBE}, but differently from \cite{HONG2}, we assume that there is no possibility of mitigating 
the effects of natural disasters.

We build on the model in \cite{HAMB21,VAN_VAN} for atmospheric carbon-temperature,
introducing a stochastic process for the pollution stock $f$, which is defined as 
$$
f=f_{pre}+\widehat{f},
$$
$f_{pre}$ being the unavoidable pollution (the pre-industrial level), and $\widehat{f}$ evolving as
\begin{equation}
\label{FOSS}
d\widehat{f}(t)=\left(e(t) - k_f u(t) -\varphi \widehat{f}(t)\right) dt+\sigma_f dW_2(t),
\end{equation}
$W_2$ being a Wiener process independent of $W_1$ and $N$.

As in \cite{HONG}, we assume that natural resource consumption (increase of pollution) is proportional to output:
\begin{equation}
\label{TECH}
e(t)=k_e e^{-gt} y(t),
\end{equation}
where $k_e e^{-gt}>0$ is the increase in pollution stock per unit of output.
We model technological advancement through the exogenous rate $g\ge 0$, i.e., as time goes the pollution intensity decreases, see \cite{VAN_VAN} for a similar assumption.

On top of the increase proportional to output, we have two additional components in the drift:
the stock of pollution decreases by $k_f u(t)$
thanks to climate adaptation expenditure, where $k_f>0$ represents
the amount of pollution removed per unit of expenditure;
moreover, pollution decays by natural sinks at the rate $\varphi > 0$. Thanks to the term $k_f u(t)$,
the model allows for climate adaptation expenditures to reduce the stock of pollution.
We can interpret $u(t)$ as replenishment of natural resources, afforestation, reforestation, reclamation of land.
The decision is at the hands of the representative agent similarly to \cite{HAMB21,NORD}, for the emission control rate, and
\cite{HONG}, where emissions are proportional to the stock of capital
and emission removals are proportional to the decarbonization stock that evolves
according to a stochastic differential equation. Notice that we model an exogenous technological advancement only for the effect of production on pollution and not on
adaptation expenditure.

Equation (\ref{FOSS}) implies that the pollution stock decays to the pre-industrial level if consumption of natural resources and adaptation expenditures cease.
 To keep the model tractable, we  assume a bound on $u$, i.e.,  $u \in [0, \bar{u}y]$, 
that is, $\bar{u}$ is the maximum output fraction devoted to climate adaption expenditure. Notice that 
$f$ is specular to the stock of natural resources as modeled in \cite{GOLO,HONG}.

In the model, we introduce State debt which plays a key role to determine the feasibility of financing climate adaptation projects, particularly in countries with limited fiscal capacity and high climate vulnerability. More precisely, let $B(t)$ be the outstanding nominal government debt, represented by a bond amortized at the rate $\lambda>0$. 
As we do not consider inflation, debt is in real term.
The continuous time evolution of the outstanding debt is:
\begin{equation}
\label{BUDG0}
dB(t)=B^{new}(t)dt-\lambda B(t)dt
\end{equation}
where $B^{new}$ is the flow of new debt issued in $t$. 

Each bond pays a proportional coupon $\delta$, which is assumed to be the risk-free rate, 
so that a default-free State will issue debt at par. Its market price is denoted by $Q(t)$.
The flow of funds for the State satisfies the budget constraint:
\begin{equation}
\label{BUDG}
Q(t)B^{new}(t)=(\lambda+\delta)B(t)+c(t)+u(t)-y(t),
\end{equation}
where $c$ is consumption.
(\ref{BUDG0}) and (\ref{BUDG}) render the following evolution of outstanding debt:
\begin{equation}
\label{BUDG1}
dB(t)= \left(\frac{1}{Q(t)}\left(\left(\lambda+\delta\right)B(t)+c(t)+u(t)-y(t)\right)-\lambda B(t) \right) dt.
\end{equation}

In this framework, the representative agent (the State) maximizes a recursive utility from consumption:
\begin{equation}
\label{UTILITY}
 V(t) = \mathbb{E} \bigg{[} \int_t^{\infty} h(c(s),V(s)) ds \bigg{]}
\end{equation}
where the continuous-time aggregator is provided by
\begingroup
\singlespacing
\begin{equation}
\label{AGGR}
    h(c,V) =
    \begin{cases*}
    \rho \theta V\left(\left(\frac{c}{\left(\left(1-\gamma\right)V\right)^{\frac{1}{1-\gamma}}}\right)^{1-1/\psi}-1\right) & if $\psi \neq 1$\\
     \rho (1-\gamma) V   \ln \left( \frac{c}{\left(\left(1-\gamma\right)V\right)^\frac{1}{1-\gamma}}\right)   & if $\psi = 1$,
    \end{cases*}
\end{equation}
\endgroup
with $\theta:=\frac{1-\gamma}{1-1/\psi}$, $\gamma >1$ measuring the degree of relative risk aversion,
$\psi$ representing the elasticity of intertemporal substitution, and $\rho$ denoting the time preference.

At each instant of time the representative agent may opt to default on debt.
As in the literature on optimal debt default, default entails two types of costs for the economy, see \cite{AGU,ARE,NUNO}: permanent endowment loss; moreover, 
the country is excluded from international capital markets temporarily, the duration of the exclusion period being a random variable $\tau$
which is distributed as an exponential distribution with an average duration $\frac{1}{\chi}$.

Therefore, after default the economy is reduced right away to an autarkic mode with a deduction of the endowment due to the institutional and economic disorder associated with the default.
Let $T$ be the default time, $y(T-)$ the output shortly before the default and
$\eta \in (0,1]$ the fraction of the output recovered after the default event, then 
\begin{equation}
\label{OUTPUT_A}
y(T)=\eta  y(T-).
\end{equation}

During the period of exclusion from international markets, the country simply consumes
the output reduced by the abatement expenditure ($c= y - u$), and debt is freezed at the nominal debt shortly before the default ($B(T-)$); 
the country is not allowed to issue new debt and does not pay interests on existing debt.
Default is not an absorbing state, the exclusion period is due to the loss of credibility of the country after default that
prevents the State from issuing public debt on the market. 
The exclusion period allows for renegotiation
between the country and bond holders on existing debt.
\cite{AGU,ARE} assume that when the country is admitted again to issue debt, the amount is set to $0$, here we 
assume that  when the country is admitted again to issue debt at time $T+\tau$ then, upon renegotiation, the amount of debt is
$B(T+\tau) = \theta_D B(T-)$,  see \cite{NUNO} for a similar assumption.
$\theta_D \in [0,1]$ is the fraction of debt recovered by investors. 

\section{Solving the model}
\label{OPT}
Neglecting the default option, the setting is characterized by three state variables $(y, B, \hat{f})$
and two admissible controls $(c, u)$, i.e., measurable and adapted to the
available information, satisfying integrability conditions, and with $u(t)$ having values in
$[0, \overline{u} y(t)],\,\forall t>0$. 

We consider three different problems: the original problem abstracting from the default decision, the optimal problem after default in the autarky regime, and the complete problem. $v^{nd}, \ v^{def}, \ v$ denote the corresponding value functions. 

First of all, we deal with the no-default option case. The optimization problem for the representative agent concerns the following value function:
\begin{equation*}
 v^{nd}(t, y, \widehat{f}, B) = \max_{c,u}\mathbb{E} \bigg{[} \int_t^{\infty} h(c(s),v^{nd}(s)) ds  \,|\, y(t)=y,\, B(t)=B,\, \widehat{f}(t)=\widehat{f} \bigg{]}
\end{equation*}
subject to (\ref{OUTPUT}), (\ref{FOSS}), (\ref{BUDG1}). 
The Hamilton-Jacobi-Bellman (HJB) equation associated with the problem is:
\begin{equation}\label{HJB}
0= v^{nd}_t +\underset{c,u}{\max} \big{[} h(c,v^{nd}) + \mathcal{L}_{c,u} \mkern9mu v^{nd}  \big{]} +
					            (\nu_0 + \nu_1 f^{\psi_f})
					            \mathbb{E} \big{[} v^{nd}(Zy) - v^{nd}(y) \big{]}
\end{equation}
where
\begin{align}\label{DIFF}
\mathcal{L}_{c,u} \mkern9mu v = &\mu \left(1 - \phi (f_{pre}+{\widehat{f}})^{1+\theta}\right) y v_y + %\nonumber \\
			          %&
			          \left(r B + \frac{c + u - y}{Q}\right) v_B + %\nonumber \\
					          %&
					          \left(k_e e^{-gt} y - k_f u - \varphi \widehat{f}\right) v_{\widehat{f}}\, + \nonumber \\
					          & 
					         \frac{1}{2} \sigma_{y}^{2} y^2 v_{yy} +
					             %\frac{1}{2} \sigma_{y}^{2} \Theta^2 v_{BB} +
					           \frac{1}{2} \sigma_{f}^{2}  v_{{\widehat{f}}{\widehat{f}}}\,,
					            %-					 \sigma_{y}^{2} y \Theta v_{yB},
\end{align}
$r$ being defined below. For ease of exposition, we may drop all or some of the arguments of the value function, i.e.,  $v^{nd}$ is to be interpreted as $v^{nd}(t,y,{\widehat{f}},B)$, and $v^{nd}(Zy)$ as $v^{nd}(t,Zy,{\widehat{f}},B)$, the same shortcut is used for other value functions to be defined below.

The right-hand side of (\ref{DIFF}) deserves some comments.
The first term takes into account the damage function derived from the pollution stock.
 The second term contains the bond yield
 \begin{equation}
\label{r}
r := \frac{\lambda+\delta}{Q} - \lambda,
\end{equation}
 which is given by the internal rate of return of the bond, i.e., the rate for which the 
 discounted future cash flow of the bond coincides with its price for a risk-neutral
investor. 
In case of a State fully committed to honor its payment obligations, the bond is priced at par by the risk-neutral investor:
\begin{equation}
\label{Qfc}
Q = \mathbb{E} \bigg{[} \int_0^\infty (\delta+\lambda) e^{-(\delta+\lambda)s} ds \bigg{]} = 1,
\end{equation}
and $r$ is constant and equal to the risk-free rate $\delta$. 
The third term takes into account the dependence of the value function from the drift of the pollution stock, with explicit time-dependence owed to the technological advancement.
The fourth and the fifth terms are the standard second order terms. 
The last term in \eqref{HJB} is associated with jumps: after a natural disaster the output drops to the recovery value $Z y(t-)$ of its pre-jump value $y(t-)$.

The first order condition for consumption is as follows:
\begin{equation}\label{FOC_c}
h_c = - \frac{1}{Q} v^{nd}_B,
\end{equation}
$h_c$ denoting the first order derivative of $h$ with respect to $c$. Condition \eqref{FOC_c} equates the marginal cost of debt to the negative
of the marginal utility of consumption. Given the utility aggregator \eqref{AGGR}, in case $\psi \ne 1$, we get:
\begin{equation*}
	c = \left(- \frac{1}{Q} v^{nd}_B \frac{\left(\left(1-\gamma\right) v^{nd} \right)^{\frac{1}{\theta}}}{\rho (1-\gamma) v^{nd}} \right)^{-\psi}	.
\end{equation*}
Dealing with the abatement expenses, the first order condition implies
\begin{equation}\label{FOC_u}
	u =
	\begin{cases*}
		0 & if $v^{nd}_B < Q k_f v^{nd}_f$ \\
		\in [0,\bar{u}y] & if $v^{nd}_B = Q k_f v^{nd}_f$\\
		\bar{u}y & if $v^{nd}_B > Q k_f v^{nd}_f$,
	\end{cases*}
\end{equation}
that is, the abatement expenditure follows a step function of the state variables, jumping to its maximum value as soon as it becomes convenient. 

Let us now consider the exclusion/autarky regime, that is, the post-default problem: new debt is set equal to zero, 
consumption is given by output minus government's expenditure in pollution abatement ($c=y-u$).
As default occurs, there is a sudden drop in the output, see \eqref{OUTPUT_A}.
The stochastic processes of the output and the pollution
stock during the exclusion phase are unchanged with respect to the no-default regime.

Let $T$ be the default time, and $T+\tau$ the time in which the country is admitted again to issue debt.
The value function at time $t$, with $T \leq  t< T+\tau$, is
\begin{align}
\label{VALUE_A}
&v^{def}(t,  y, {\widehat{f}}, B) = \underset{u}{\max} \quad \mathbb{E} \bigg{[}  \int_t^{T+\tau} h(y(s) - u(s),v^{def}(s)) ds \, +\, \int_{T+\tau}^{\infty} h(c(s),v(s)) ds \nonumber \\
 							    & \quad
							    |\quad y(t)=y,\, %y(T)=\eta y(T-),\,
                                B(s)= B(T-)=B\,\forall s\in[t,T+\tau),\, B(T+\tau)=\theta_D B,\,{\widehat{f}}(t)={\widehat{f}}
							    \bigg{]},
\end{align}
where the integration in the first term is across the exclusion period $(t, T+\tau)$
and the second term is associated with
 the general problem after renegotiation of debt, whose value function is denoted by $v = v(t, y, f, B)$ and will be defined thereafter.

The HJB equation associated with the optimization problem \eqref{VALUE_A} for $t \ge T$ during the autarky period is:
\begin{align}\label{HJB_A}
0= v^{def}_t + &\underset{u}{\max} \big{[} h( y -u,v^{def}) + \mathcal{L}^{def}_{u} \mkern9mu v^{def}  \big{] }+
					            (\nu_0 + \nu_1 (f_{pre}+{\widehat{f}})^{\psi_f})
					            \mathbb{E} \big{[} v^{def}(Z y) - v^{def}( y) \big{]} + \nonumber \\
					            &\chi \big{[} v( \theta_D B) - v^{def}(B)\big{]},
\end{align}
where
we have introduced the operator associated to the diffusion processes \eqref{OUTPUT} and \eqref{FOSS} during the autarky period:
\begin{align}\label{DIFF_A}
\mathcal{L}^{def}_{u} \mkern9mu v = &\mu (1 - \phi (f_{pre}+{\widehat{f}})^{1+\theta}) y v_y + 
			        	          (k_e e^{-gt} y - k_f u - \varphi {\widehat{f}}) v_{\widehat{f}} + 
					          \frac{1}{2} \sigma_{y}^{2} y^2 v_{yy} +
					           \frac{1}{2} \sigma_{f}^{2}  v_{{\widehat{f}}{\widehat{f}}}.
\end{align}
The last term on the right-hand side of \eqref{HJB_A}
represents the change in value owed to jumping back into the no-default regime, with the country's resumption of access to capital markets at time $t>T$ with reduced
debt ($B(t) = \theta_D B(T-)$)
 and output $y(t) = y(t-)$. Without loss of generality, we assign zero probability to the simultaneous occurrence of the two jumps, i.e., climate disaster and recovery to the normal regime.

The first order condition of the HJB problem during the exclusion phase is:
\begin{equation}\label{FOC_u_autarky}
h_c \biggr\rvert_{y-u} = - k_f v^{def}_{\widehat{f}}.
\end{equation}
This condition implicitly defines the optimal climate risk expenditure in the range $u \in [0, \bar{u}y]$ based on the derivative of the
value function with respect to the pollution stock. In case of $\psi \ne 1$ we get:
\begin{equation*}
	u = \max  \big{\{}  \min  \big{\{} \bar{u}y ; y- c^{def} \big{\}} ; 0\big{\}} \text{ where } c^{def} = \left( - k_f  v^{def}_{\widehat{f}} \frac{[(1-\gamma) v^{nd} ]^{\frac{1}{\theta}}}{\rho (1-\gamma) v^{nd}} \right)^{-\psi}.
	\end{equation*}

The value function $v$ in the last term of the right-hand side of \eqref{VALUE_A} and \eqref{HJB_A}
is the solution of a HJB variational inequality, see \cite{NUNO,PHAM},
i.e., the optimal value including the default option, modeled here as an optimal stopping time. More precisely, $v=v(t, y, \hat{f}, B)$ is determined
 solving the following HJB variational inequality:
\begin{eqnarray}\label{HJB_VAR}
0&=&\max \bigg{\{} v^{def}(\eta y) - v(y),\\
&&v_t +\underset{c,u}{\max} \big{[} h(c,v) + \mathcal{L}_{c,u} \mkern9mu v  \big{]} +
					            (\nu_0 + \nu_1 (f_{pre}+{\widehat{f}})^{\psi_f})
					            \mathbb{E} \big{[} v(Zy) - v(y) \big{]} \bigg{\}}\nonumber
\end{eqnarray}
where the operator $\mathcal{L}_{c,u} $ associated to the processes \eqref{OUTPUT}, \eqref{FOSS} and \eqref{BUDG1} is
provided in \eqref{DIFF} and the value function $v^{def}$ solves the HJB equation \eqref{HJB_A}.
The solution of the variational inequality \eqref{HJB_VAR} provides at the same time two distinct
outcomes: the value function $v$ and the default barrier at time $t$,
i.e., the boundary in the state variables space $(y, {\widehat{f}}, B)$ for which the condition determining the threshold for the optimal default 
$$
v^{def}(\eta y) = v(y),
$$
implied by the assumption $y(T)=\eta y(T-)$, occurs.

Notice that the solution of the optimization problem in the autarky regime is intrinsically coupled to the solution of the original 
problem, since both $v$ and $v^{def}$ appear in Equation \eqref{HJB_A} and \eqref{HJB_VAR}. In this perspective, our approach is different from the one followed by \cite{REBE}, where the default
of the country is induced by a rare disaster. 
 In that setting, it is possible to solve two de-coupled optimization problems in the two separate regimes (\emph{full commitment} and \emph{autarky}). Moreover, in our framework the solution of the HJB \eqref{HJB_A} and \eqref{HJB_VAR} is also coupled with the computation of the bonds price $Q$, which is addressed in the next section. 

\section{Pricing bonds}
\label{BOND}
The price of the sovereign bond is set in a market populated by (international) risk-neutral investors under the assumption of no arbitrage as in \cite{REBE}.
The price is strictly connected with the default option. 
In case of a defaulted State, the debt is priced by a risk-neutral investor at \emph{market recovery}, i.e., $Q^{def}$ is the value of the bond after reentry into
the capital market under the risk-neutral measure $\mathbb{Q}$ discounted at the risk-free rate across the autarky regime period. 
In case of a deterministic exclusion period ($\hat{\tau}$), the value of the bond would be:
\begin{equation}
\label{Qdef_naive}
Q^{def}(T) = e^{-\delta \hat{\tau}} \theta_D \mathbb{E}^{\mathbb{Q}} \big{[} Q(T+\hat{\tau}) \big{]}.
\end{equation}
The right-hand side is provided by the product of the discount factor across the autarky regime period and the expected risk-neutral price of risky bonds at time $T+\hat{\tau}$ scaled by the factor $\theta_D$.
$T$ is the optimal default time that solves the variational problem \eqref{HJB_VAR},
i.e., $T$ is the first time in which the inequality  $v^{def}(t, \eta y, {\widehat{f}}, B) \ge  v(t, y, {\widehat{f}}, B)$ holds true.

As we assume that the exclusion time is a random variable $\tau$ distributed according to
an exponential distribution with intensity $\chi$, the market recovery is
expressed by the following expected value of the price $Q$:
\begin{equation}\label{Qdef}
Q^{def}(T) = \mathbb{E}^{\mathbb{Q}} \bigg{[} \int_T^\infty e^{-\delta (s-T)} \chi e^{-\chi (s-T)} \theta_D Q(s,y(s), {\widehat{f}}(s), \theta_D B(T-)) ds \bigg{]},
\end{equation}
where we recall that $y(T)=\eta y(T-)$. The integral on the right-hand side of \eqref{Qdef} is over the exclusion time period with probability of reentry in the capital market during the time 
interval $(T+\tau, T+\tau + d\tau)$ assigned by $\chi e^{-\chi \tau} d\tau$. $Q(s, y(s), {\widehat{f}}(s), \theta_D B(T-))$ denotes the price of the bond at time $s$ in case the country is readmitted to capital markets at that time.

We follow \cite{NUNO} and evaluate the risk-neutral price of the risky bond before default time $T$ discounting
the cash flow at the risk-free discount rate until the default time $T$ and the final single cash flow arising from market recovery $Q^{def}$ at time $T$ in (\ref{Qdef}).
Therefore, the price of the bond before default is:
\begin{equation}\label{Q}
Q(t)= \mathbb{E}^{\mathbb{Q}} \bigg{[} \int_t^T (\delta+\lambda) e^{-(\delta+\lambda)(s-t)} ds
					+ e^{-(\delta+\lambda)(T-t)} Q^{def}(T,\eta y(T), {\widehat{f}}(T), B(T)) \bigg{]}.
\end{equation}
The relationship between $Q$ and $Q^{def}$ shows the continuity of the risk-neutral price of the debt in the space $(y, {\widehat{f}}, B)$ as the system approaches the default barrier.

So far we have derived two interlinked pricing formulas for the government bond.
The pricing formula \eqref{Q} applies if, at time $t$, the variables $(y,{\widehat{f}},B)$ lie in the
 {\em no-default region} of the state space, formula \eqref{Qdef}
holds if we are in the {\em default region}, the two regions being separated by the
default boundary. To handle the bond price, it is convenient to introduce the optimal default policy function $d$
defined on the state space, which may take only two values: $1$ (default regime) or $0$ (no-default regime):

\begingroup
\singlespacing
\begin{equation}\label{d}
	d(t,y,f,B) =
	\begin{cases*}
		1 & if $v(t,y,{\widehat{f}},B) = v^{def}(t,\eta y,{\widehat{f}},B)$ \\
		0 & otherwise.
	\end{cases*}
\end{equation}
\endgroup

To simplify the analysis, 
we assume that $W_1$, the realization of the recovery fraction, and the exit time from the autarky are independent. 
These risks can be hedged by risk-neutral investors, see \cite{REBE} for a similar assumption.
On the contrary, we assume that the jump-arrival timing risk owed to the pollution stock cannot be hedged away by diversification and is priced by the market. Therefore, to move to risk-neutral pricing, we introduce the risk-neutral disaster frequency $\nu^\mathbb{Q}(f)$ in \eqref{Qdef} and \eqref{Q}, see \cite{REBE}. 
We assume that the risk-neutral expectation of the disaster arrival time under the risk-neutral probability 
is shorter than under the physical measure, i.e., $\nu^\mathbb{Q}(f)>\nu(f)$. This assumption captures the idea that well-diversified
investors demand a premium for natural disasters. 
To keep the model tractable under the risk-neutral probability measure, 
we assume the same nonlinear dependence of the jump frequency from the pollution stock laid down in \eqref{NU_T}, with a market-implied sensitivity to the pollution $\nu_1^\mathbb{Q}$ to be calibrated from the market prices of the bonds:
\begin{equation}
\label{NU_T_mkt}
\nu^\mathbb{Q} = \nu_0 + \nu_1^\mathbb{Q}  f^{\psi_f},
\end{equation}
where the pollution stock is given by the overall level 
$f=f_{pre}+{\widehat{f}}$.

In this framework, starting from the no-default regime,  we apply the Feynman-Kac formula to recast \eqref{Q} into a partial differential equation:
\begin{align}\label{Q_FK}
			(\delta + \lambda) Q &= Q_t + (\delta + \lambda) + \mathcal{L}_{c,u} \mkern9mu Q  +
					            \nu^\mathbb{Q}(f)
					            \mathbb{E} \big{[} Q(Zy) - Q(y) \big{]} \ &\mathrm{if} \ d=0 \nonumber \\					
 Q(t, y, {\widehat{f}}, B)&= Q^{def}(t, \eta y, {\widehat{f}}, B)  \  &\mathrm{if} \  d=1,
\end{align}
where the operator $\mathcal{L}_{c,u,}$, associated to the processes \eqref{OUTPUT}, \eqref{FOSS} and \eqref{BUDG1}, is provided
by \eqref{DIFF} and is evaluated for the optimal $c$ and $u$ solving the optimization problem \eqref{HJB_VAR}. 

In turn, we apply the Feynman-Kac formula to \eqref{Qdef}:
\begin{equation}\label{Qdef_FK}
	\chi \left(Q^{def}(B) - \theta_D Q(\theta_D B)\right) + \delta Q^{def} =
	Q_t^{def} + \mathcal{L}^{def}_{u} Q^{def} + \nu^\mathbb{Q}(f)
					            \mathbb{E} \big{[} Q^{def}(Z y) - Q^{def}(y) \big{]},
\end{equation}
where the operator $\mathcal{L}^{def}_{u}$ is provided by \eqref{DIFF_A} and is
evaluated along the optimal strategy $u$  which defines the value function $v^{def}$.

The risk-neutral price of the risky bond is therefore determined by the simultaneous solution of two interconnected problems, i.e., \eqref{Q_FK} and \eqref{Qdef_FK}. These problems are also coupled with the solution of the HJB \eqref{HJB_A} and \eqref{HJB_VAR} through the definition of the optimal default policy function $d$ in \eqref{d}. We refer to Appendix \ref{SCHEME} for details on the solution algorithm.

\section{Estimation and model calibration}
\label{CALIBR}

The model includes 23 parameters. It is impossible to estimate all of them on a country specific basis because some pieces of information are missing, therefore we rely on different sources to estimate/calibrate our model: parameters taken from the literature, parameters estimated for macro-regions (taking the median value as a reference), and parameters estimated/calibrated country by country. In what follows, we present a summary of the procedure and the main results, for a detailed discussion we refer to Appendix \ref{CALIB}. 

Parameters concerning agent's preferences, renegotiation on public debt and default are not estimated/calibrated ex-novo, we refer to those provided in the literature. In Table \ref{tbaseline} we report the parameter values and sources between brackets.

\begin{table}[tp]
\centering
\begin{tabular}{c|c|c}
  parameter & \mc{1}{c}{value} & \mc{1}{|l}{meaning}\\\hline
   \mc{1}{c|}{$\psi$} 			& \mc{1}{l|}{$2/3$} & \mc{1}{l}{Elasticity of intertemporal substitution \cite{VAN_VAN}} \\
   \mc{1}{c|}{$\gamma$} 		& \mc{1}{l|}{$2.0$} & \mc{1}{l}{Coefficient of relative risk aversion \cite{REBE}} \\
   \mc{1}{c|}{$\rho$} 			& \mc{1}{l|}{$0.052$} & \mc{1}{l}{Rate of time preference \cite{REBE}} \\
   \mc{1}{c|}{$\delta$} 		& \mc{1}{l|}{$0.040$} & \mc{1}{l}{Risk-free rate \cite{REBE}} \\
\mc{1}{c|}{$\nu_0$} 		& \mc{1}{l|}{$0.36\%$} & \mc{1}{l}{Basic frequency of environmental events \cite{KAR2019}} \\
   \mc{1}{c|}{$\theta_D$} 		& \mc{1}{l|}{$0.50$} & \mc{1}{l}{Fraction of debt recovered after recovery form autarky \cite{NUNO}} \\ %\cite{NUNO}: $\theta_D=0.5$
   \mc{1}{c|}{$\lambda$} 		& \mc{1}{l|}{$0.264$} & \mc{1}{l}{Bond amortization rate \cite{NUNO}} \\
   \mc{1}{c|}{$\eta$} 			& \mc{1}{l|}{$0.75$} & \mc{1}{l}{Fraction of the output recovered after the default event \cite{de2006costs}} \\
   \mc{1}{c|}{$\chi$} 			& \mc{1}{l|}{$0.25$} & \mc{1}{l}{Intensity of the exponential distribution describing exit from autarky \cite{REBE}} \\ 
    \mc{1}{c|}{$\beta$} 			& \mc{1}{l|}{$6.27$} & \mc{1}{l}{Power law exponent of the fraction of output recovered in case of a disaster \cite{BAR}} \\
    
   \end{tabular}
\caption{Parameter values taken from the literature (source in brackets).}\label{tbaseline}
\end{table}

As we are able to collect full information only for a limited set of developing and emerging countries, we opt to proceed with the estimation of some parameters for three clusters of homogeneous countries. 
Developing and emerging countries are identified as those classified middle-income or low-income by the International Monetary Fund. The three macro-regions are: Sub-Saharan Africa (SSA), Latin America \& Caribbean (LATAM); Asia (South \& East) \& Pacific (SA-APAC) corresponding to 
the IMF macro-regions {\em South Asia}, {\em East Asia}, and {\em Pacific}. The clusters are made up respectively of  42, 29, and 29 countries\footnote{For the complete list of countries, we refer to Table \ref{COUNTRIES} of Appendix \ref{CALIB}.}.
The dataset containing all information necessary for the estimation of parameters spans from 2013 to 2023. The values of the parameters obtained as medians of estimated parameters across countries at the macro-region level are reported in Table \ref{tmean_SSA}, \ref{tmean_LATAM}, and \ref{tmean_SA_APAC}.

\begin{table}[tp]
\centering
\begin{tabular}{c|c|c}
  parameter & \mc{1}{c}{value} & \mc{1}{|l}{meaning}\\\hline
   \mc{1}{c|}{$\mu$} 			& \mc{1}{l|}{$0.047$} & \mc{1}{l}{Output drift} \\ %See file: "2024-07-07 Results Calibration Output Process"
   \mc{1}{c|}{$\theta$} 		& \mc{1}{l|}{$0.52$} & \mc{1}{l}{Nonlinearity in pollution stock of the damage to output } \\ %See file: "Jump_diffusion Process - calibration for SSA Countries"
   \mc{1}{c|}{$\varphi$} 		& \mc{1}{l|}{$0.31$} & \mc{1}{l}{Annual pollution decay rate by natural sinks} \\ %See file: "2024-07-28 Results Calibration Pollutions Stock Process"
   \mc{1}{c|}{$\sigma_y$} 		& \mc{1}{l|}{$0.051$} & \mc{1}{l}{Output volatility} \\ %See file: "2024-07-07 Results Calibration Output Process"
   \mc{1}{c|}{$\sigma_f$} 		& \mc{1}{l|}{$0.025$} & \mc{1}{l}{Pollution stock volatility} \\  %See file: "2024-07-28 Results Calibration Pollutions Stock Process"
   \mc{1}{c|}{$g$} 			& \mc{1}{l|}{$0.014$} & \mc{1}{l}{Improvement rate of output cleanliness by technological advancement} \\  %See file: "2024-07-28 Results Calibration Pollutions Stock Process"
   \mc{1}{c|}{$\nu_1$} 		& \mc{1}{l|}{$0.16$} & \mc{1}{l}{Climate disaster frequency: increase by unit pollution stock} \\  %See file: "Jump_diffusion Process - calibration for SSA Countries"
   \mc{1}{c|}{$\psi_f$} 		& \mc{1}{l|}{$2.1$} & \mc{1}{l}{Nonlinearity in pollution stock of the climate disaster frequency} \\ %See file: "Jump_diffusion Process - calibration for SSA Countries"
   \mc{1}{c|}{$f_{pre}$} 		& \mc{1}{l|}{$0.27$} & \mc{1}{l}{Unavoidable pollution} \\ %Average 2013-2023 of Cabo Verde: the country in the lowest income class with minimum pollution stock in 2023
   \mc{1}{c|}{$\phi$} 			& \mc{1}{l|}{$0.93$} & \mc{1}{l}{Amount of damage to output for unit $f$} \\	 %See file: "Jump_diffusion Process - calibration for SSA Countries"
   \mc{1}{c|}{$k_e$} 			& \mc{1}{l|}{$0.11$} & \mc{1}{l}{Increase in pollution stock per unit of output} \\ %See file: "2024-07-28 Results Calibration Pollutions Stock Process"
   \end{tabular}
\caption{Parameters: median across the pool of countries, SSA macro-region.}\label{tmean_SSA}
\end{table}

\begin{table}[tp]
\centering
\begin{tabular}{c|c|c}
  parameter & \mc{1}{c}{value} & \mc{1}{|l}{meaning}\\
    \hline
   \mc{1}{c|}{$\mu$} 			& \mc{1}{l|}{$0.050$} & \mc{1}{l}{Output drift} \\ %See file: "2024-07-07 Results Calibration Output Process"
   \mc{1}{c|}{$\theta$} 		& \mc{1}{l|}{$0.73$} & \mc{1}{l}{Nonlinearity in pollution stock of the damage to output } \\ %See file: "Jump_diffusion Process - calibration for SSA Countries"
   \mc{1}{c|}{$\varphi$} 		& \mc{1}{l|}{$0.34$} & \mc{1}{l}{Annual pollution decay rate by natural sinks} \\ %See file: "2024-07-28 Results Calibration Pollutions Stock Process"
   \mc{1}{c|}{$\sigma_y$} 		& \mc{1}{l|}{$0.054$} & \mc{1}{l}{Output volatility} \\ %See file: "2024-07-07 Results Calibration Output Process"
   \mc{1}{c|}{$\sigma_f$} 		& \mc{1}{l|}{$0.028$} & \mc{1}{l}{Pollution stock volatility} \\  %See file: "2024-07-28 Results Calibration Pollutions Stock Process"
   \mc{1}{c|}{$g$} 			& \mc{1}{l|}{$0.028$} & \mc{1}{l}{Improvement rate of output cleanliness by technological advancement} \\  %See file: "2024-07-28 Results Calibration Pollutions Stock Process"
   \mc{1}{c|}{$\nu_1$} 		& \mc{1}{l|}{$0.12$} & \mc{1}{l}{Climate disaster frequency: increase by unit pollution stock} \\  %See file: "Jump_diffusion Process - calibration for SSA Countries"
   \mc{1}{c|}{$\psi_f$} 		& \mc{1}{l|}{$2.5$} & \mc{1}{l}{Nonlinearity in pollution stock of the climate disaster frequency} \\ %See file: "Jump_diffusion Process - calibration for SSA Countries"
   \mc{1}{c|}{$f_{pre}$} 		& \mc{1}{l|}{$0.27$} & \mc{1}{l}{Unavoidable pollution} \\ %Average 2013-2023 of Cabo Verde: the country in the lowest income class with minimum pollution stock in 2023
   \mc{1}{c|}{$\phi$} 			& \mc{1}{l|}{$0.98$} & \mc{1}{l}{Amount of damage to output for unit $f$} \\	 %See file: "Jump_diffusion Process - calibration for SSA Countries"
   \mc{1}{c|}{$k_e$} 			& \mc{1}{l|}{$0.045$} & \mc{1}{l}{Increase in pollution stock per unit of output} \\ %See file: "2024-07-28 Results Calibration Pollutions Stock Process"
   \end{tabular}
\caption{Parameters: median across the pool of countries, LATAM macro-region.}\label{tmean_LATAM}
\end{table}

\begin{table}[tp]
\centering
\begin{tabular}{c|c|c}
  parameter & \mc{1}{c}{value} & \mc{1}{|l}{meaning}\\\hline
   \mc{1}{c|}{$\mu$} 			& \mc{1}{l|}{$0.058$} & \mc{1}{l}{Output drift} \\ %See file: "2024-07-07 Results Calibration Output Process"
   \mc{1}{c|}{$\theta$} 		& \mc{1}{l|}{$0.48$} & \mc{1}{l}{Nonlinearity in pollution stock of the damage to output } \\ %See file: "Jump_diffusion Process - calibration for SSA Countries"
   \mc{1}{c|}{$\varphi$} 		& \mc{1}{l|}{$0.35$} & \mc{1}{l}{Annual pollution decay rate by natural sinks} \\ %See file: "2024-07-28 Results Calibration Pollutions Stock Process"
   \mc{1}{c|}{$\sigma_y$} 		& \mc{1}{l|}{$0.039$} & \mc{1}{l}{Output volatility} \\ %See file: "2024-07-07 Results Calibration Output Process"
   \mc{1}{c|}{$\sigma_f$} 		& \mc{1}{l|}{$0.022$} & \mc{1}{l}{Pollution stock volatility} \\  %See file: "2024-07-28 Results Calibration Pollutions Stock Process"
   \mc{1}{c|}{$g$} 			& \mc{1}{l|}{$0.021$} & \mc{1}{l}{Improvement rate of output cleanliness by technological advancement} \\  %See file: "2024-07-28 Results Calibration Pollutions Stock Process"
   \mc{1}{c|}{$\nu_1$} 		& \mc{1}{l|}{$0.22$} & \mc{1}{l}{Climate disaster frequency: increase by unit pollution stock} \\  %See file: "Jump_diffusion Process - calibration for SSA Countries"
   \mc{1}{c|}{$\psi_f$} 		& \mc{1}{l|}{$2.1$} & \mc{1}{l}{Nonlinearity in pollution stock of the climate disaster frequency} \\ %See file: "Jump_diffusion Process - calibration for SSA Countries"
   \mc{1}{c|}{$f_{pre}$} 		& \mc{1}{l|}{$0.27$} & \mc{1}{l}{Unavoidable pollution} \\ %Average 2013-2023 of Cabo Verde: the country in the lowest income class with minimum pollution stock in 2023
   \mc{1}{c|}{$\phi$} 			& \mc{1}{l|}{$0.99$} & \mc{1}{l}{Amount of damage to output for unit $f$} \\	 %See file: "Jump_diffusion Process - calibration for SSA Countries"
   \mc{1}{c|}{$k_e$} 			& \mc{1}{l|}{$0.064$} & \mc{1}{l}{Increase in pollution stock per unit of output} \\ %See file: "2024-07-28 Results Calibration Pollutions Stock Process"
    \end{tabular}
\caption{Parameters: median across the pool of  countries, SA-APAC macro-region.\\
	}\label{tmean_SA_APAC}
\end{table}

To estimate process \eqref{FOSS}, we use the Climate-driven INFORM Risk indicator as a proxy for pollution stock. We map indicator values from negligible to high pollution levels. To enhance statistical robustness, a two-step estimation approach is applied: first, we estimate technological progress in pollution reduction in (\ref{TECH}), i.e, parameter $g$, using $CO_2$ emissions data as a proxy of physical pollution. We observe a stronger technology advancement for LATAM than for the SA-APAC and the SSA macro-region. Then we estimate the remaining parameters through the maximum likelihood method.

To estimate the endowment process \eqref{OUTPUT}, we use the CRED data on climate-related events (7,598 climate events across 229 countries). We focus on pollution-driven events, such as floods and extreme temperatures, excluding geophysical events like earthquakes. 
The dataset comprises many disasters with a wide variability of output loss. The output model \eqref{OUTPUT} requires to disentangle the economic effects of frequent, low-severity events (in a sense systematic) from those of rarer, severe disasters. The first type is captured by the damage function, the second one by the jump process.

To incorporate both types of events and to calibrate the key parameters, we assume that the recovery distribution in case of a rare disaster follows a power law as in \eqref{Z} with $\beta=6.27$, as estimated in \cite{BAR} based on a large sample of rare disasters. The baseline frequency of a rare disaster $\nu_0$ is taken from \cite{KAR2019}, representing the average environmental event frequency from 1916 to 1955. 
Then, by momentum matching with respect to the output process, we derive the parameters of the damage function $(\phi, \theta)$ and the rare-disaster frequency $(\nu_1,\psi_f)$ from the output data of the countries for each of the macro-region in scope. The median values of the parameters are reported in Table \ref{tmean_SSA}, \ref{tmean_LATAM}, and \ref{tmean_SA_APAC}.

The calibration reveals that the SSA macro-region is the weakest, exhibiting the lowest growth rate alongside the highest pollution intensity per unit of output. Furthermore, the SSA region shows the lowest rate of output cleanliness improvement from technological advancement. Conversely, the SA-APAC macro-region displays the most favorable output parameters, with both a high growth rate and low volatility. However, the SA-APAC region appears to be the most vulnerable to climate disasters, as shown by the high frequency of $\nu_1$. In summary, the LATAM macro-region seems to be the one least exposed to environmental risk, showing intermediate values for output growth and volatility, combined with the lowest pollution intensity per unit of output and the lowest $\nu_1$ value. Additionally, the LATAM region achieves the highest rate of output cleanliness improvement from technological advancements.

We are left with two parameters that are not obtained neither from the literature nor from data: $k_f$ (decrease of pollution stock per unit of expenditure) and $\overline{u}$ (cap on climate adaptation expenditure as a percentage of the output). We set the two parameters to be $0.20$. 

Country specific information concerns output, pollution, public debt and market data.
We obtain market bond yields corrected for exchange rate (country-US dollar) from LSEG Data\&Analytics by calculating the monthly, volume-weighted average market yield as of December 2023 for medium to long-term bonds issued by countries belonging to the three macro-regions. Sovereign debt is traded on the secondary market for only a subset of these countries. In Table \ref{Q_SSA}, \ref{Q_LATAM}, and \ref{Q_SA_APAC}, for each macro-region, we report the information for countries with available market data. In the fifth column, we report the market yield labeled as $r^{mkt}$. The second, third, and fourth columns show the values of the three state variables of the model $(y,f,B)$ corresponding to the year 2023 observation; output and debt are normalized by the average output of the macro-region. 

\begin{table}[tp]
\centering
\begin{tabular}{c|ccc|c||cc|cc}
Country 		& $y$ & $f$ & $B$ &$r^{mkt}$	& $r^{mod}$		      &Error&$d$&$u/y$\\\hline
Median 		& 1.58& 0.42& 0.75& 0.1004	&   0.1004 &   0.0000& 0&0\\\hline	
Uganda		& 0.56& 0.70& 0.27& 0.1597	&  0.1309  &  0.0288 &0  &0\\
Zambia		& 0.71& 0.39& 0.70& 0.2425	&      0.1178   & 0.1247&0  &0\\
Nigeria		& 1.08& 0.66& 0.42& 0.1040	&  0.1295  & -0.0255&0  &0\\
Kenya		& 1.13& 0.66& 0.79& 0.1787	&      0.2276  & -0.0489&0  &0\\
Namibia		& 2.03& 0.38& 1.37& 0.0968	&   0.1353  & -0.0385&0  &0\\
South Africa	& 2.88& 0.45& 2.12& 0.0954	&  0.2029 &  -0.1075&0  &0\\
Botswana		& 3.36& 0.26& 0.63& 0.0867	&      0.0667  &  0.0200&0 &0.2\\
Mauritius		& 4.93& 0.21& 3.93& 0.0446	&     0.1756  & -0.1310&0 &0.2\\
\end{tabular}
\caption{Data and model results of the countries of the SSA macro-region for which bond data are available. Debt and output are normalized by the average output, equal to $5,460$ US Intl. dollars.}\label{Q_SSA}
\end{table}

\begin{table}[tp]
\centering
\begin{tabular}{c|ccc|c||cc|cc}
Country 		& $y$ & $f$ & $B$ &$r^{mkt}$	& $r^{mod}$		      &Error&$d$&$u/y$\\\hline
Median		& 1.15& 0.47& 0.69& 0.0849	&  0.0849   & 0.0000 & 0&0\\\hline
Peru			& 0.94& 0.47& 0.32& 0.0637	&     0.0630 &   0.0007 &0 &0\\
Colombia		& 1.14& 0.52& 0.63& 0.0849	&      0.0933  & -0.0084&0 &0\\
Brazil		& 1.15& 0.46& 1.02& 0.1040	&   0.1101  & -0.0061&0  &0\\
Mexico		& 1.44& 0.49& 0.76& 0.0914	&   0.0878   & 0.0036&0 &0\\
Chile			& 1.79& 0.32& 0.69& 0.0478	&  0.0590 &  -0.0112&0 &0\\ 
\end{tabular}
\caption{Data and model results of the countries of the LATAM macro-region for which bond data are available. Debt and output are normalised by the average output, equal to $16,367$ US Intl. dollars.}\label{Q_LATAM}
\end{table}

\begin{table}[tp]
\centering
\begin{tabular}{c|ccc|c||cc|ccc}
Country 		& $y$ & $f$ & $B$ &$r^{mkt}$	& $r^{mod}$		      &Error&$d$&$u/y$\\\hline
Median		& 1.54& 0.45& 0.72& 0.0556	&     0.0556  &  0.0000   &      0       &    0\\\hline
Pakistan		& 0.75& 0.61& 0.57& 0.1071	&      0.0781   & 0.0290    &     0    &     0\\
Bangladesh	& 0.89& 0.57& 0.35& 0.0766	&   0.0564   & 0.0202     &    0    &      0\\
India			& 0.94& 0.53& 0.77& 0.0705&	 0.0726  & -0.0021   &      0          &   0\\
Philippines	& 1.17& 0.53& 0.68& 0.0584	& 0.0632  & -0.0048   &      0         &    0\\
Sri Lanka		& 1.59& 0.31& 1.84& 0.0528	& 0.0687  & -0.0159   &      0    &     0\\
Viet Nam		& 1.48& 0.37& 0.50& 0.0261	&  0.0499  & -0.0238   &      0    &      0\\
Indonesia		& 1.64& 0.46& 0.64& 0.0678	& 0.0540  &  0.0138   &      0    &       0\\
Thailand		& 2.36& 0.44& 1.45& 0.0340	&   0.0631  & -0.0291   &      0     &        0\\
China		& 2.39& 0.30& 1.99& 0.0351	&   0.0626  & -0.0275   &      0    &         0\\
Malaysia		& 3.89& 0.29& 2.60& 0.0422	&   0.0639  & -0.0217   &      0       &0.2\\
\end{tabular}
\caption{Data and model results of the countries of the SA-APAC macro-region for which bond data are available. Debt and output are normalised by the average output, equal to $8,946$ US Intl. dollars.}\label{Q_SA_APAC}
\end{table}

The top row of each table presents the median values for the three state variables at the macro-region level, alongside the corresponding median rates. These values are used to calibrate the risk-adjusted frequency ($\nu_1^{\mathbb{Q}}$) identified as the parameter that, when incorporated in Equation \eqref{NU_T_mkt}, aligns the the model bond yield with the market bond yield for the median values of the three macro-regions:
\begin{equation}
    r^{mod}(Q) = r^{mkt}
\end{equation}
where $r^{mod}$ is derived from the bond price $Q$ by Equation \eqref{r}. 
Table \ref{nu_mkt} (second column) shows the values of $\nu_1^{\mathbb{Q}}$ for each of the three macro-regions. Notice that the risk adjusted frequency $\nu_1^{\mathbb{Q}}$ is much higher than the historical one $\nu_1$ in all the three macro-regions.

\begin{table}[tp]
    \centering
    \begin{tabular}{l|c|cc}
         & $\nu_1^{\mathbb{Q}}$ & $\Tilde{f}$  & Mkt. freq. $\nu^{\mathbb{Q}}(\Tilde{f})$\\\hline
         SSA&           $19.5$ & $0.42$ & $3.2$\\
         LATAM&         $24.6$ & $0.47$ & $3.7$\\
         SA-APAC&    $10.1$ & $0.45$ & $1.9$\\
    \end{tabular}
    \caption{Risk-adjusted $\nu_1^{\mathbb{Q}}$, calibrated on the median across countries with traded debt (second column). Fourth column: current market-implied expected frequency of climate-driven  disasters $\nu^{\mathbb{Q}}(\Tilde{f})$ calculated from Equation \eqref{NU_T_mkt} for the median country with traded debt, given the current pollution stock (third column).}
    \label{nu_mkt}
\end{table}

\section{Main results: bond spread, renegotiation, climate action}
\label{NUM}

In the sixth column of 
Table \ref{Q_SSA}, \ref{Q_LATAM}, and \ref{Q_SA_APAC}, we present the model rate $r^{mod}(Q)$ for each country. The rate is derived from the bond price using Equation \eqref{r} evaluated for the median risk neutral density at the macro-region level, see Table \ref{nu_mkt}, and the current state variables $(y,f,B)$ of the country as reported in the second, third, and fourth columns, respectively. The seventh column reports the difference between the model-implied rate and the observed market rate. The model fits quite well the market rate with the exception of the SSA macro-region where the errors can be significant. The results are particularly good for the LATAM region (error below 100 basis points).

The last two columns 
report the optimal default decision and the optimal investment in pollution abatement. We observe that no country should default on debt according to our model showing that there is not a strong conflict between climate risk and debt sustainability.   
As far as climate risk action is concerned, we show that only Malaysia, Botswana and Mauritius should invest in reducing the pollution stock.

\begin{figure}[tp]
\centering

 \includegraphics[width=0.3\textwidth]{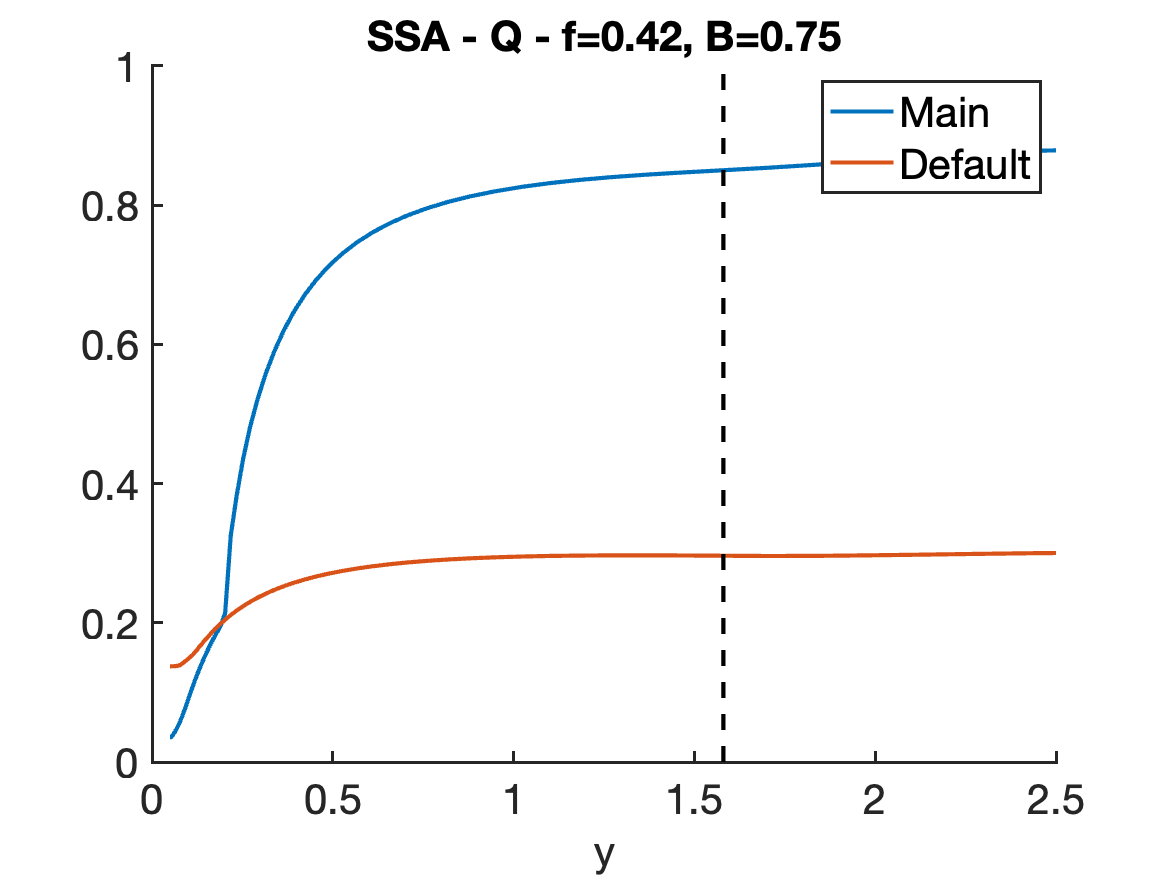}
 \includegraphics[width=0.3\textwidth]{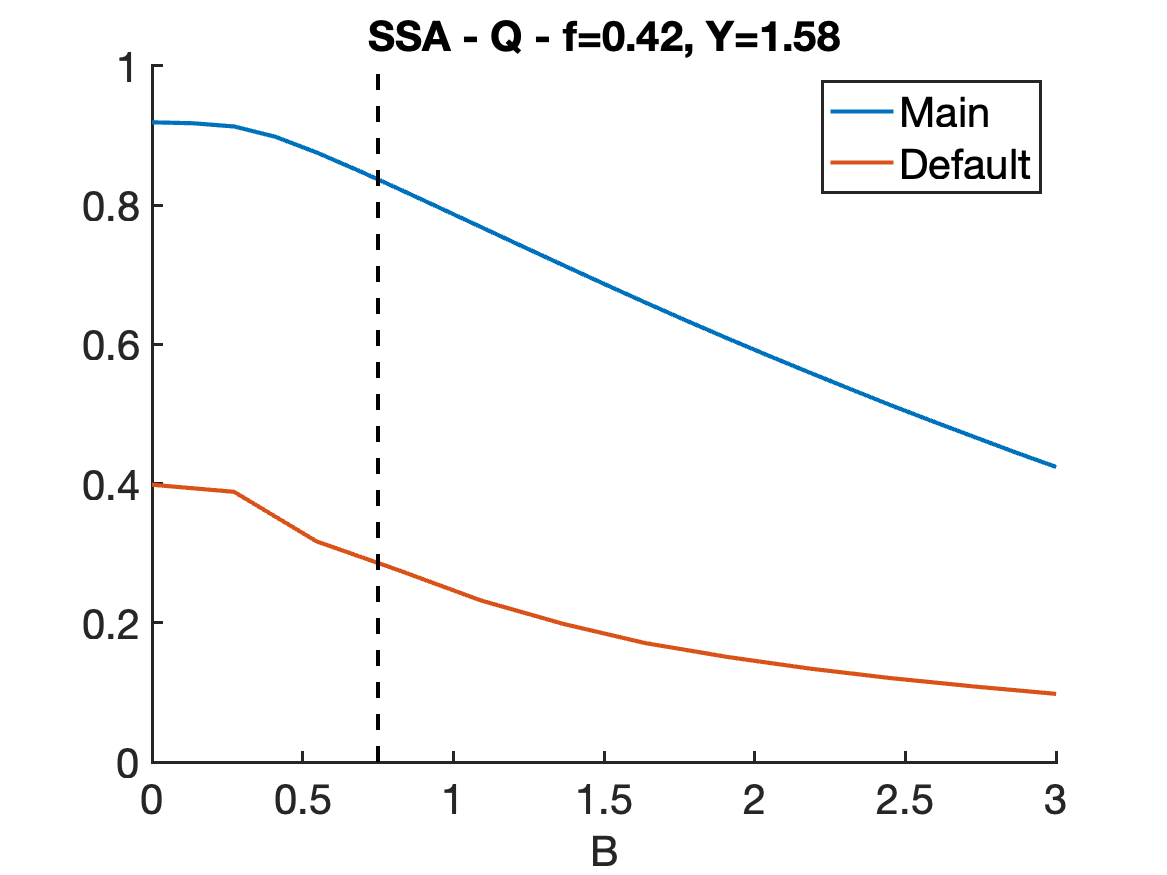}
 \includegraphics[width=0.3\textwidth]{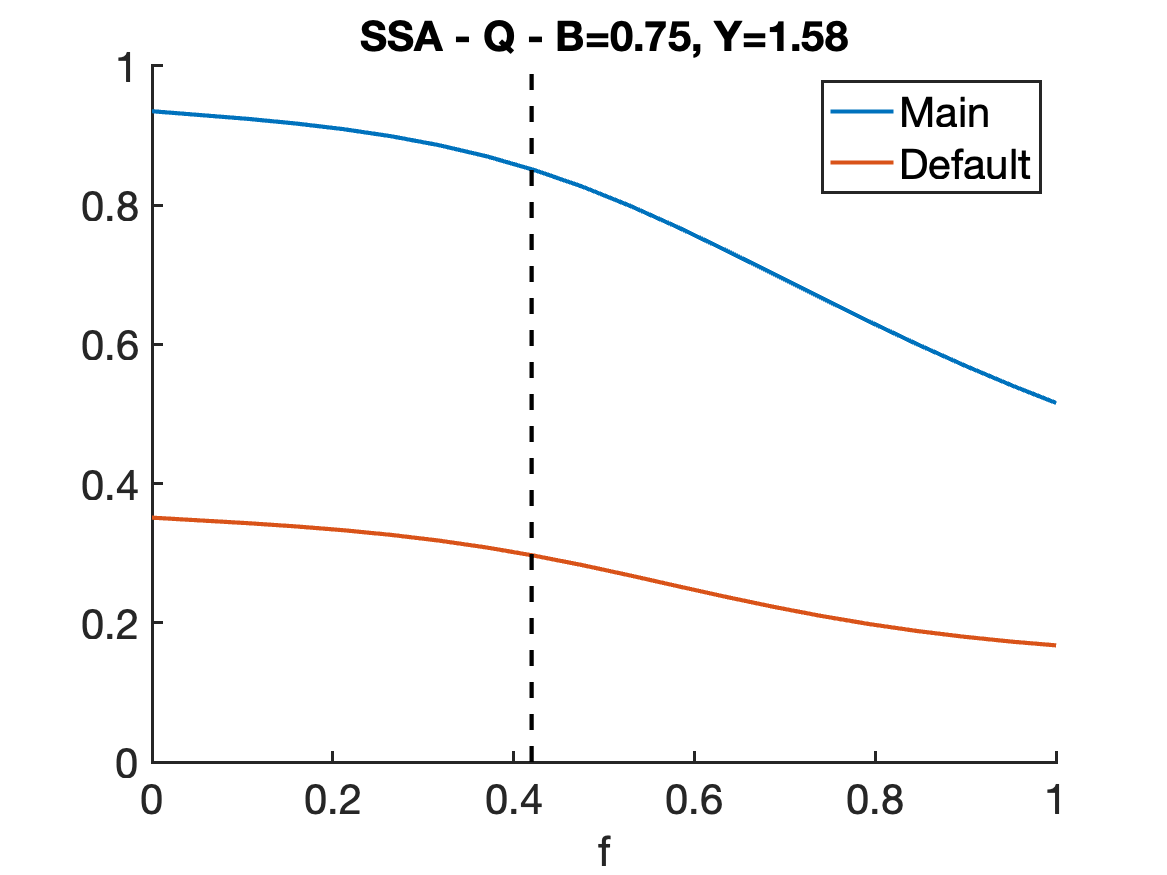}
 
 \includegraphics[width=0.3\textwidth]{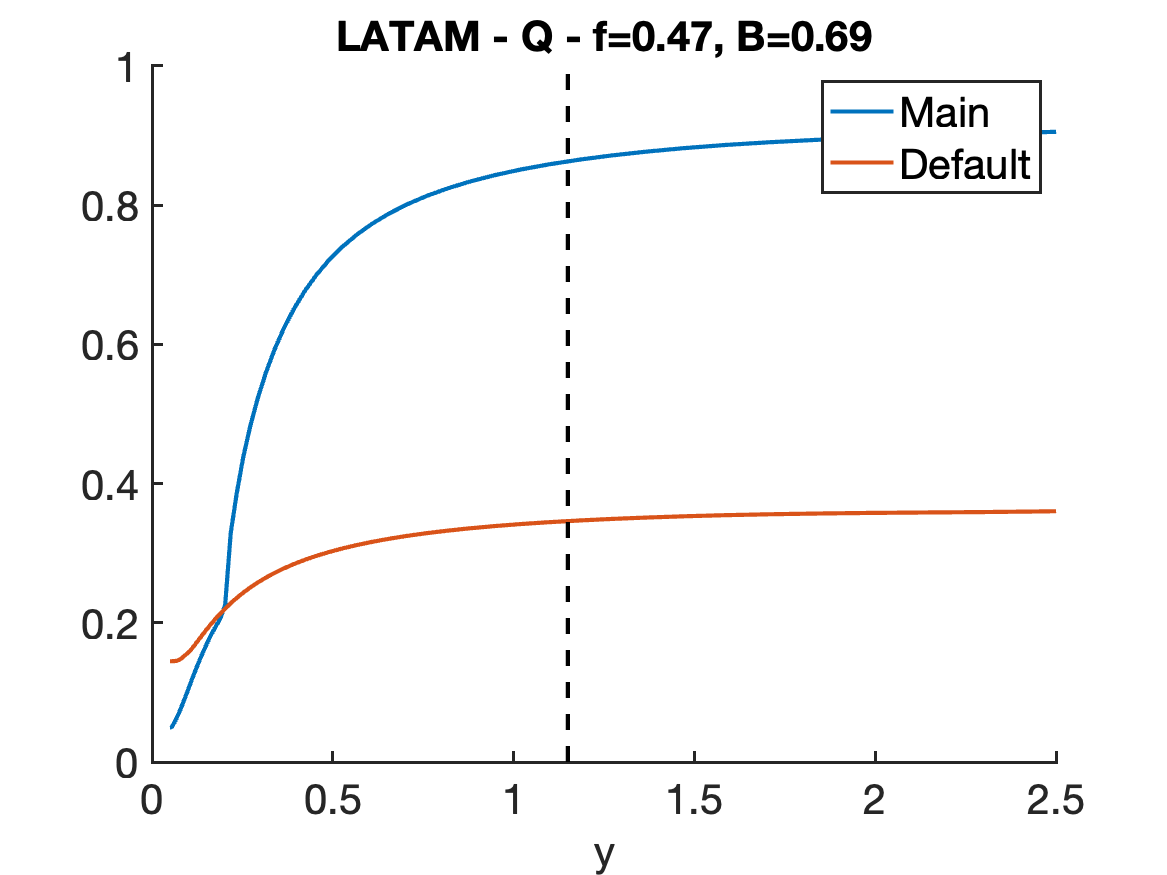}
 \includegraphics[width=0.3\textwidth]{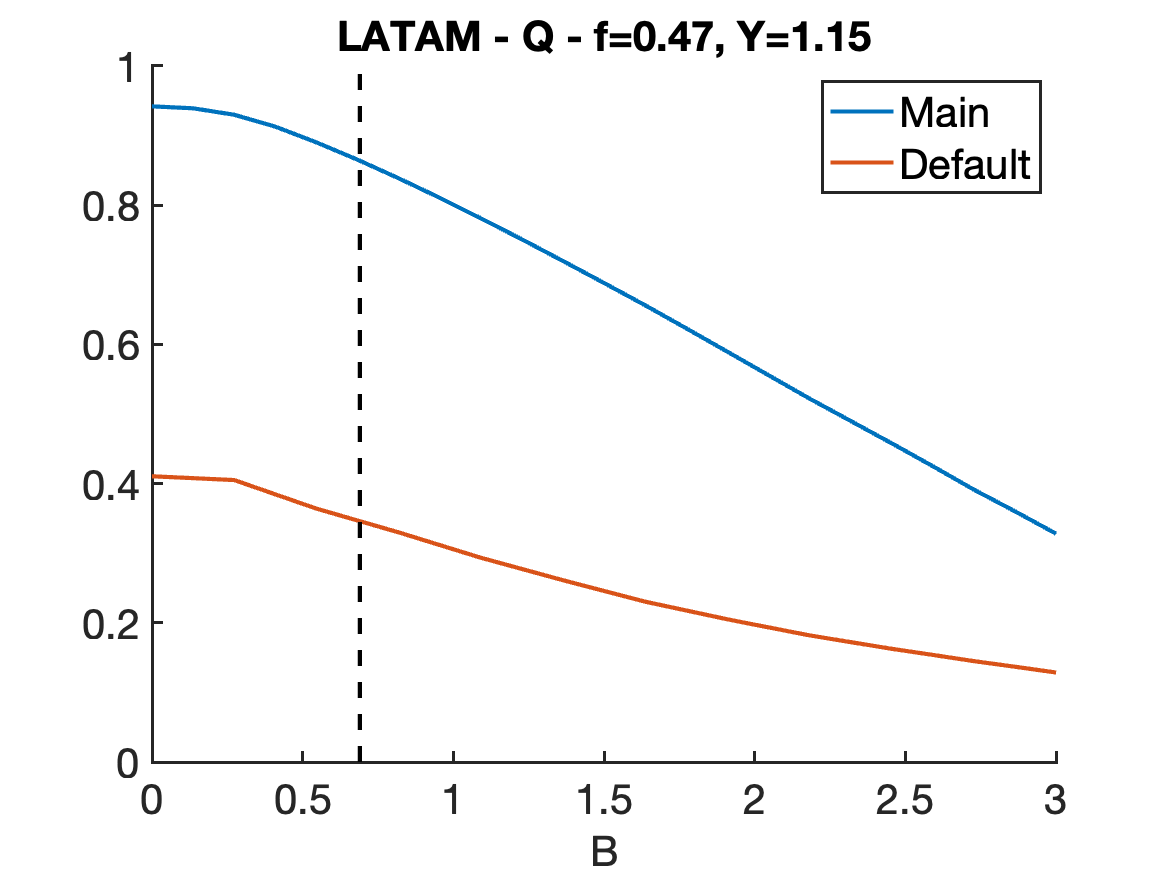}
 \includegraphics[width=0.3\textwidth]{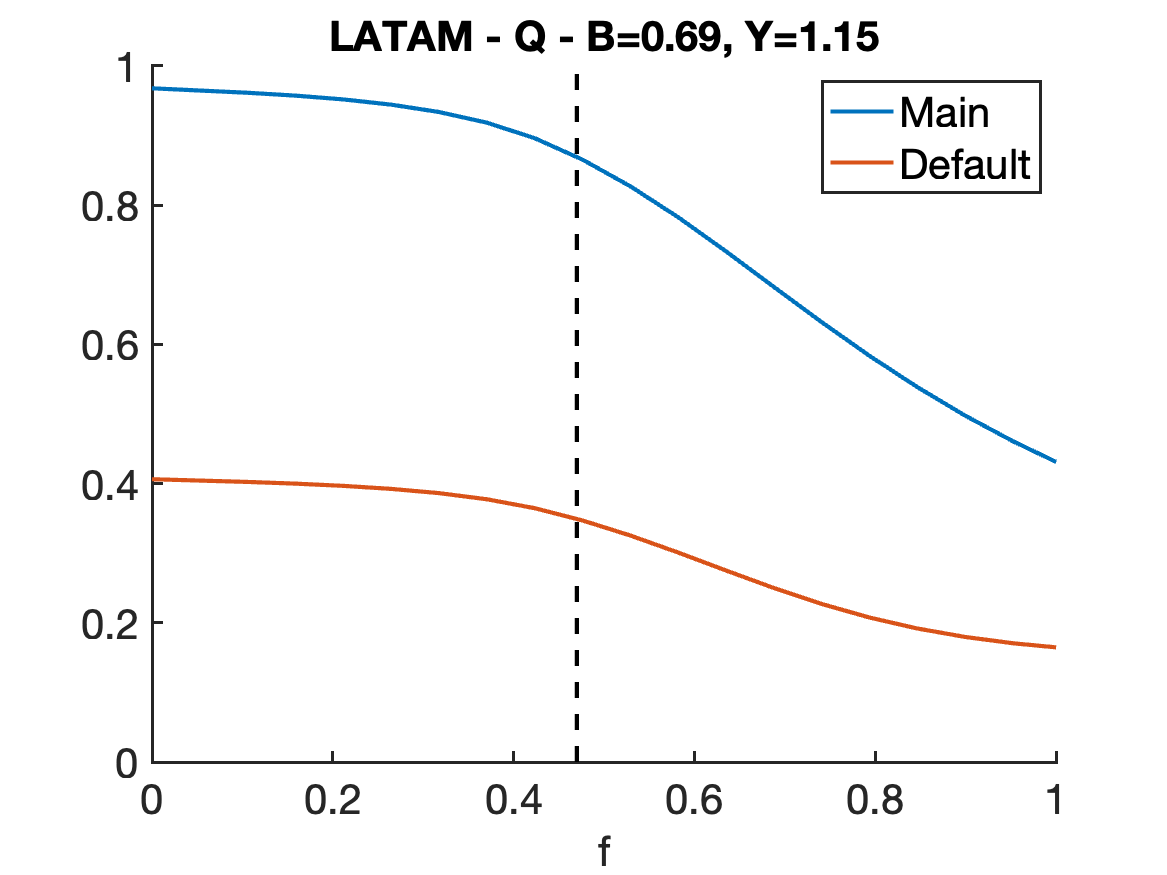}
 
 \includegraphics[width=0.3\textwidth]{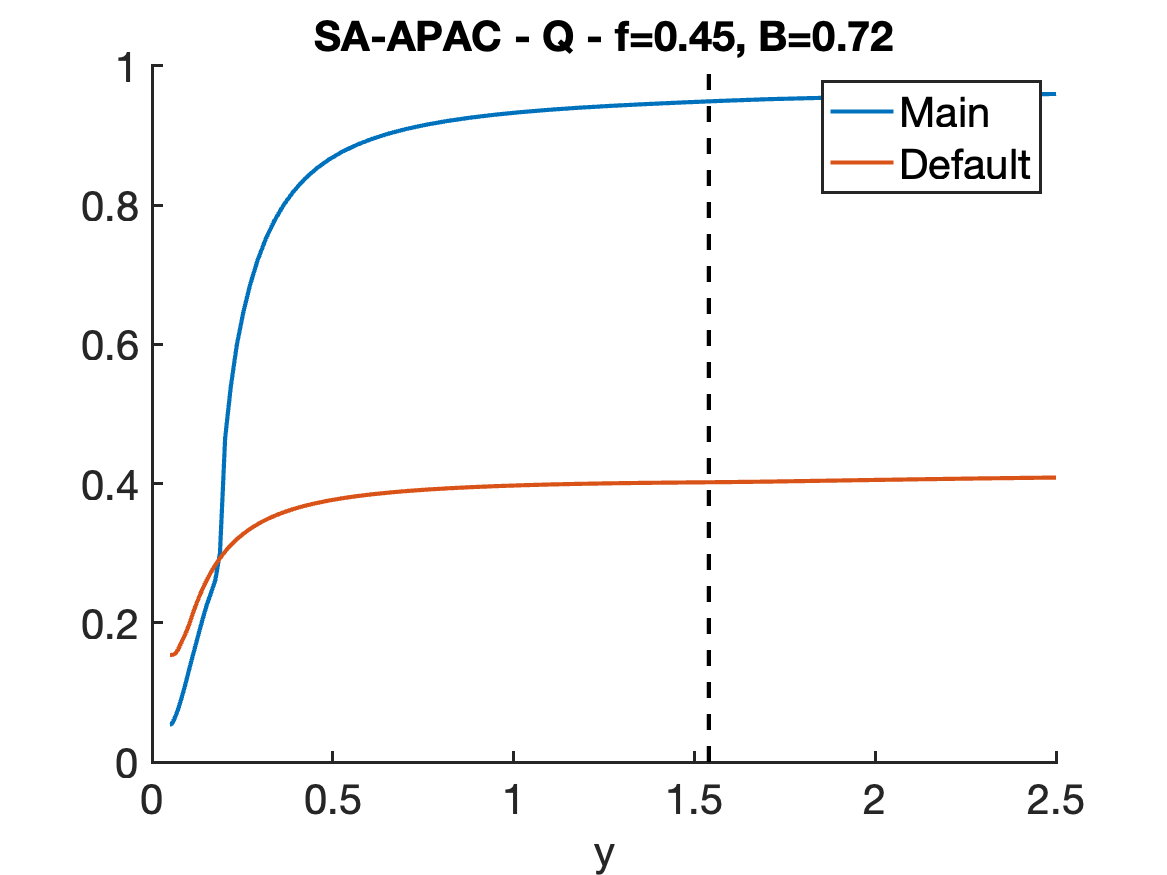}
 \includegraphics[width=0.3\textwidth]{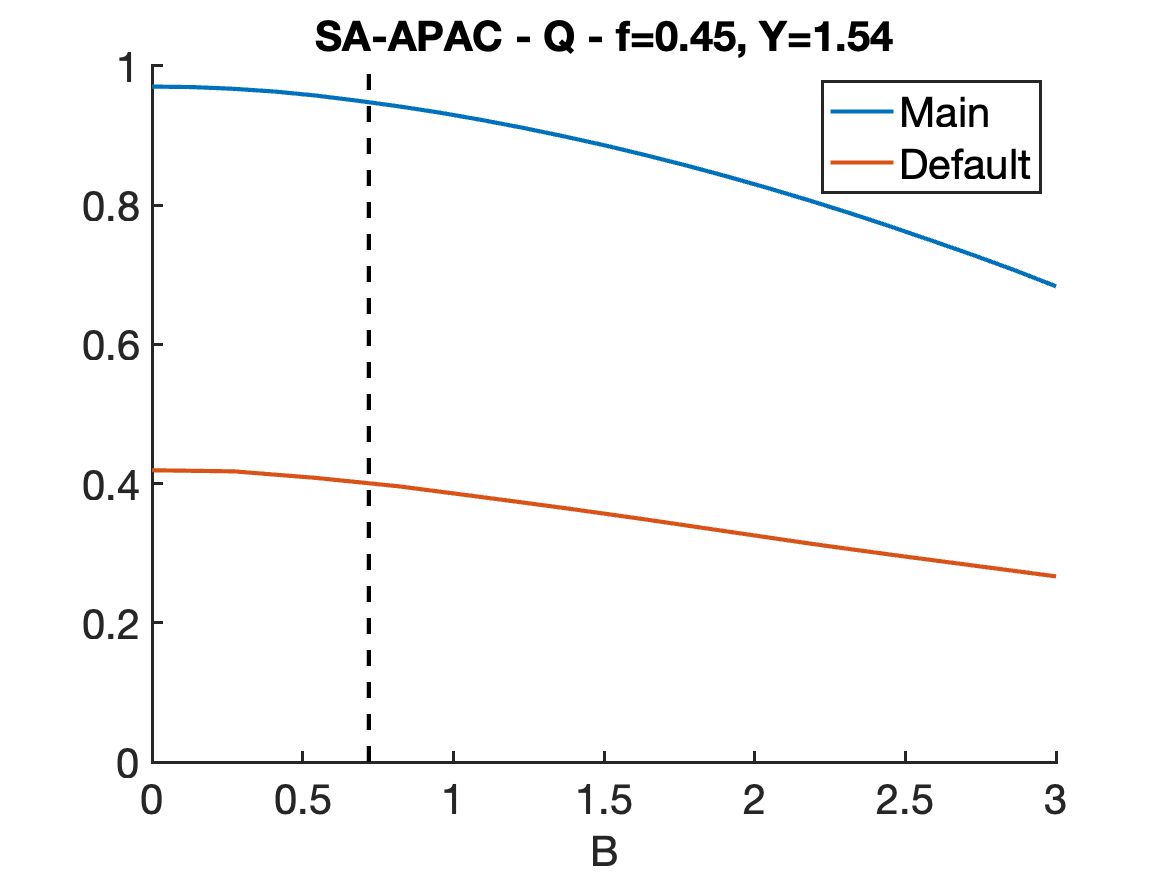}
 \includegraphics[width=0.3\textwidth]{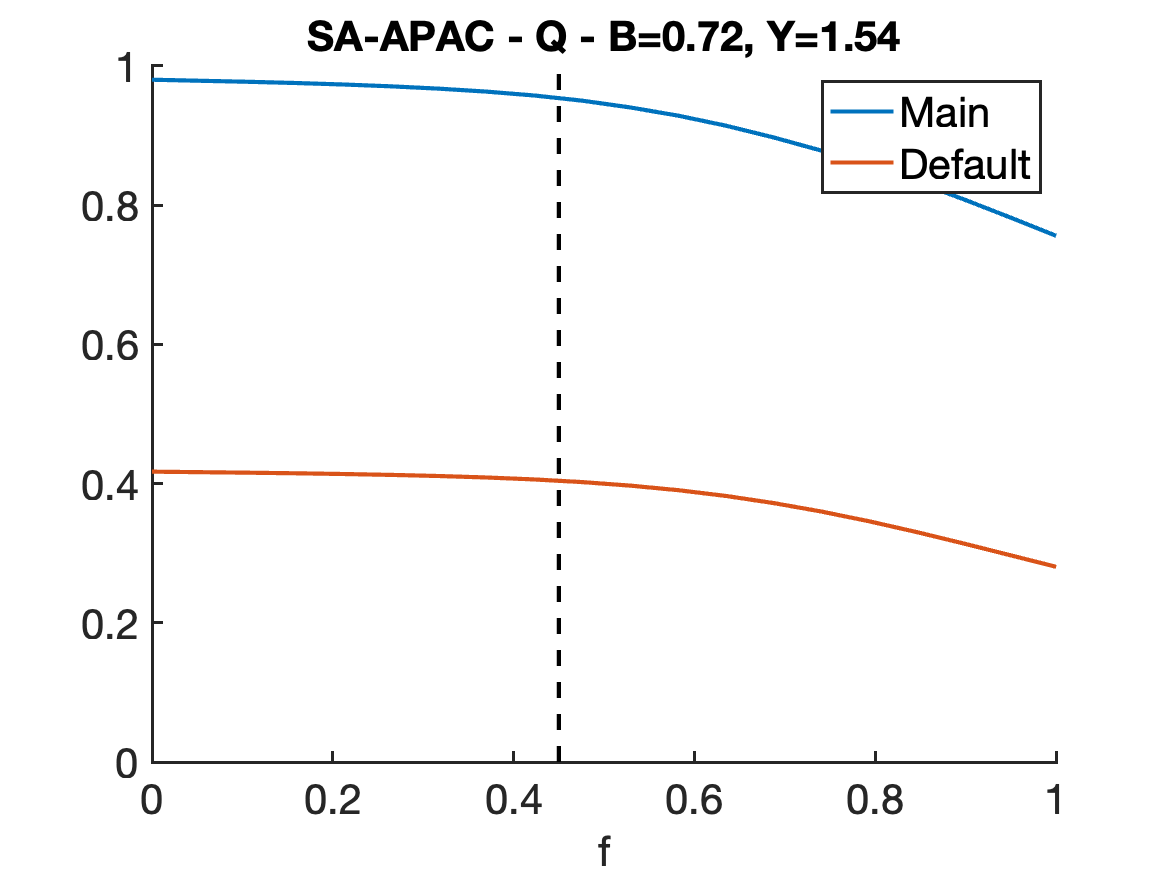}

\caption{$Q$ as a function of $y$ (first column), $B$ (second column) and $f$ (third column): SSA (first line), LATAM (second line) and SA-APAC (third line). On top of each picture we report the other two state variables centered on median values. The dotted line represents the median value.}\label{M1}
 \end{figure}

In Figure \ref{M1}, for each macro-region, we show the price of bonds $Q$ as a function of one of the three state variables ($y,B,f$), centering the other two at the median values. We consider both the default and no-default case. 
As expected, the price of bonds is increasing in the output, decreasing in public debt, and decreasing in pollution. The reverse is observed for the bond return. For the interval of values of $f$ observed in our analysis  ($0.21-0.70$ in the SSA, $0.32-0.52$ in the LATAM and $0.29-0.61$ in the SA-APAC macro-region) we observe a limited sensitivity of bond prices with respect to the pollution. Increasing the level of pollution in the three macro regions of 10\% (leaving output and debt at the median level) we go from $0.0556$ to $0.0580$ in the SA-APAC macro-region, from $0.1004$ to $0.1099$ in the SSA macro-region and from $0.0849$ to $0.0984$ in the LATAM region. The estimates show a rather limited effect of pollution on government bond spreads.

\begin{figure}[tp]
\centering
 \includegraphics[width=0.3\textwidth]{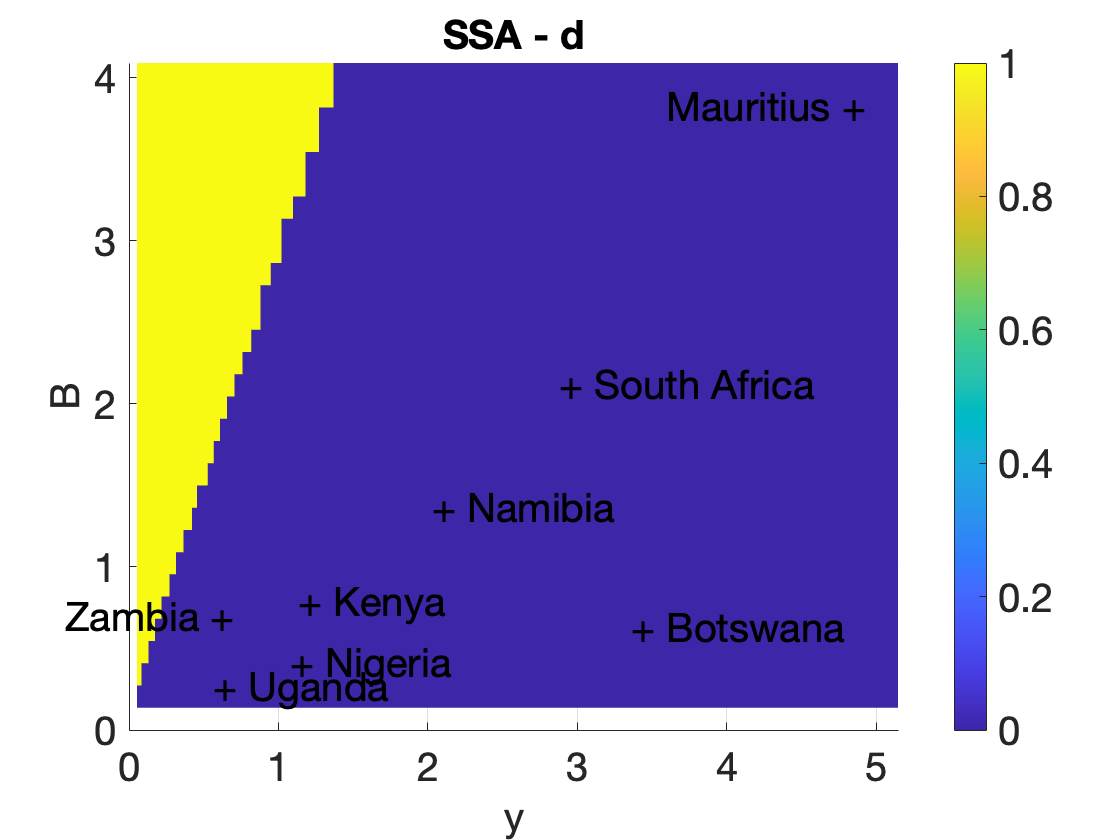}
 \includegraphics[width=0.3\textwidth]{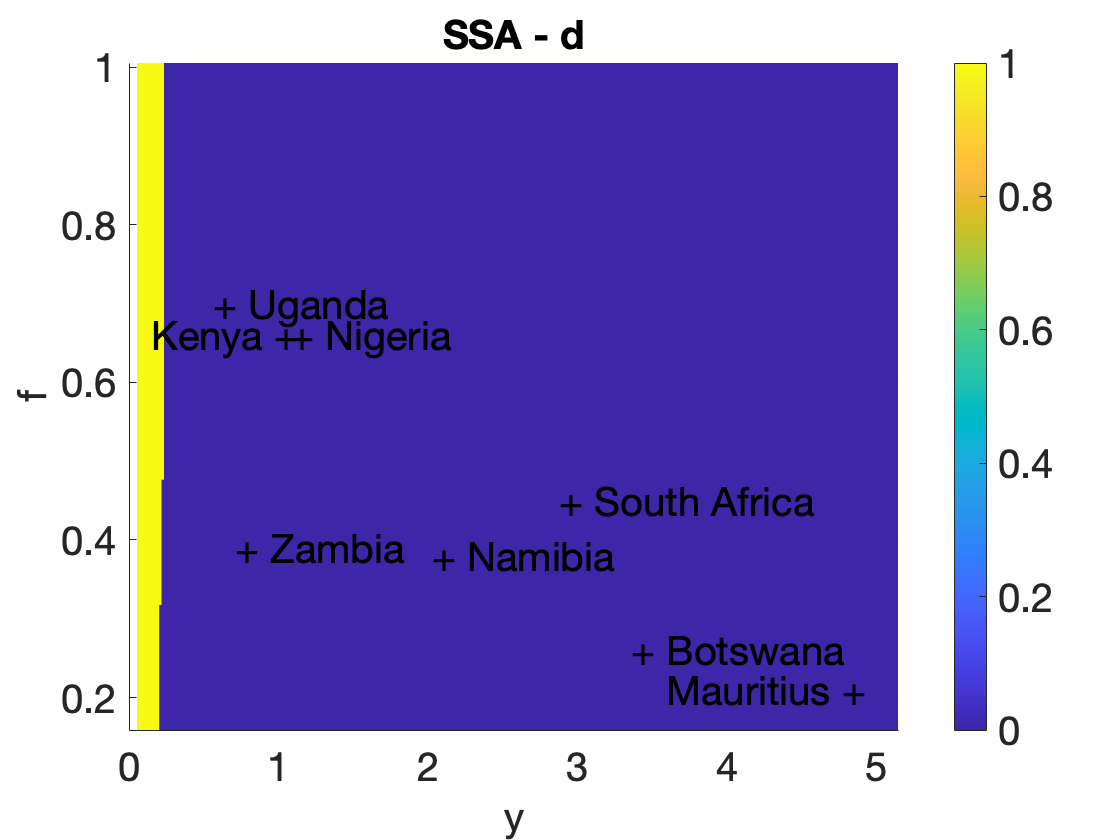}
 \includegraphics[width=0.3\textwidth]{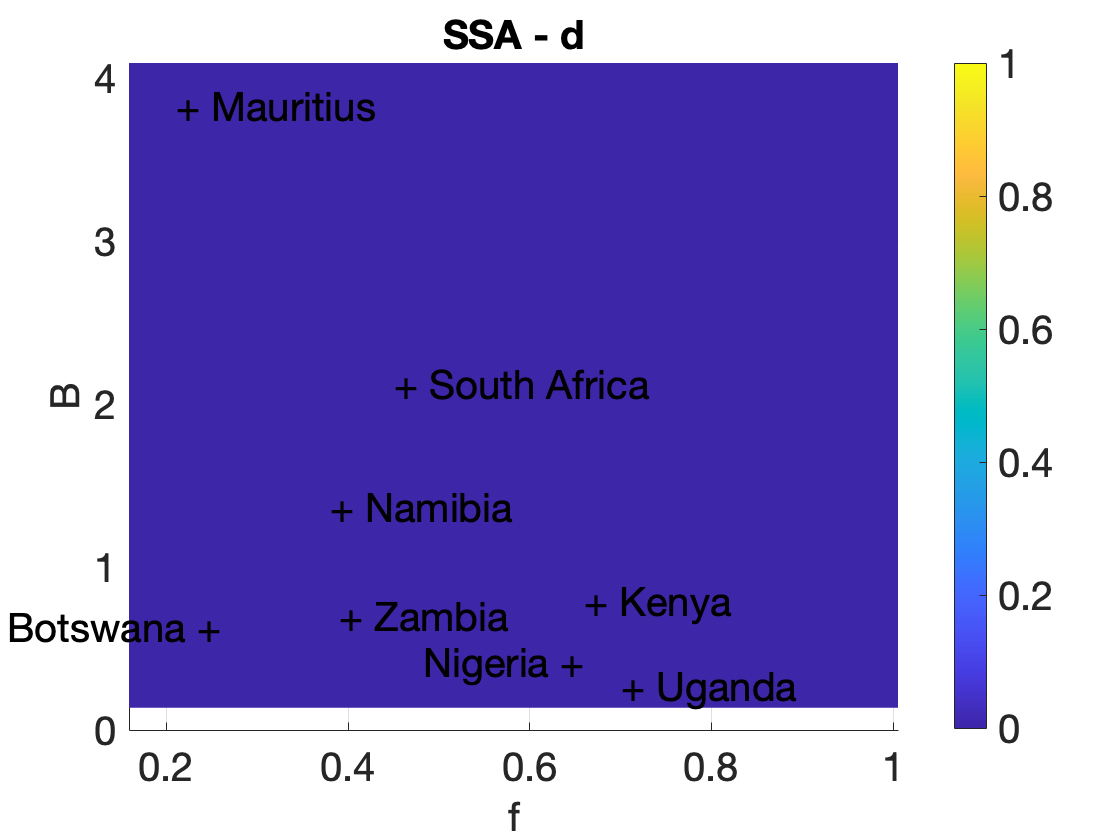}
 
 \includegraphics[width=0.3\textwidth]{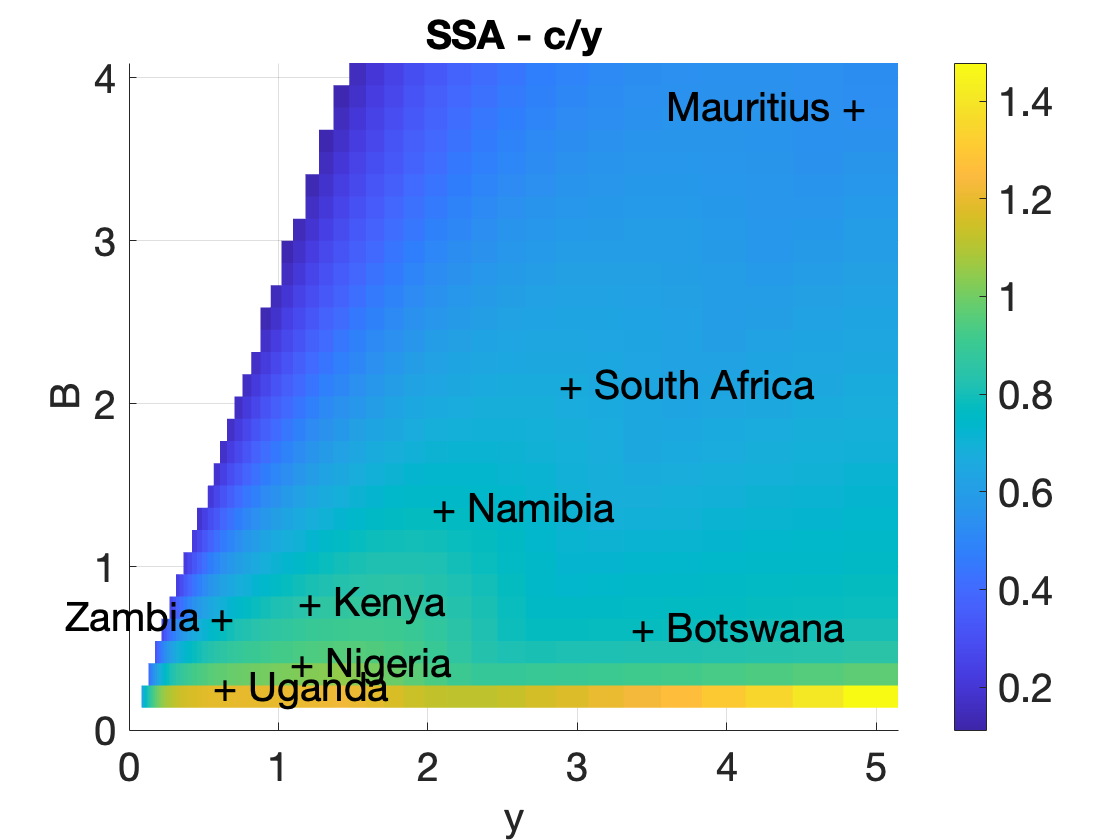}
 \includegraphics[width=0.3\textwidth]{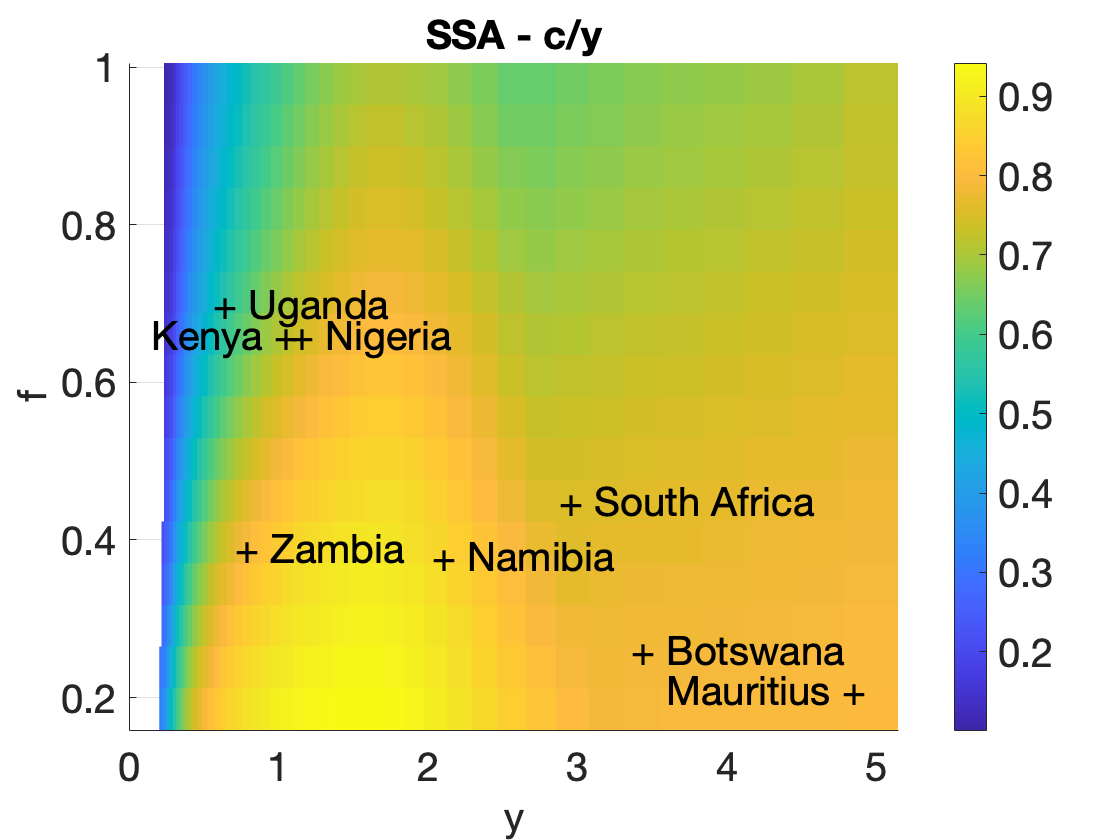}
 \includegraphics[width=0.3\textwidth]{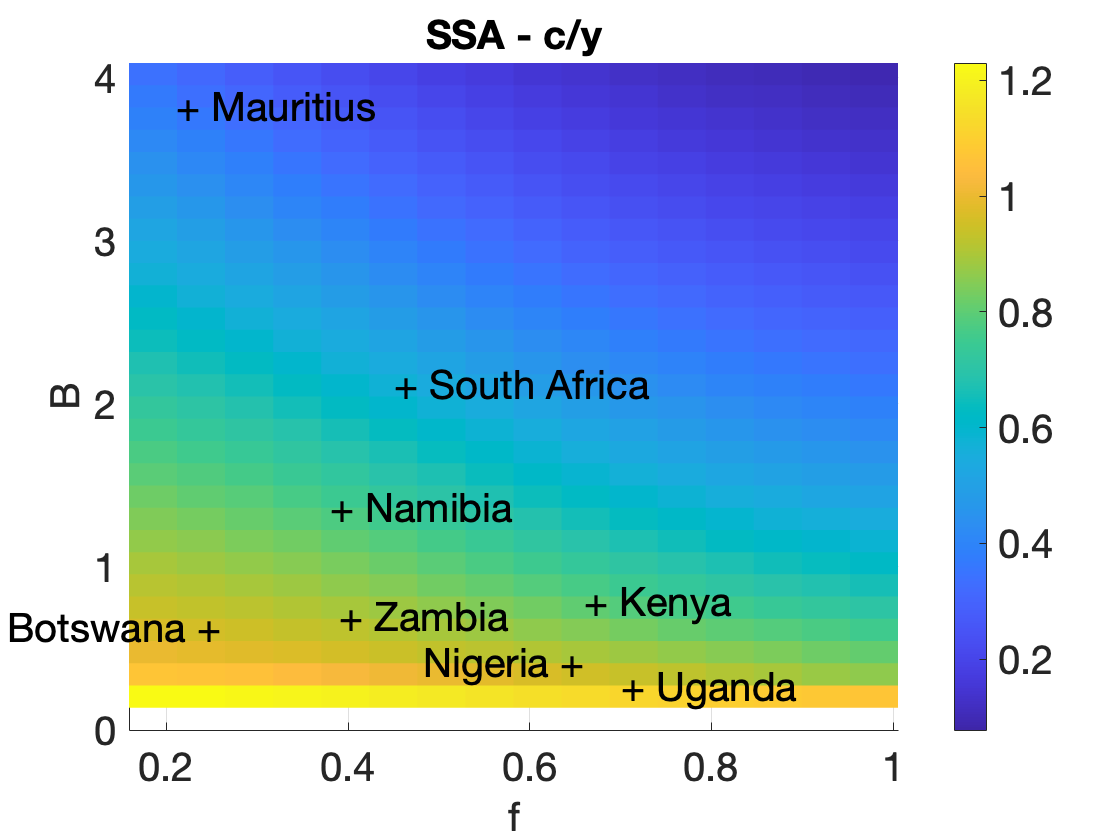}
 
 \includegraphics[width=0.3\textwidth]{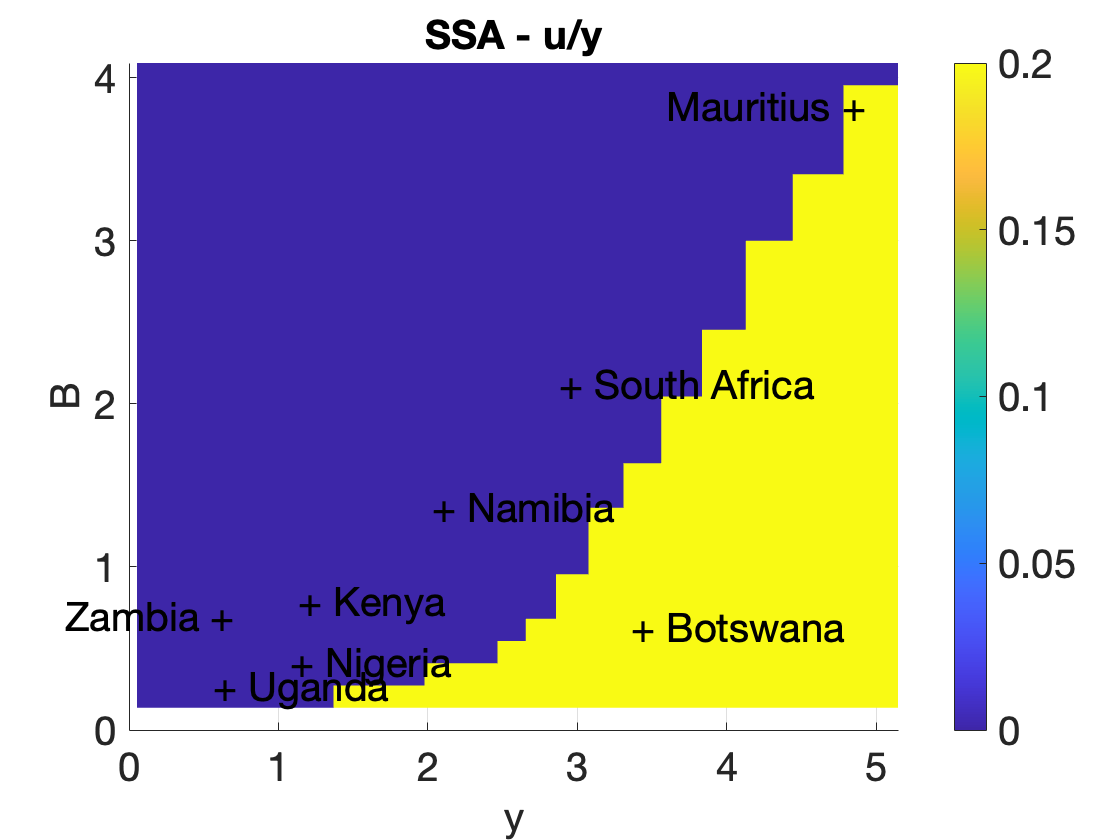}
 \includegraphics[width=0.3\textwidth]{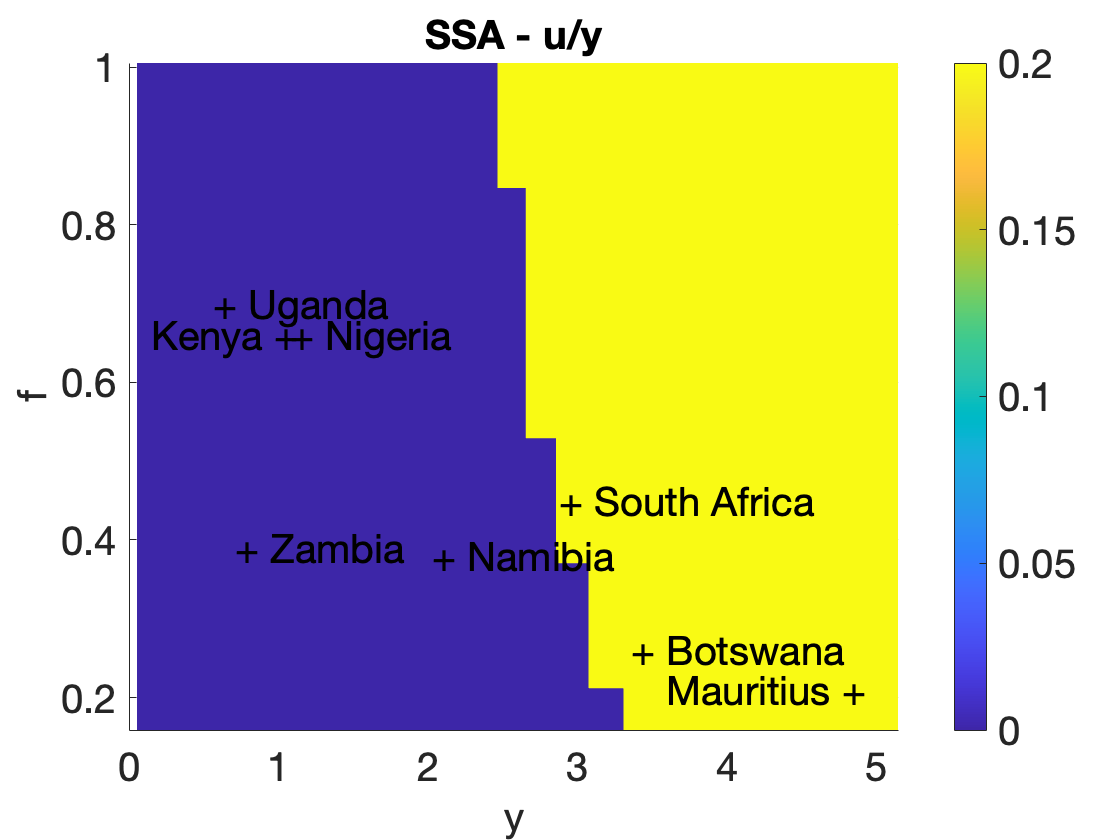}
 \includegraphics[width=0.3\textwidth]{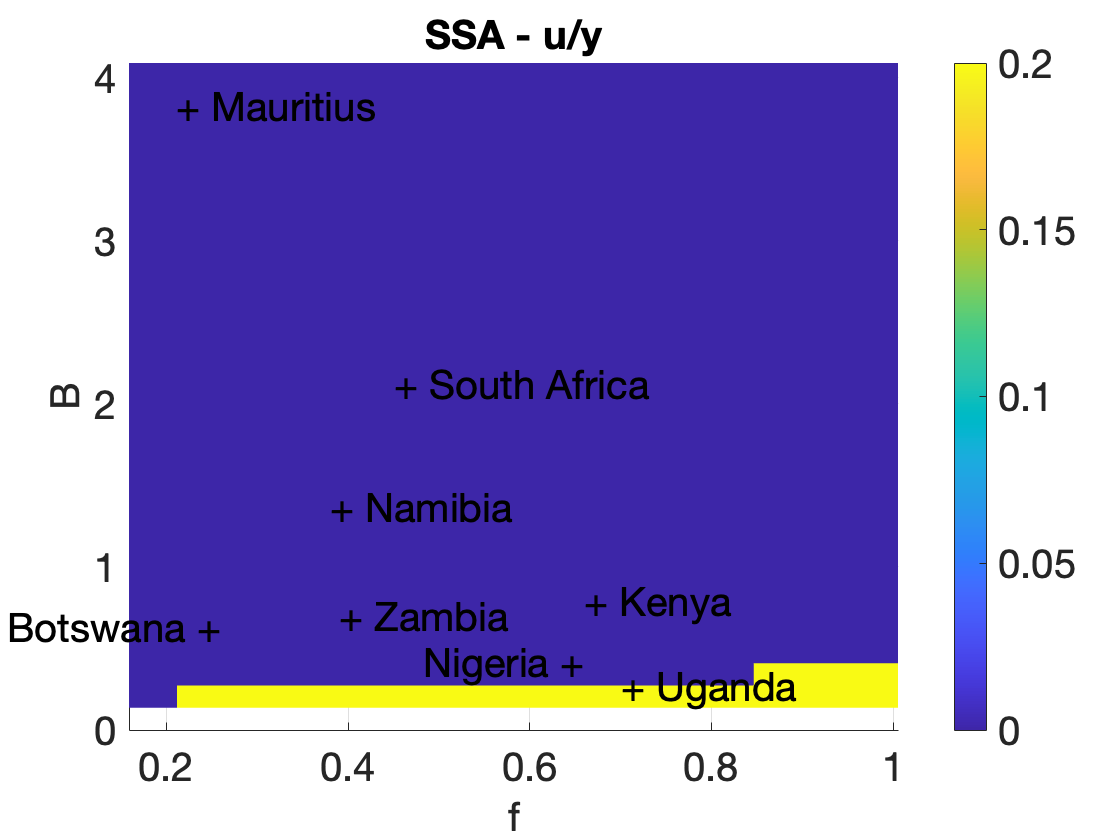}

\caption{SSA region: 
decision to default ($d$), consumption over output ($c/y$) adaptation expenditure over output ($u/y$).
When a variable is fixed, it is set equal to the median value for the macro-region, see Table \ref{Q_SSA}: plane $(y,B)$, $f=0.42$; plane $(y,f)$, $B=0.75$; plane $(f,B)$, $y=1.58$. The white region in the picture for $c/y$ corresponds to the default region. The positioning of the country labels in the pictures is purely indicative because, for example, in the $(y, B)$ plane, the values of $d$, $c/y$, and $u/y$ correspond to the median value of $f$ (specifically, $f = 0.42$).}\label{M2-SSA}
 \end{figure}

\begin{figure}[tp]
\centering
 \includegraphics[width=0.3\textwidth]{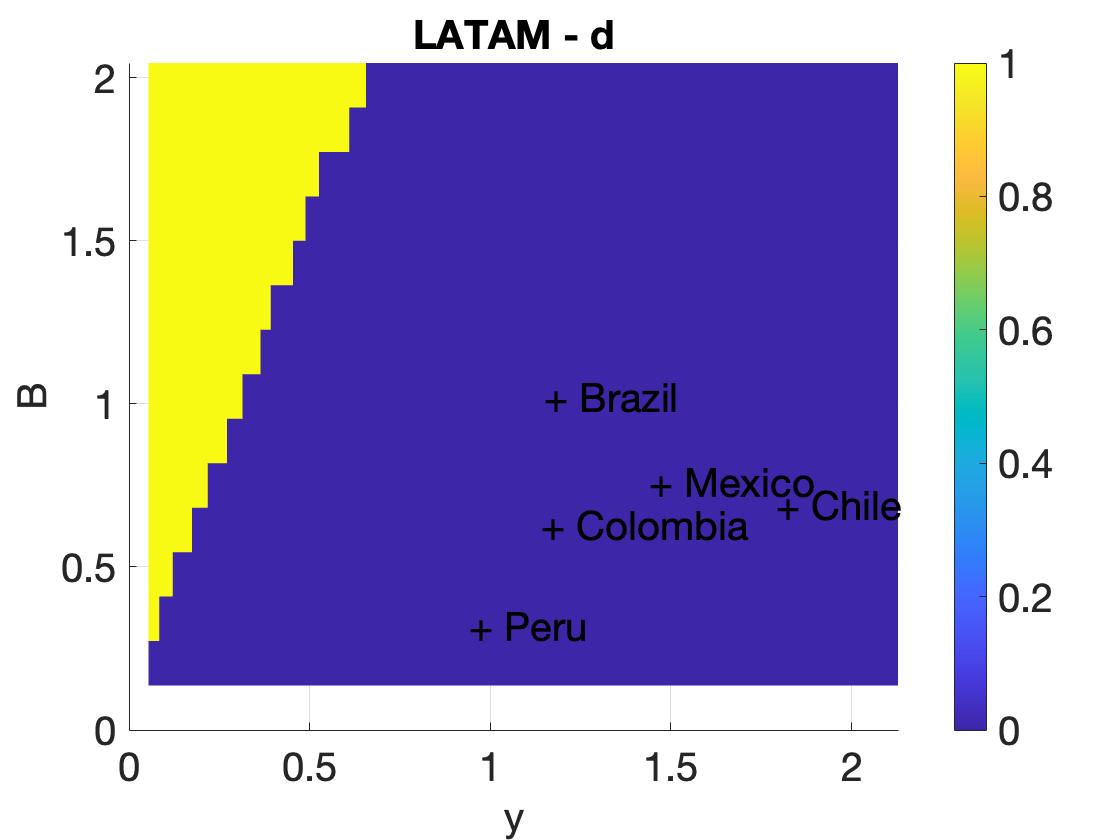}
 \includegraphics[width=0.3\textwidth]{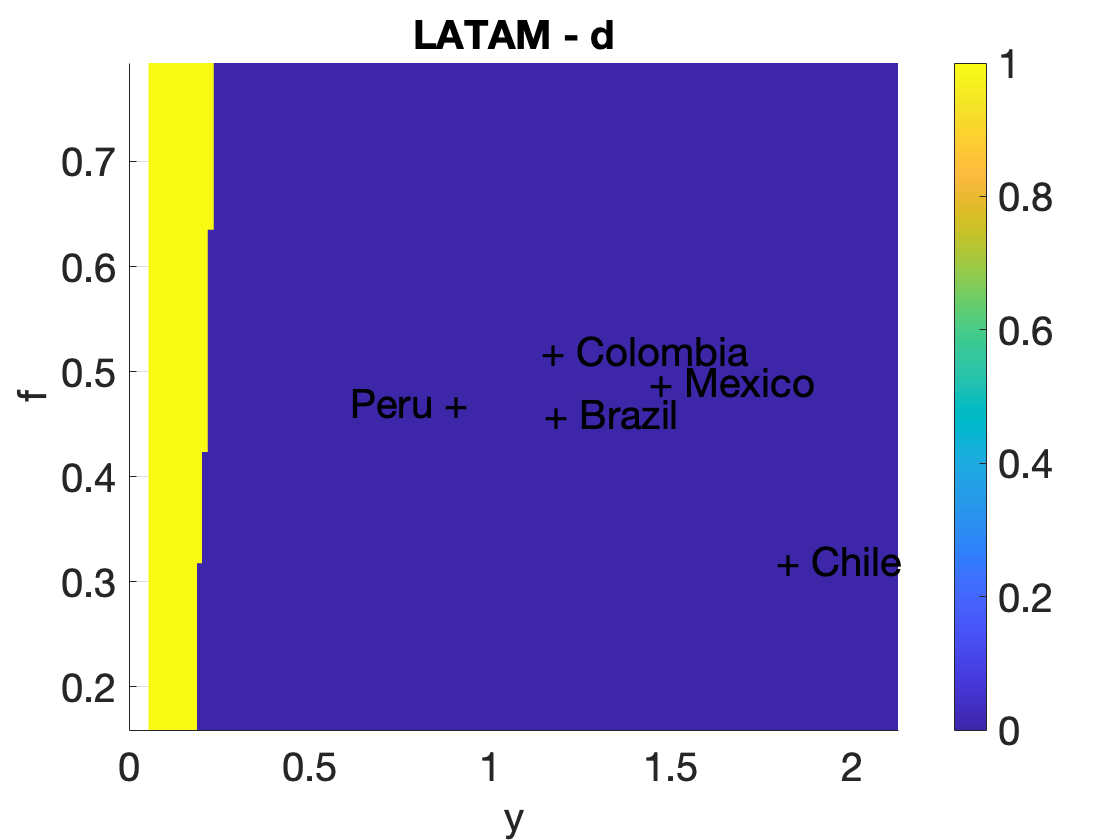}
 \includegraphics[width=0.3\textwidth]{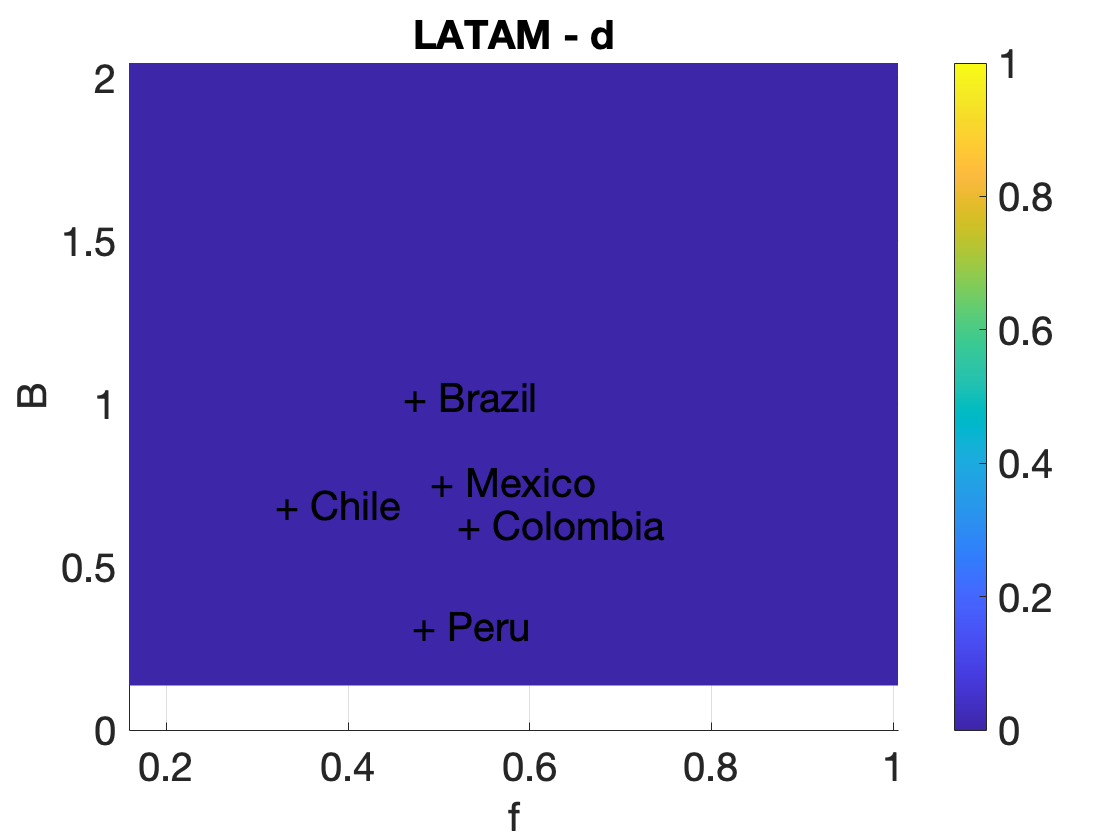}
 
 \includegraphics[width=0.3\textwidth]{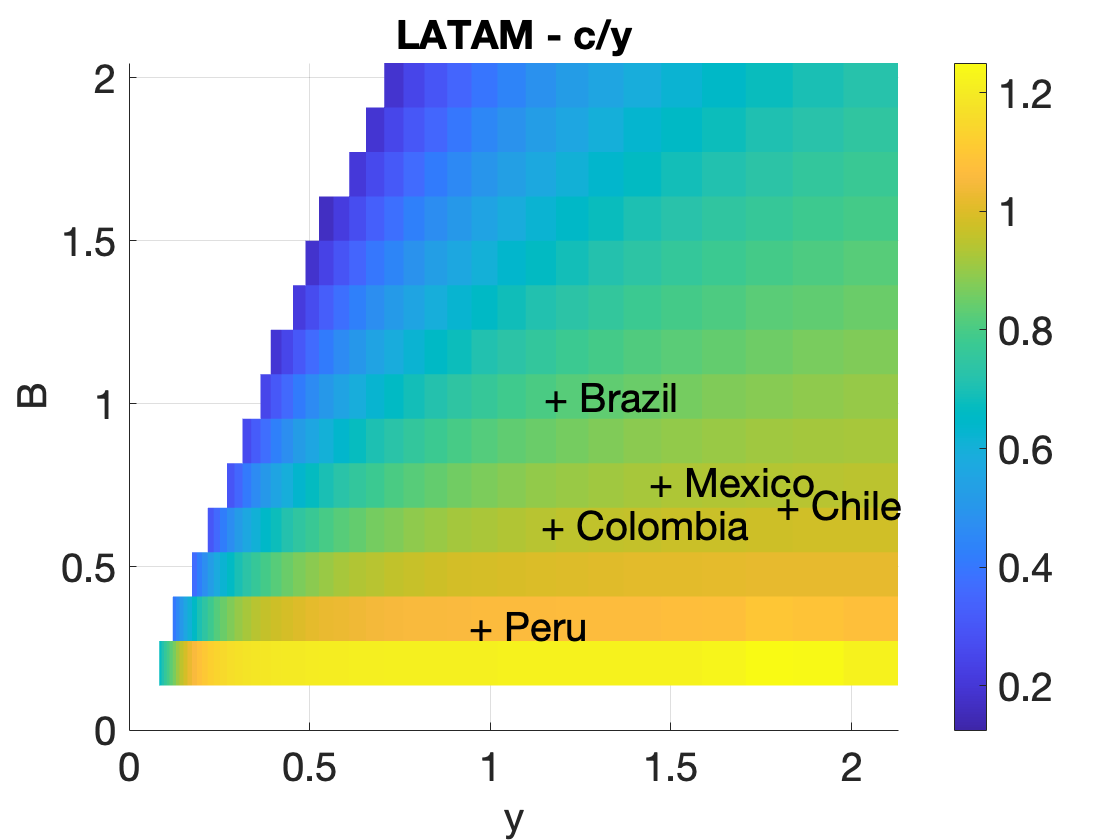}
 \includegraphics[width=0.3\textwidth]{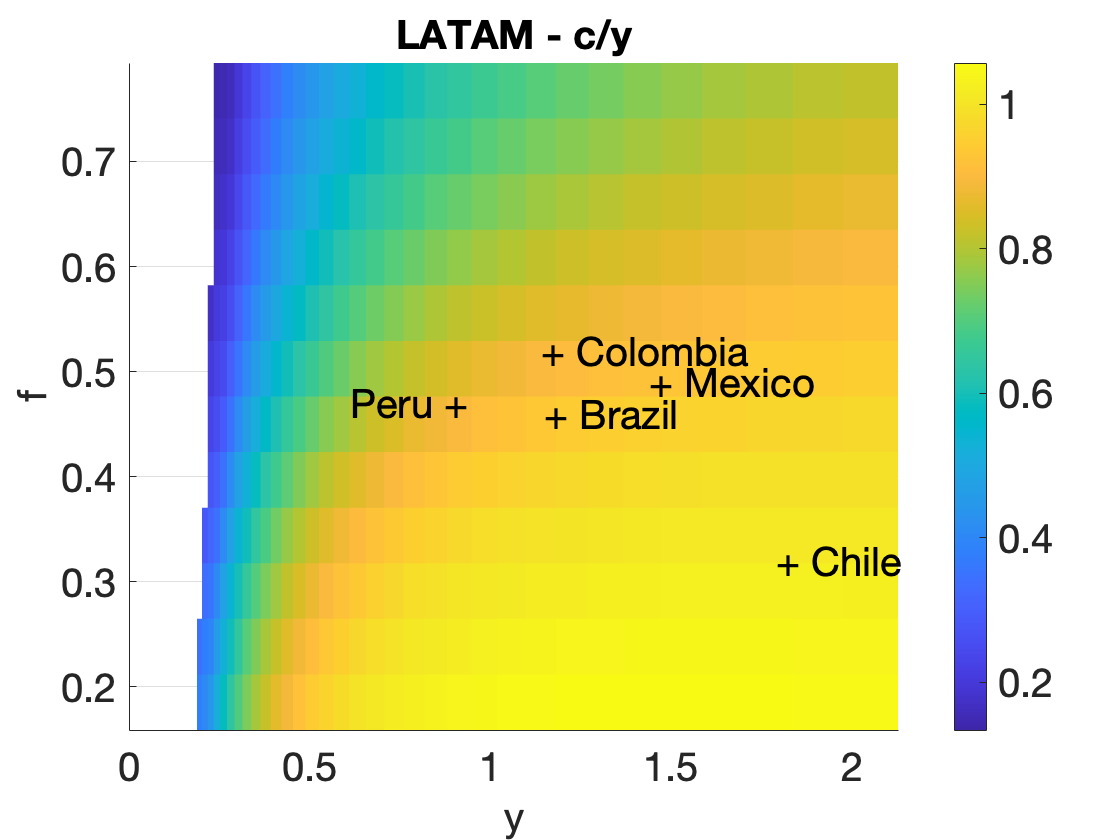}
 \includegraphics[width=0.3\textwidth]{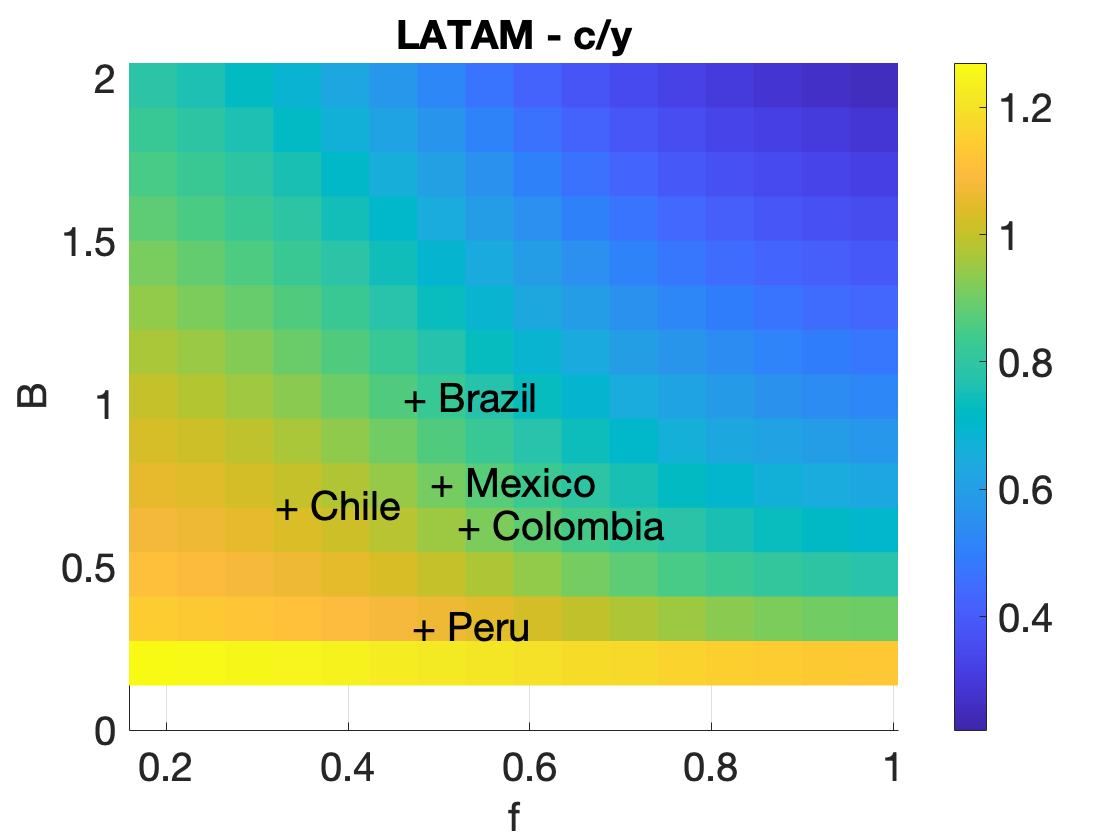}
 
 \includegraphics[width=0.3\textwidth]{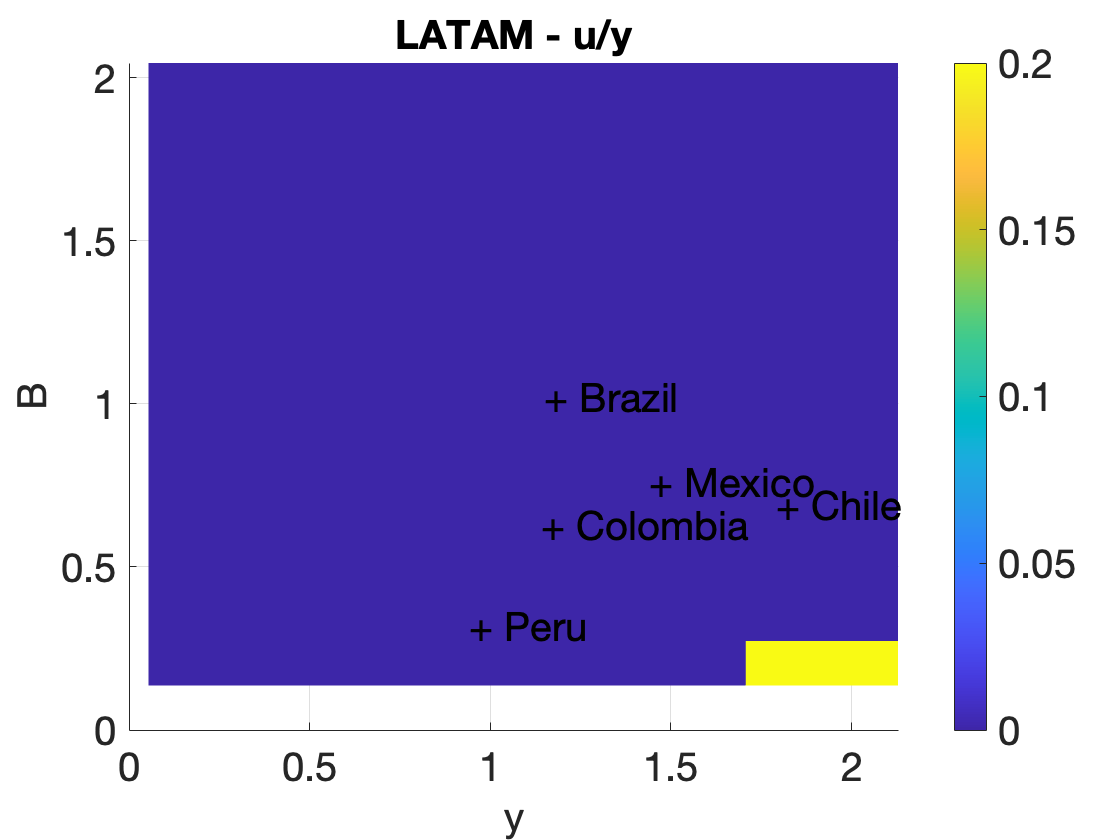}
 \includegraphics[width=0.3\textwidth]{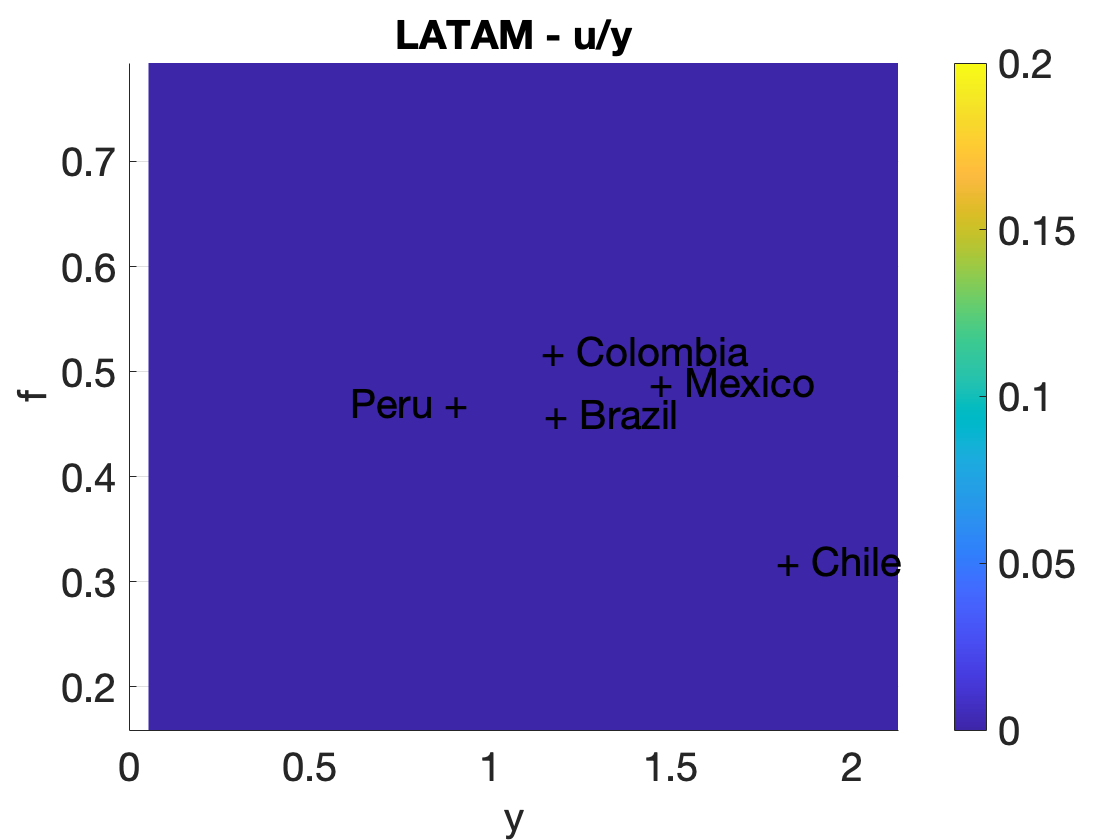}
 \includegraphics[width=0.3\textwidth]{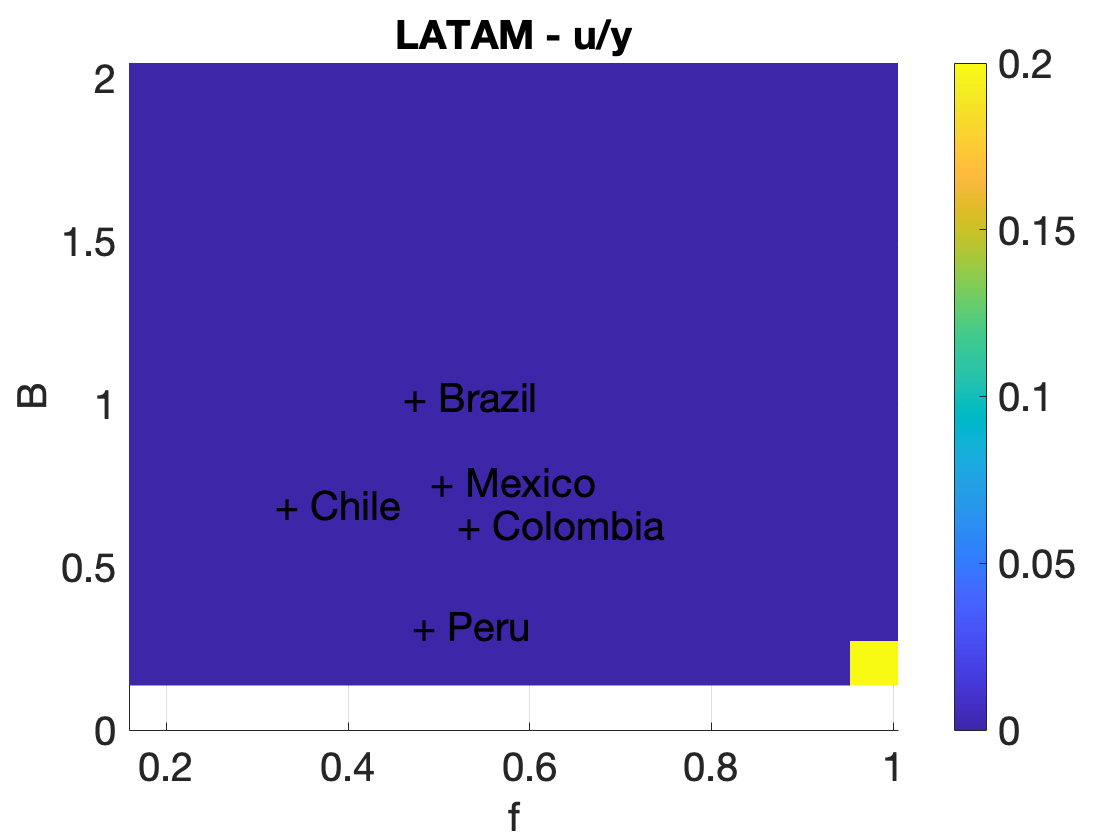}

\caption{LATAM region: 
 decision to default ($d$), consumption over output ($c/y$) adaptation expenditure over output ($u/y$).
When a variable is fixed, it is set equal to the median value for the macro-region, see Table \ref{Q_LATAM}: plane $(y,B)$, 
$f=0.47$; plane $(y,f)$, $B=0.69$; plane 
$(f,B)$, $y=1.15$. The white region in the picture for $c/y$ corresponds to the default region. The positioning of the country labels in the pictures is purely indicative because, for example, in the $(y, B)$ plane, the values of $d$, $c/y$, and $u/y$ correspond to the median value of $f$ (specifically, $f = 0.47$). }\label{M2-LATAM}
 \end{figure}

\begin{figure}[tp]
\centering
 \includegraphics[width=0.3\textwidth]{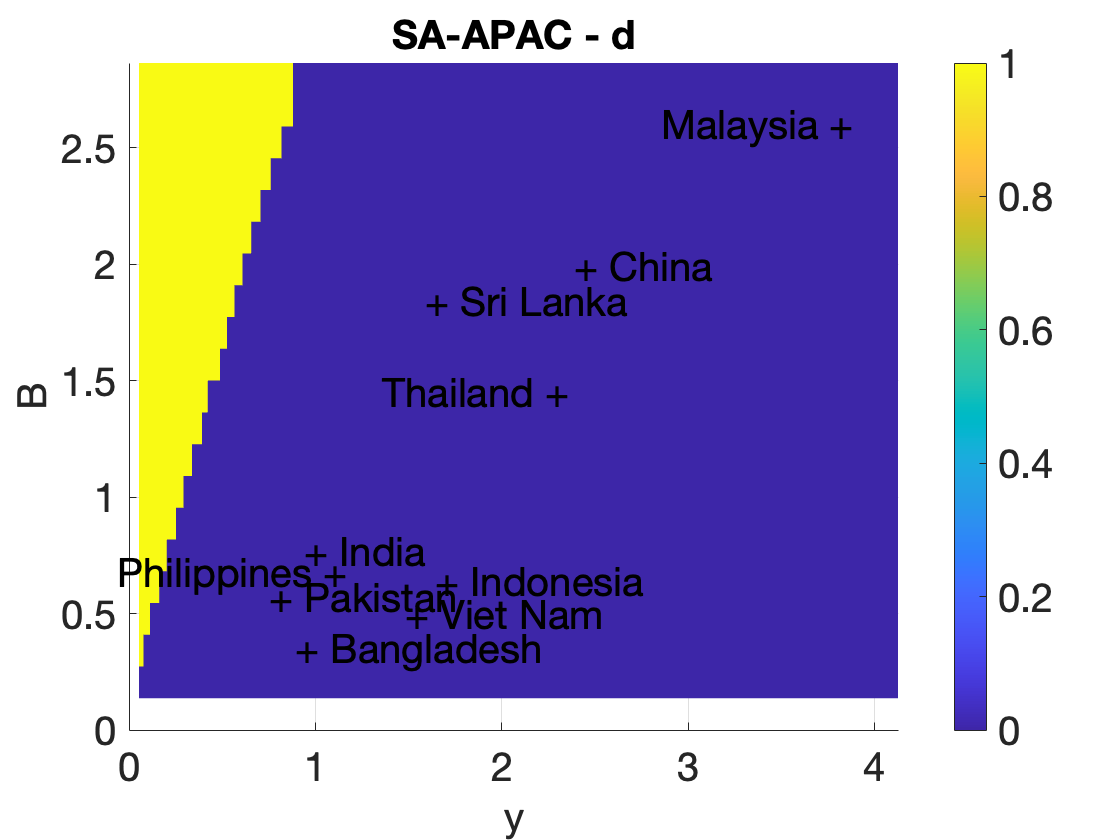}
 \includegraphics[width=0.3\textwidth]{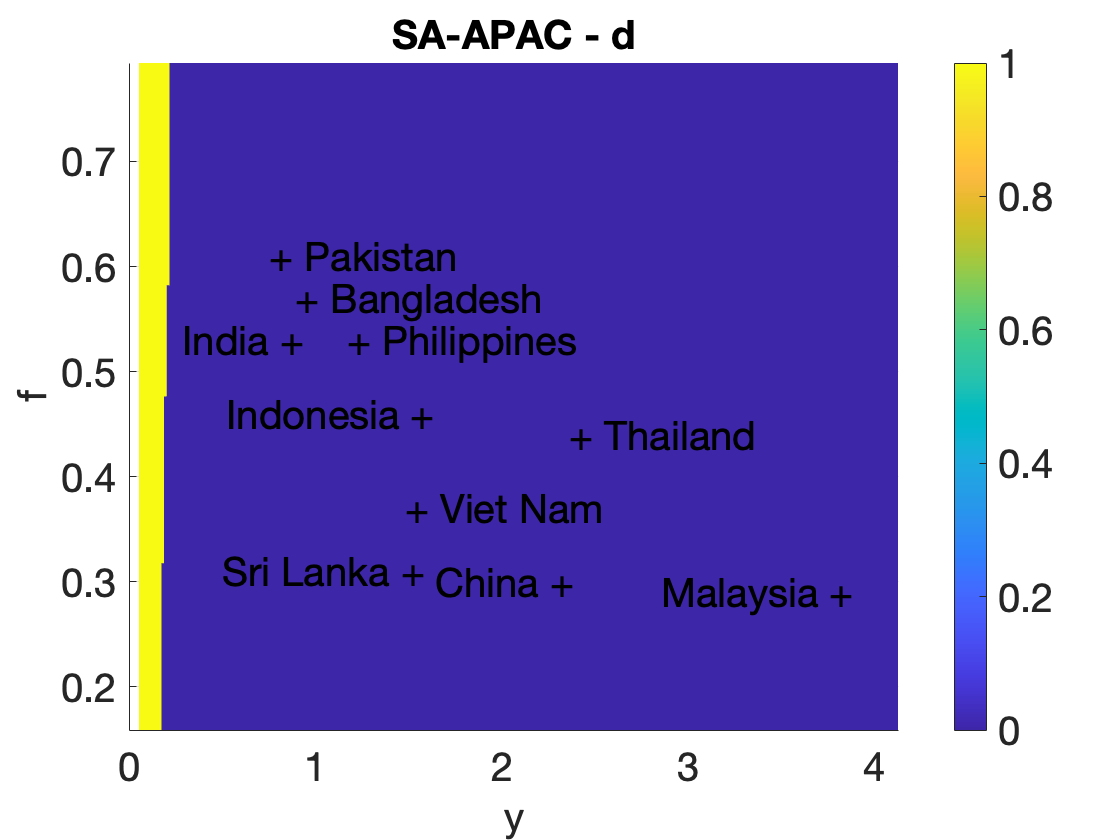}
 \includegraphics[width=0.3\textwidth]{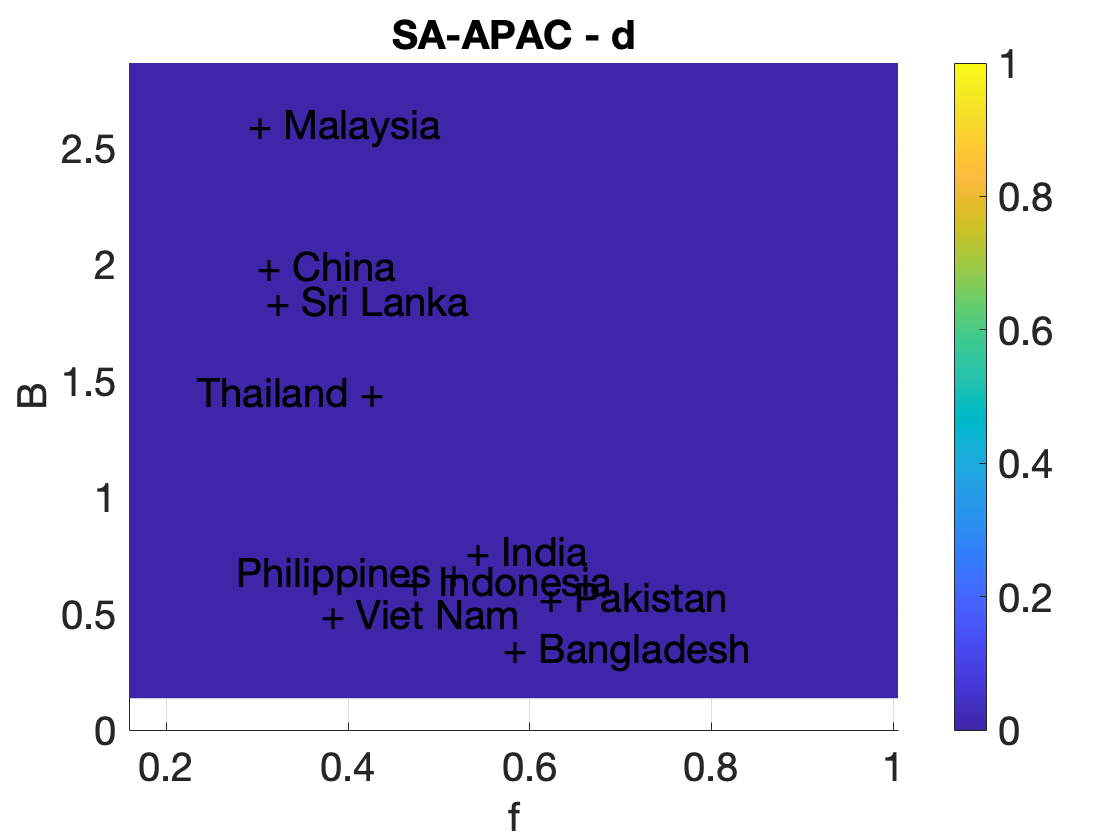}
 
 \includegraphics[width=0.3\textwidth]{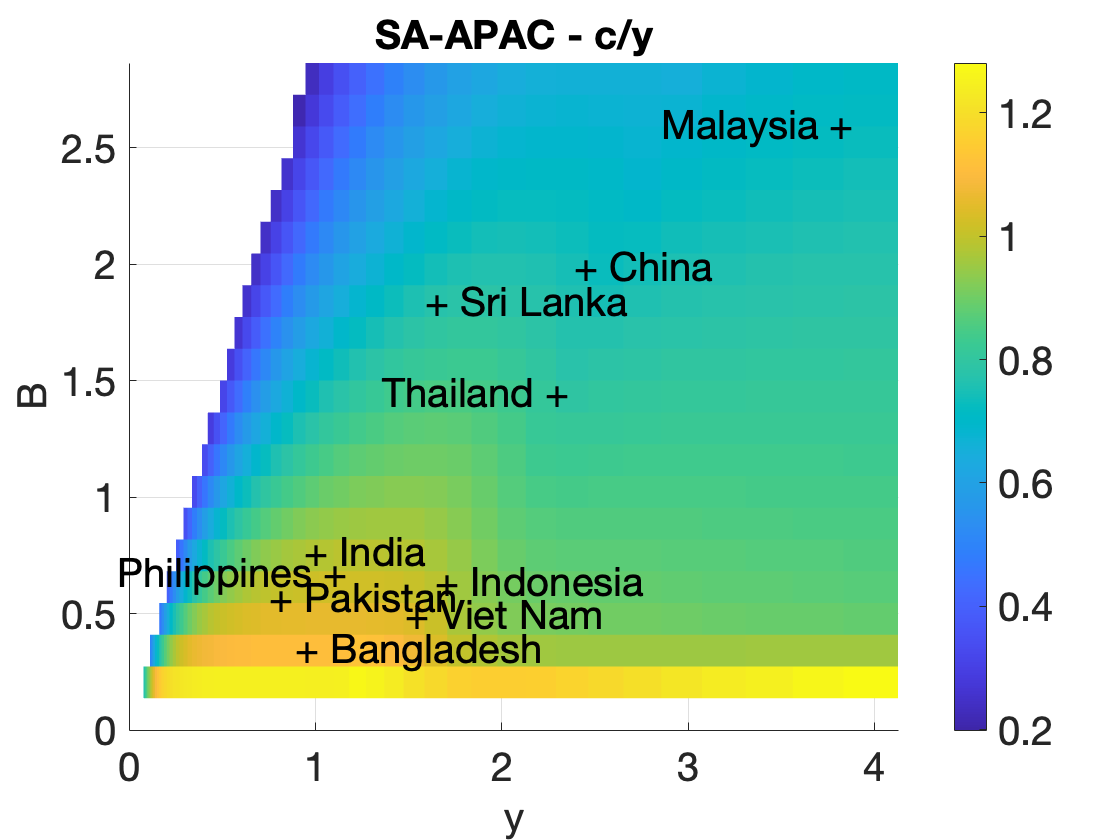}
 \includegraphics[width=0.3\textwidth]{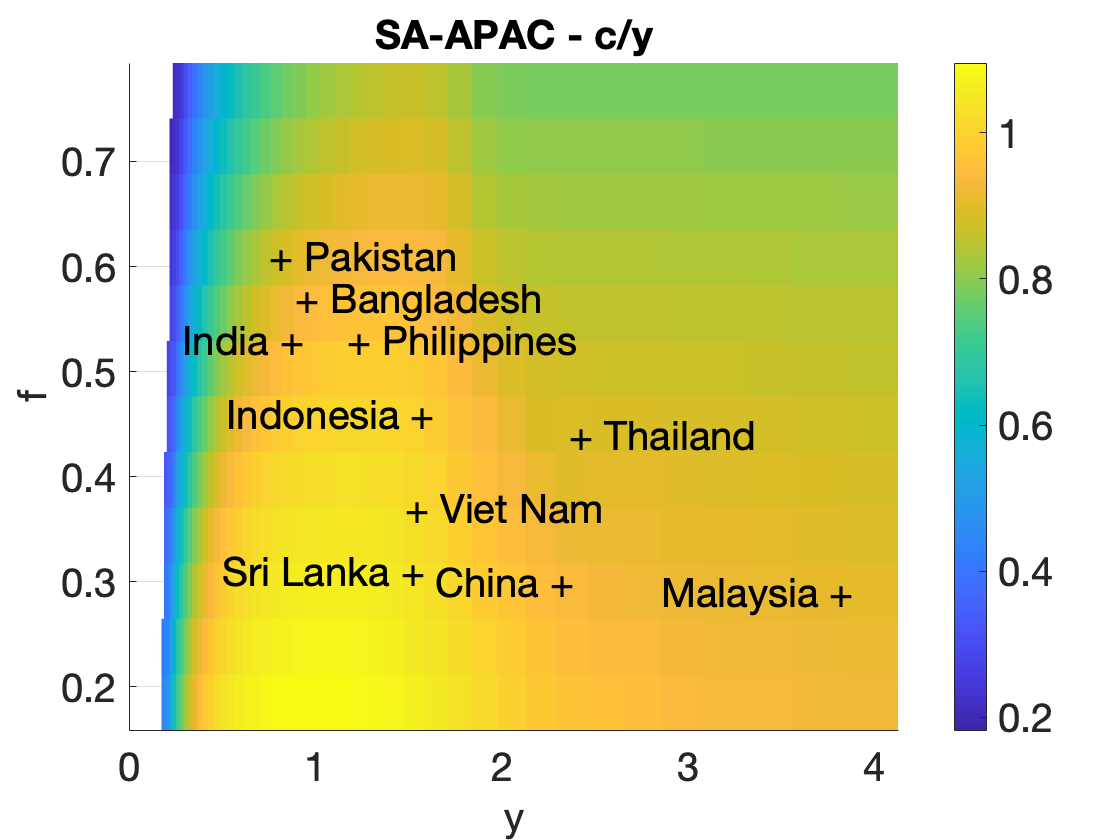}
 \includegraphics[width=0.3\textwidth]{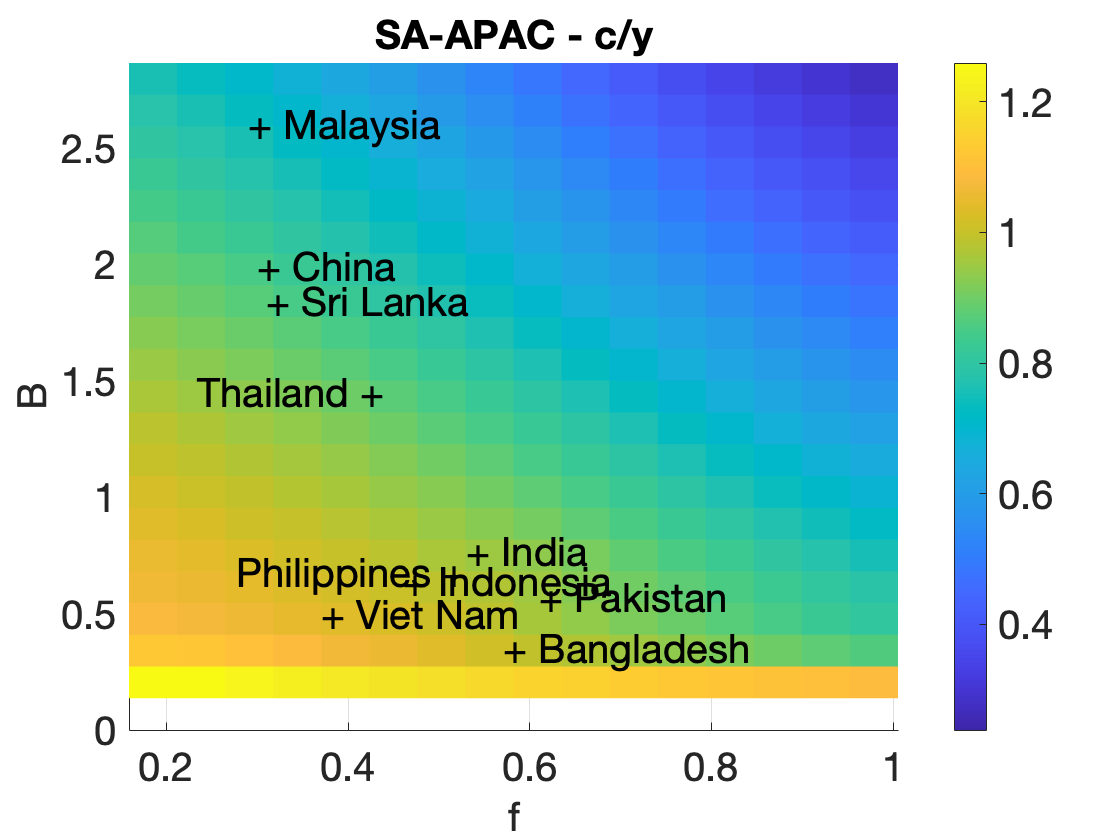}
 
 \includegraphics[width=0.3\textwidth]{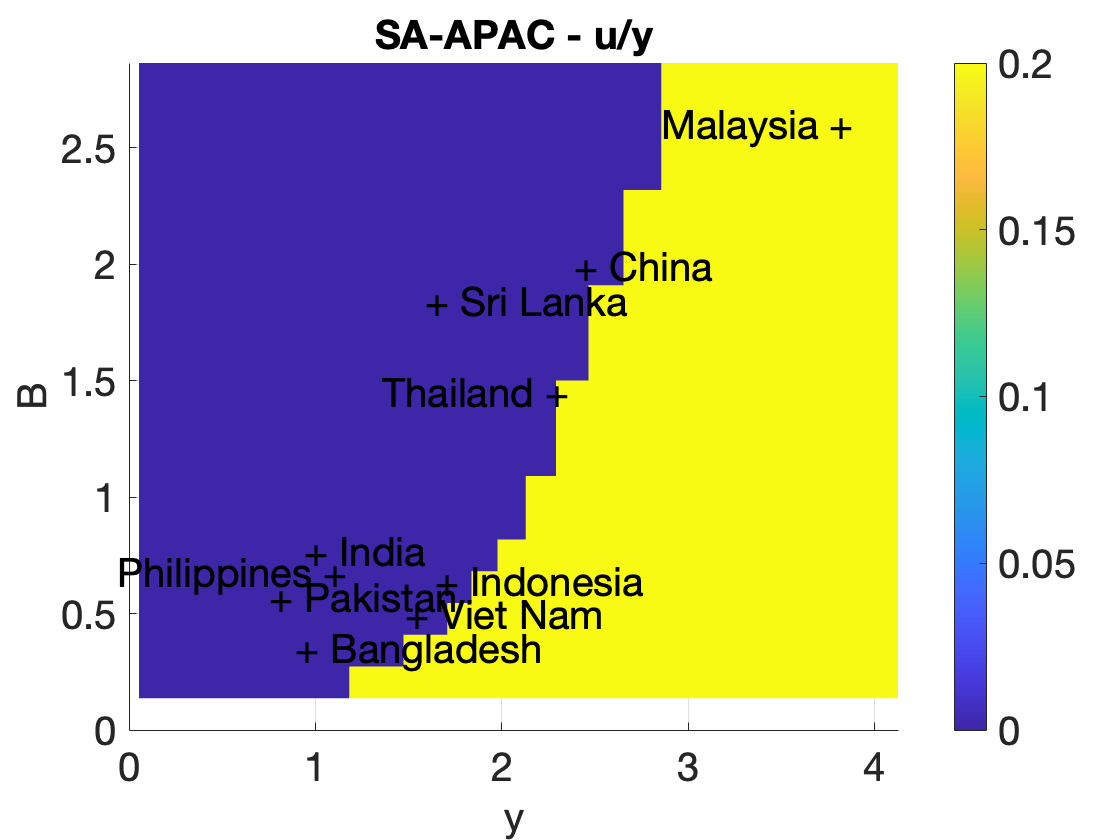}
 \includegraphics[width=0.3\textwidth]{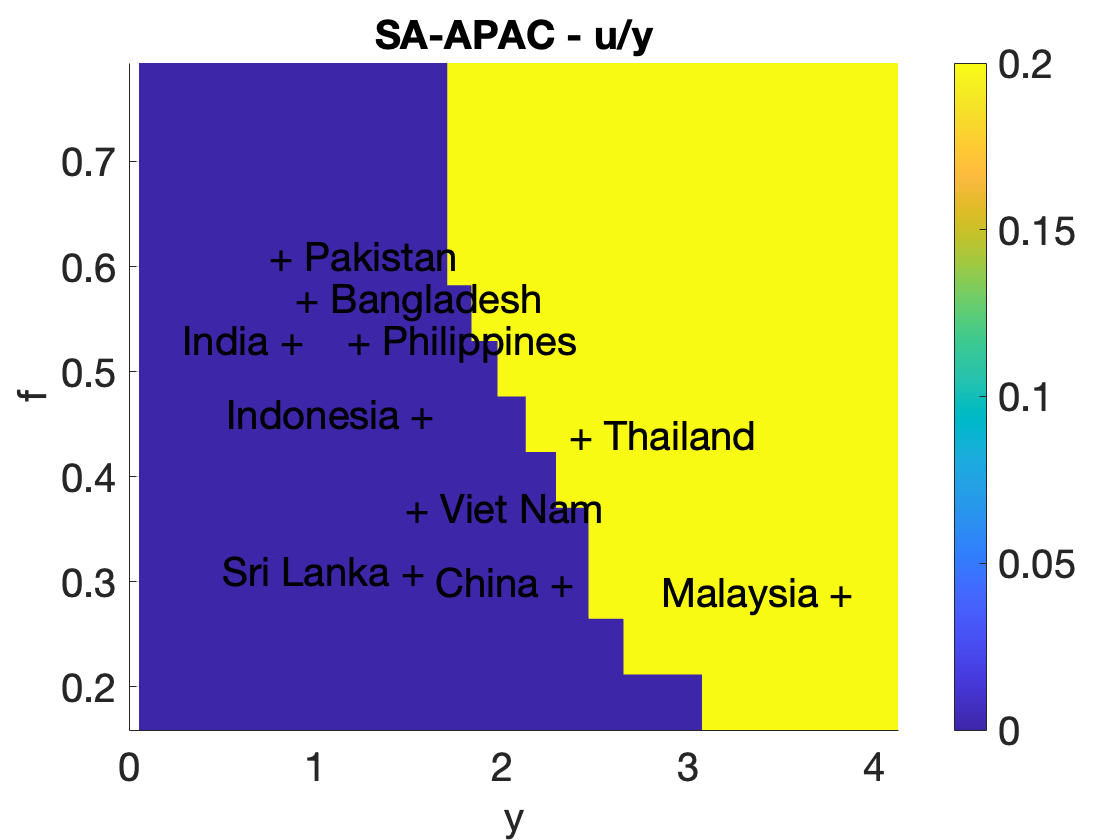}
 \includegraphics[width=0.3\textwidth]{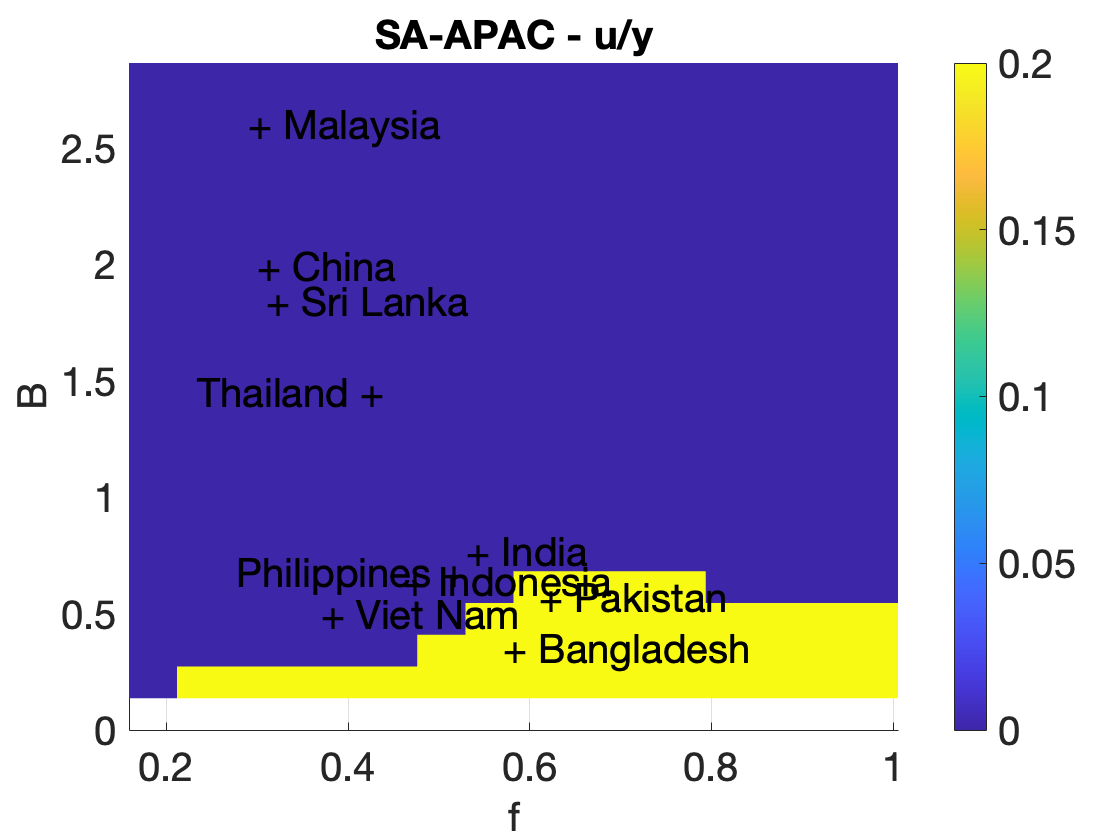}

\caption{SA-APAC region: 
decision to default ($d$), consumption over output ($c/y$) adaptation expenditure over output ($u/y$).
When a variable is fixed, it is set equal to the median value for the macro-region, see Table
\ref{Q_SA_APAC}: plane $(y,B)$, 
$f=0.45$; plane $(y,f)$, $B=0.72$; plane $(f,B)$, $y=1.54$. The white region in the picture for $c/y$ corresponds to the default region. The positioning of the country labels in the pictures is intended to be indicative because, for example, in the $(y, B)$ plane, the values of $d$, $c/y$, and $u/y$ correspond to the median value of $f$ (specifically, $f = 0.45$).}\label{M2-SAAPAC}
 \end{figure}

We investigate the decision to default ($d$), the consumption over output ($c/y$), and the adaptation expenditure over output ($u/y$) through three sets of figures:
Figure \ref{M2-SSA} for the SSA macro-region, Figure \ref{M2-LATAM} for the LATAM macro-region, and Figure \ref{M2-SAAPAC} for the SA-APAC macro-region. Each figure plots the variables on the output-public debt ($y,B$) plane, output-pollution ($y,f$) plane, pollution-public debt ($f,B$) plane taking the median value at the macro-region level for the variable left out, e.g. $B$ for the first figure. We have also reported the position of the countries which is to be considered as purely indicative because the control/decision of a country is obtained for the country specific values of the two state variables under consideration and the median value for the third one, e.g., in the $(y,B)$ plane the point representing Botswana is obtained considering its output and debt $(3.36, \ 0.63)$ and the median value for $f$, i.e., $0.42$, which is different from its own value ($0.26$).

In all the macro-regions, the decision to default only occurs for a very high debt to GDP ratio (around 250\% for the SSA region, and 300\% for SA-APAC and LATAM macro-region). On the output-pollution plane we observe that for a very low endowment the countries may opt to default, the default region enlarges as pollution increases but the effect is very limited.
Climate risk doesn't lead to default on sovereign debt, in the pollution-debt plane, there is no space for the decision to default on debt.
Notice that according to our model, no country should opt for default as shown in Tables \ref{Q_SSA}-\ref{Q_SA_APAC}. 

As the stock of pollution increases, consumption over output decreases in all the three macro-regions. The relationship between consumption over output and output is bell shaped for most of the values of the state variables $B$ and $f$ in the the SA-APAC and SSA macro-region and is increasing in the LATAM macro-region. As far as dependence on public debt is concerned, we observe a negative relationship in all three macro-regions.  

As far as adaptation expenditure is concerned, we observe very little evidence that countries take action to adapt to climate risks. Positive investment to address climate risks is observed only when $y \gg B$ or $y \gg f$ for the SA-APAC and the SSA macro-region, while in the LATAM macro-region there is little space for adaptation expenditure.
This outcome is due to the fact that, as already observed, LATAM countries are less exposed to environmental risk. An increase in $f$ reduces the activation threshold for $y$, that is, the minimum level of $y$ at which a country begins to take action on climate change. This effect is more pronounced in the SA-APAC macro-region compared to the SSA macro-region.

Note that there is no country in the LATAM region that should invest in climate adaptation ($u/y$ is always 0, see Table \ref{Q_LATAM}, in line with Figure \ref{M2-LATAM}), instead there are some countries with positive investment in the SSA and SA-APAC macro-region (Botswana and Mauritius, see Table \ref{Q_SSA}, and Malaysia, see Table \ref{Q_SA_APAC}), the main driver for the decision being the high values of $y$, as also shown in Figure \ref{M2-SSA} and \ref{M2-SAAPAC}.

\section{Sensitivity analysis and policy options}
\label{SENS}
In Appendix \ref{Sensitivity} we provide a sensitivity analysis with respect to the main parameters of the model: $\chi, \ \eta, \ k_f, \ \lambda, \ \sigma_f, \ \theta_d, \ \varphi, \ g, \ \nu_0$. We perform the analysis showing the difference in $d, \ c/y, \ u/y$ between the value obtained for the median parameter augmented by $20\%$ and the value obtained decreasing it by $20\%$. We provide the analysis for the SA-APAC macro-region, results for the other regions are similar and available upon request.

Increasing $\chi$ (decreasing the average duration of exclusion from issuing public debt) does not affect neither the decision to default on debt nor to abate pollution, see Figure \ref{S_chi}. A very low positive effect is observed for the consumption ratio.

Increasing the parameter $\eta$ (the fraction of the output recovered after the default event), the region where it is optimal to default increases, as expected, while the region where it is optimal to have a non null adaptation expenditure decreases, see Figure \ref{S_eta}. 
It seems that a higher recovering rate leads the country to care less of climate risk. Mild renegotiation conditions lead to less action against climate risk. 
The effect on the consumption ratio is negative.
%EB: perché?

As the amount of pollution removed per unit of expenditure ($k_f$) increases, the region with a positive adaptation expenditure significantly increases, see Figure \ref{S_kf}. No effect is observed on the default decision, the effect on the consumption ratio is mixed. It is positive in general, but it becomes negative when climate risk action becomes positive. 

The default region slightly increases as $\lambda$ increases (debt maturity becomes shorter), see Figure \ref{S_lambda}. No effect is observed on the adaptation expenditure decision, the effect on the consumption ratio is negative the rationale being that short maturity doesn't provide hedge to output fluctuations, see \cite{ARE3}.

As the debt fraction after default $\theta_D$ increases, no effect is detected on the default and risk adaptation expenditure decision, see Figure \ref{S_thetaD}, the effect of the consumption ratio is limited.

The parameter $\varphi$ models how the pollution decays by natural sinks. As this parameter increases, the region with a positive adaptation expenditure shrinks with a positive effect on the consumption ratio, no effect on the default decision is observed, see Figure \ref{S_varphi}.

Several parameters do not affect the outcome of the model: 
uncertainty surrounding pollution evolution, technology improvement ($\sigma_f, \ g$), see Figure \ref{S_sigmaf}, \ref{S_g}.
 The baseline frequency for a climate disaster $\nu_0$ has no effect on the default and adaptation risk expenditure, the effect on the consumption ratio is negative but rather small, see Figure \ref{S_nu0}. This confirms the weak nexus between debt management and climate risk.

We consider two policy options to promote adaptation risk expenditure: carbon offset incentive and readmission to financial markets dependent on pollution reduction.

Thanks to a carbon offset incentive, the country is reimbursed a certain fraction of the expenditure to decrease the stock of pollution. To model this incentive we add term $-pu(t)$ in (\ref{BUDG}), that is, the budget constraint becomes:
\begin{equation*}
Q(t)B^{new}(t)=(\lambda+\delta)B(t)+c(t)+(1-p)u(t)-y(t).
\end{equation*}
The idea comes from the UN REDD+ initiative, which is a global program developed by the United Nations to combat climate change by reducing emissions from deforestation and forest degradation in developing countries\footnote{https://unfccc.int/topics/land-use/workstreams/redd/what-is-redd.}. 

\begin{figure}[tp]
\centering
 \includegraphics[width=1\textwidth]{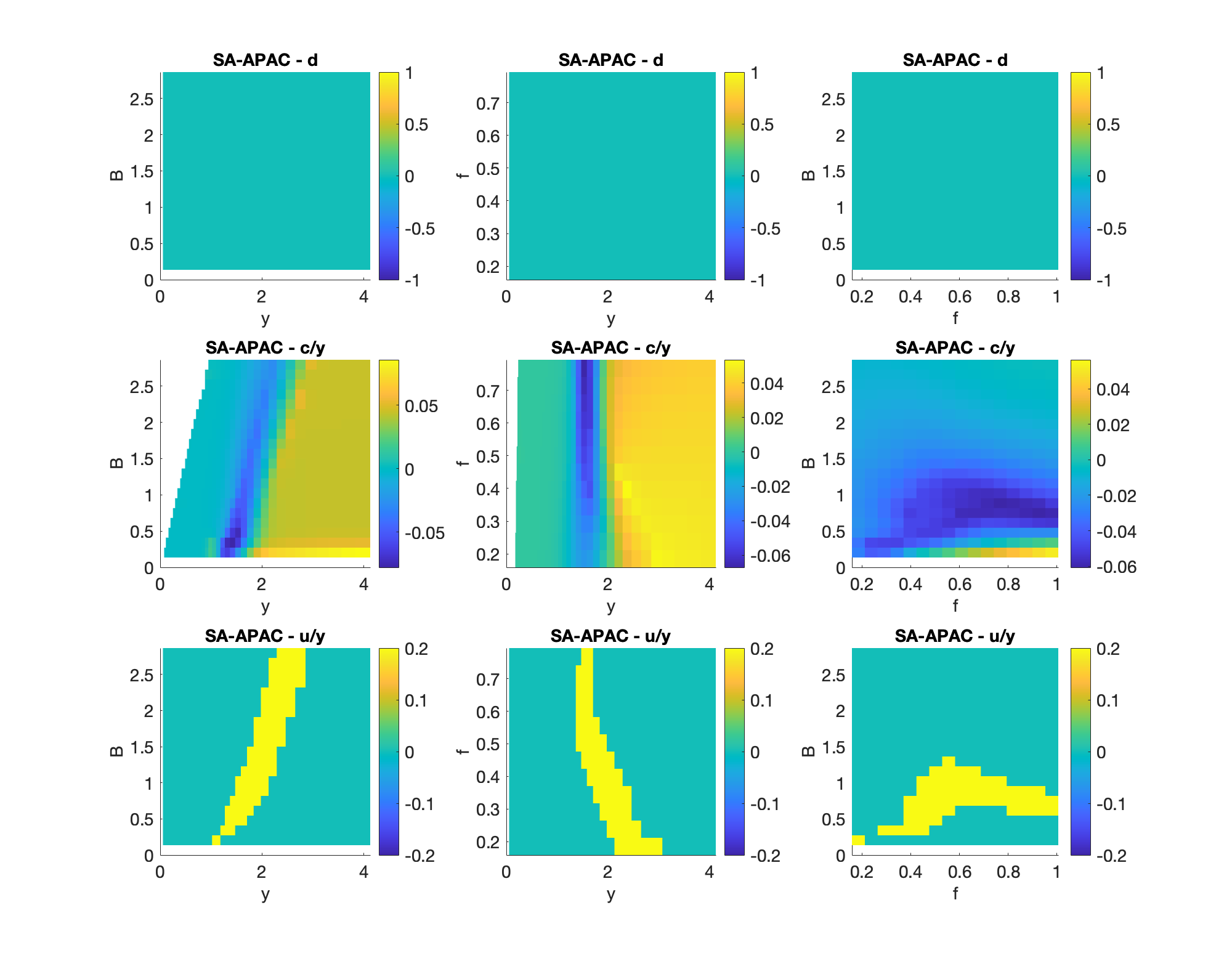}
\caption{Carbon offset incentive. Differences of the optimal control values obtained considering the incentive and the baseline model, that is,  dealing for example with $d$ we show  $d_{incentive}-d_{baseline}$ for the SA-APAC region, see Figure \ref{M2-SAAPAC} for details on the baseline analysis.}\label{Opt6}
 \end{figure}

We investigate this option assuming a 20\% incentive, i.e., $p=0.2$. Results are reported in Figure \ref{Opt6} for the SA-APAC macro-region, results for the other regions are similar and are available upon request.

Comparing it with Figure \ref{M2-SAAPAC}, we notice that the incentive has no effect on the default decision, instead the region with a positive abatement expenditure enlarges: the yellow color in the last row of Figure \ref{Opt6} (dealing with $u/y$) corresponds to the region where a positive abatement expenditure is optimal if $p=0.2$, but not in the $p=0$ case. 
As far as consumption ratio is concerned, we observe an improvement only for a high output because for intermediate values the consumption rate is already high. 

Considering Table \ref{Q_SA_APAC}, where optimal default and abatement expenditure are reported for each country, we observe that only 
Malaysia would invest in climate risk adaptation. Results 
not reported for sake of brevity show that the 20\% incentive would also lead Thailand, China, as well as the median case, to invest in climate risk abatement. As expected, there is no change in the default decision. 

\begin{figure}[tp]
\centering
 \includegraphics[width=1\textwidth]{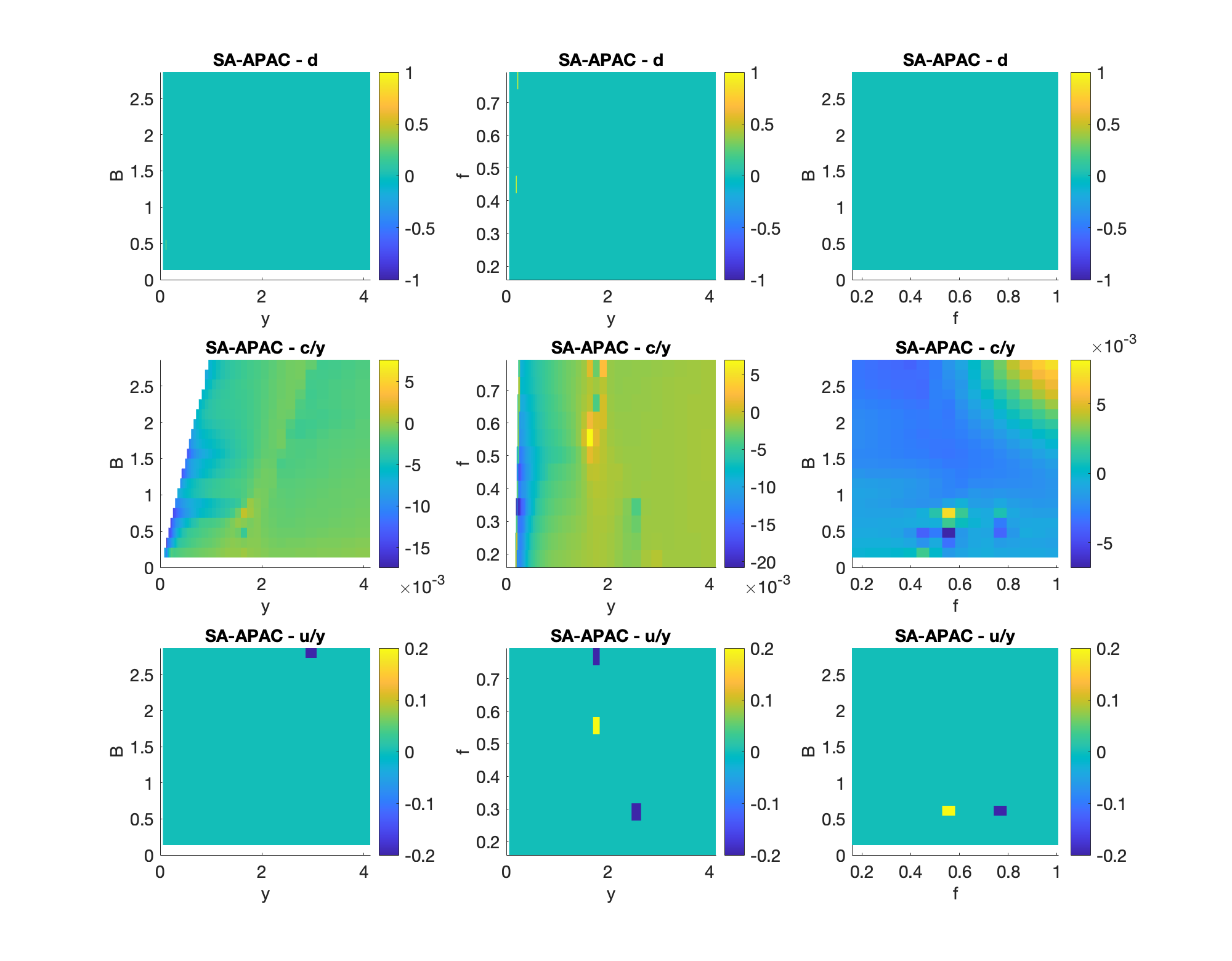}
\caption{Readmission to the market dependent on the pollution stock. 
Differences of the optimal control values obtained considering the modified probability of readmission and the baseline model, that is,  dealing for example with $d$ we show  $d_{modified}-d_{baseline}$ for the SA-APAC region, see Figure \ref{M2-SAAPAC} for details on the baseline analysis.}\label{Opt3}
 \end{figure}

We now assume that the probability of being readmitted to the market after the default - and to issue new debt - depends on the stock of pollution, that is:
$$
\chi=\chi(f)= \chi_b e^{-\alpha_{\chi} f},
$$
$\chi_b$ being the baseline parameter. In this case, increasing the pollution level, the value of $\chi$, and so the probability of exiting from the autarky, decreases, approaching zero when the pollution is high. Therefore climate risk enters in the debt restructuring and in the readmission to the market and the country should have an incentive to invest in pollution abatement. This 
opportunity seems to have a negligible impact both on the default and the risk adaptation expenditure; the effect on the consumption ratio is mixed but still small, see Figure \ref{Opt3}, where we set $\alpha_{\chi}=0.5$. Results with $\alpha_{\chi}=1$, therefore increasing the sensitivity of $\chi$ with respect to $f$, here not reported, show a similar behavior. 

\section{Conclusions}
\label{CONC}
There is empirical evidence of a positive relationship between sovereign debt rates and climate risk but, up to now, no theoretical foundation for it has been established. The model presented in this paper aims to shed light on this relationship considering climate risk and endogenous default on public debt. 

The analysis shows that the connection is rather weak. The model calibrated on developing and emerging countries show that sovereign bond rates are weakly affected by the physical pollution of the country. Countries have very little incentive to take actions to abate pollution and climate risk/country vulnerability has limited effect on the decision to default. We may conclude that climate risk is not a key issue in the management of public debt for developing and emerging countries. 
Climate risk does not seem to be an issue for these countries. 
To foster investment to abate pollution it may be useful to subsidize such expenditure, while renegotiation upon default related to climate risk does not provide enough incentives.

\section*{Acknowledgment}
This research has been partially funded by the European Union - Next Generation EU, Mission 4-Component 1-CUP D53D23017690001 - Project PRIN PNRR 2022 P20228CHNL “Measuring, managing and hedging indirect climate-transition risk”, and by Italian MUR, grant Dipartimento di Eccellenza 2023-2027. Daniele Marazzina is members of the Gruppo Nazionale Calcolo Scientifico-Istituto Nazionale di Alta Matematica (GNCS-INdAM). The simulations discussed in this work were, in part, performed on the HPC Cluster of the Department of Mathematics of Politecnico di Milano which was funded by MUR grant Dipartimento di Eccellenza 2023-2027.

\clearpage
\appendix
\renewcommand\thefigure{\thesection.\arabic{figure}}    
\renewcommand\thetable{\thesection.\arabic{table}}    
\setcounter{figure}{0}  
\setcounter{table}{0}  

\section{Model calibration and estimation}
\label{CALIB}
In Table \ref{COUNTRIES} we report the countries considered in our analysis.

The output of each country is provided by the gross domestic product per capita, measured by purchasing power parity (GDP PPP)  as reported by the International Monetary Fund (IMF)\footnote{Source: \url{https://www.imf.org/external/datamapper/PPPPC@WEO/OEMDC/ADVEC/WEOWORLD}.}. The data analysis reveals significant disparities in output levels across countries. To enhance comparability across the countries of the three macro-regions, we normalize both output and debt relative to the regional sample average for the period 2013-2023, expressed in dollars (SSA: $5,460\$ $,  LATAM: $16,367\$ $, SA-APAC: $8,946\$ $).

\begin{table}[tp]
    \centering
    \begin{tabular}{l|l|l}
         SSA &   LATAM &  SA-APAC\\\hline
        Angola& Antigua and Barbuda& Bangladesh \\
        Benin& Argentina& Bhutan \\
        Botswana& Bahamas, The& Cambodia \\
        Burkina Faso& Barbados& China \\
        Burundi& Belize& Fiji \\
        Cabo Verde& Bolivia& India \\
        Cameroon& Brazil& Indonesia \\
        Central African Republic& Chile& Kiribati \\
        Comoros& Colombia& Lao PDR \\
        Congo, Dem. Rep.& Costa Rica& Malaysia \\
        Congo, Rep.& Dominica& Maldives \\
        Cote d'Ivoire& Dominican Republic& Marshall Islands \\
        Eswatini& Ecuador& Micronesia, Fed. Sts. \\
        Ethiopia& El Salvador& Mongolia \\
        Gabon& Grenada& Myanmar \\
        Gambia, The& Guatemala& Nepal \\
        Ghana& Guyana& Pakistan \\
        Guinea& Haiti& Palau \\
        Guinea-Bissau& Honduras& Papua New Guinea \\
        Kenya& Jamaica& Philippines \\
        Lesotho& Mexico& Samoa \\
        Liberia& Nicaragua& Solomon Islands \\
        Madagascar& Panama& Sri Lanka \\
        Malawi& Paraguay& Thailand \\
        Mali& Peru& Timor-Leste \\
        Mauritania& St. Lucia& Tonga \\
        Mauritius& Suriname& Tuvalu \\
        Mozambique& Trinidad and Tobago& Vanuatu \\
        Namibia& Uruguay& Viet Nam \\
        Niger& &  \\
        Nigeria& &  \\
        Rwanda& &  \\
        Senegal& &  \\
        Seychelles& &  \\
        Sierra Leone& &  \\
        South Africa& &  \\
        Sudan& &  \\
        Tanzania& &  \\
        Togo& &  \\
        Uganda& &  \\
        Zambia& &  \\
        Zimbabwe& &  \\
    \end{tabular}
    \caption{List of developing and emerging countries split by macro-region}
    \label{COUNTRIES}
\end{table}

\subsection{Pollution process}
As a measure of pollution stock, we consider the Climate-driven INFORM Risk indicator, computed by the IMF for a panel of 188 countries on an annual basis since 2013\footnote{Source: \url{https://climatedata.imf.org}.}. 
The indicator has three dimensions (climate-driven hazard \& exposure, vulnerability, and lack of coping capacity).  
We assume a linear mapping from the indicator to the level of pollution stock $f$, ranging from $f=0$ (INFORM Risk indicator: $0$, negligible pollution stock) to $f=1$ (INFORM Risk indicator: $10$, very high pollution stock). 
The level of pre-industrial pollution is approximated by the time average of the least polluted country in the sample (Cape Verde), setting $f_{pre}=0.27$. 

The Climate-driven INFORM Risk dataset provides eleven annual observations, subtracting $f_{pre}$ we obtain $\{\widehat{f}(t_{i})\}_{i=1}^{11}$. We assume that no country in the panel allocated a significant portion of its GDP to climate adaptation during the observation period. Therefore, the sample can be viewed as a realization of the process in \eqref{FOSS} with $u=0$.

Annual realizations $\widehat{f}(t_i)$ are normally distributed, with conditional expected value and unconditional variance:
\begin{eqnarray}\label{OU}
\mathbb{E}[\widehat{f}(t_{i+1})|\widehat{f}(t_i)]&=& e^{-\varphi} \widehat{f}(t_i) + k_e \int_{t_i}^{t_{i+1}} e^{-g s} y(s) e^{-\varphi (t_{i+1} - s)} ds \\
Var[\widehat{f}(t_i)] &=& \frac{\sigma_f^2}{2 \varphi} (1-e^{-2 \varphi}).\nonumber
\end{eqnarray}

The four parameters in \eqref{OU} ($\varphi, \sigma_f, k_e, g$) could be directly estimated using a Maximum Likelihood (ML) estimator on a country-by-country basis. However, to enhance the statistical significance, we prefer to adopt a two-step estimation strategy. 

First, we calibrate the effect of technological advancement on the relationship between output and pollution in (\ref{TECH}). As there is no measure of physical pollution, we use $CO_2$ emissions as a proxy. This information is gathered from The World Bank that reports $CO_2$ emissions (kg per PPP\$ of GDP) for a panel of 229 countries\footnote{\url{https://data.worldbank.org/indicator/EN.ATM.CO2E.PP.GD}.}. The dataset spans from 1990 to 2020. To estimate the rate of technological progress ($g$) in the pollution stock process, we fit the time series of $CO_2$ emission intensity (emissions divided by GDP) to an exponential decay model.
Availability of information narrows the clusters to 31 countries in the SSA region, 27 countries in the LATAM region, and 25 countries in the SA-APAC region. 
In Table \ref{tmean_SSA}, \ref{tmean_LATAM}, and \ref{tmean_SA_APAC} we provide the estimate of $g$
for each of the three macro-regions (median values across the sample of the parameters). 
Summary statistics of the estimated parameters across countries in each macro-region are shown in Table \ref{G}.

\begin{table}[tp]
    \centering
    \begin{tabular}{l|r|r|r|r|r|r|r}
        & Mean &   Std. Dev. &  Median & Min. & $25^{th}$ Pctile & $75^{th}$ Pctile & Max. \\\hline
         SSA & $0.011$ & $0.024$ & $0.014$ & $-0.056$ & $0.000$ & $0.029$ & $0.040$ \\
         LATAM  & $0.021$ & $0.018$ & $0.028$ & $-0.023$ & $0.011$ & $0.031$ & $0.044$ \\
         SA-APAC  & $0.019$ & $0.010$ & $0.021$ & $-0.001$ & $0.014$ & $0.027$ & $0.035$ \\
    \end{tabular}
    \caption{Technology advancement factor $g$, summary statistics (1990-2020).}
    \label{G}
\end{table}

Given the estimate of $g$, we estimate the remaining parameters ($k_e$, $\varphi$, $\sigma_f$) using a ML estimator at the macro-region level.
Results 
(median values across the sample of the parameters) are provided in 
the second column of Table \ref{tmean_SSA}, \ref{tmean_LATAM}, and \ref{tmean_SA_APAC} for each of the three macro-regions. 
Summary statistics of the estimated parameters across countries in each macro-region are shown in Table \ref{K_E}, \ref{VARPHI}, \ref{SIGMA_F}, respectively.
\begin{table}[tp]
    \centering
    \begin{tabular}{l|r|r|r|r|r|r|r}
        & Mean &   Std. Dev. &  Median & Min. & $25^{th}$ Pctile & $75^{th}$ Pctile & Max. \\\hline
         SSA & $0.208$ & $0.250$ & $0.111$ & $0.003$ & $0.027$ & $0.305$ & $0.890$ \\
         LATAM  & $0.087$ & $0.265$ & $0.045$ & $0.001$ & $0.011$ & $0.120$ & $0.483$ \\
         SA-APAC  & $0.115$ & $0.152$ & $0.064$ & $0.003$ & $0.009$ & $0.164$ & $0.481$ \\
    \end{tabular}
    \caption{Increase in pollution stock per unit output $k_e$, summary statistics (2013-2023).}
    \label{K_E}
\end{table}
\begin{table}[tp]
    \centering
    \begin{tabular}{l|r|r|r|r|r|r|r}
        & Mean &   Std. Dev. &  Median & Min. & $25^{th}$ Pctile & $75^{th}$ Pctile & Max. \\\hline
         SSA & $0.38$ & $0.27$ & $0.31$ & $0.04$ & $0.19$ & $0.54$ & $0.85$\\
         LATAM & $0.40$ & $0.27$ & $0.34$ & $0.02$ & $0.27$ & $0.52$ & $0.81$\\
         SA-APAC  & $0.40$ & $0.22$ & $0.35$ & $0.07$ & $0.25$ & $0.61$ & $0.72$\\
    \end{tabular}
    \caption{Naturals sink rate $\varphi$, summary statistics (2013-2023).}
    \label{VARPHI}
\end{table}
\begin{table}[tp]
    \centering
    \begin{tabular}{l|r|r|r|r|r|r|r}
        & Mean &   Std. Dev. &  Median & Min. & $25^{th}$ Pctile & $75^{th}$ Pctile & Max. \\\hline
        SSA & $0.029$ & $0.019$ & $0.025$ & $0.002$ & $0.013$ & $0.039$ & $0.065$ \\
         LATAM  & $0.030$ & $0.014$ & $0.028$ & $0.010$ & $0.019$ & $0.042$ & $0.053$\\
         SA-APAC  & $0.022$ & $0.017$ & $0.022$ & $0.004$ & $0.009$ & $0.027$ & $0.058$ \\
    \end{tabular}
    \caption{Volatility of the pollution stock process $\sigma_f$, summary statistics (2013-2023).}
    \label{SIGMA_F}
\end{table}

\subsection{Endowment process}

To estimate the endowment process \eqref{OUTPUT}, we start with the analysis of natural disaster frequency using climate-related event data from 1980 to 2022, collected by the IMF\footnote{\url{https://climatedata.imf.org/search?collection=Dataset}.}. These data come from the Emergency Events Database (EM-DAT) which is maintained by the Centre for Research on the Epidemiology of Disasters (CRED) at the Université Catholique de Louvain\footnote{\url{https://doc.emdat.be/docs/data-structure-and-content/emdat-public-table/}.}. To focus on pollution-related events, we only include those related to floods, extreme temperatures, storms, mass movements, wildfires, droughts, and glacial lake  outburst/floods.
CRED’s database covers a vast number of events ($7,598$ across $229$ countries), but many records lack the detailed information to evaluate the severity of the event. In Table \ref{DISASTER} we report some descriptive statistics on the dataset.
 
\begin{table}[tp]
    \centering
    \begin{tabular}{l|r|r|r|r|r|r|r}
         & Mean &   Std. Dev. &  Median & Min. & $25^{th}$ Pctile & $75^{th}$ Pctile & Max. \\\hline
         SSA & $0.84$ & $0.56$ & $0.70$ & $0.05$ & $0.37$ & $1.28$ & $2.35$ \\
         LATAM  & $1.38$ & $1.28$ & $1.09$ & $0.09$ &  $0.37$ & $1.85$ &  $4.93$ \\
         SA-APAC  & $2.84$ & $4.25$ & $0.93$ & $0.09$ & $0.23$ & $3.62$ & $17.16$ \\
    \end{tabular}
    \caption{Natural events' count per country per year, summary statistics (1980-2022).}
    \label{DISASTER}
\end{table}

The frequency of natural events is notably higher than expected for rare disasters. In our setting,  the endowment process (\ref{OUTPUT}) aims to capture significant declines associated with rare natural disasters through the jump process and frequent, smaller reductions (somehow structural) through the damage function. Therefore, 
we have to distinguish rare, severe climate disasters from more frequent minor events.
To this end, we extract $\{k(t_i)\}_{i=1}^{10}$  from the CRED database, where $k(t_i)$ represents the number of climate-related events in the {\em i-th} year for a selected country in scope. We then discretize the endowment process with yearly time steps ($\Delta t=1$) as follows:

\begin{equation}\label{OUT_DISC}
 \Delta \ln y(t_{i+1}) = \ln\left(y(t_{i+1})\right) - \ln\left(y(t_i)\right) = \varepsilon(t_{i+1}) - \eta(t_{i+1}) 
\end{equation}
where, $\varepsilon(t_{i+1})$ represents the part of the endowment process \eqref{OUTPUT} (white noise component) unaffected by climate risk, drawn from a Normal distribution with mean  $\mu - \sigma^2_{y}/2$ and variance $\sigma^2_{y}$. The impact of climate risk is captured by the term $-\eta(t_{i+1})$.
We assume that each climate-related event in the CRED data is independent of the others, with recovery following an exponential distribution with a country-specific scale parameter $\lambda_f$.  Through some algebra, we find that:
\begin{equation}\label{ERLANG}
 \eta(t_i) = 1 - e^{-\xi(k(t_i),\lambda_{f})}
\end{equation}
where $\xi(k(t_i),\lambda_{f})$ is sampled from an Erlang-k distribution with event count $k(t_i)$ and scale parameter $\lambda_f$.
We estimate the parameters $\mu$, $\sigma_{y}$ and $\lambda_{f}$ for each country in the sample from IMF output data using a ML estimator under the assumption that $\eta(t_i)$ is distributed as in \eqref{ERLANG}. Table \ref{MU}, \ref{SIGMA_Y}, and \ref{LAMBDA_F}  shows summary statistics for the parameters across the three macro-regions, with median values for $\mu$ and $\sigma_y$ detailed in Table \ref{tmean_SSA}, \ref{tmean_LATAM}, and \ref{tmean_SA_APAC}. $\lambda_f$ represents the average annual log-output reduction per climate-related event, with median values of $0.27\%$ in the SSA, $0.31\%$ in the LATAM, and $0.27\%$ in the SA-APAC macro-region.

The above estimates are obtained considering all the climate events, we want now to disentangle rare/severe disasters from frequent/minor disasters.
According to \eqref{OUTPUT}, the impact of climate risk on the log-output $\eta(t_{i+1})$ depends on pollution stock and is given by:
\begin{equation}
    \eta(t_{i+1}) = \mu D(f(t_i)) + \big(1-Z(t_i)\big)N(t_i)
\end{equation}
where 
$f(t_i)$ is the pollution stock of the country under consideration in year $t_i$, 
$N(t_i)\in \{0,1\}$ indicates a rare disaster in year $i$, and $Z(t_i)$ is recovery if $N(t_i)=1$\footnote{We assume that the occurrence of two or more rare disasters in a single year is a low-probability event which can be neglected during the calibration of the model parameters.}. As far as the the recovery distribution \eqref{Z} for rare climate events is concerned, we assume that it follows a power law characterized by a parameter of $\beta=6.27$ as estimated in \cite{BAR} based on a large sample of rare disasters. The damage function $D$ and rare-disaster frequency $\nu$ depend only on pollution stock value $\overline{f}$, so the expected value and variance of $\eta(t_{i+1})$ are:
\begin{align}\label{POLL_REAL}
    <\eta> &= \mu D(\bar{f}) + \frac{\nu(\bar{f})}{\beta+1} \\
    Var(\eta) &= \frac{2 \nu(\bar{f})}{(\beta+1)(\beta+2)} + \frac{\beta}{\beta+2} \big[ \frac{\nu(\bar{f})}{\beta+1}\big]^2 \nonumber
\end{align}
where $D(\bar{f})$ is given by \eqref{OUTPUT2}. 

The rare-disaster frequency $\nu(\bar{f})$ is provided by \eqref{NU_T}, where we set the baseline frequency $\nu_0$ to $0.36\%$, representing the average environmental event frequency according to \cite{KAR2019}. Then we calibrate the remaining parameters of the damage function $(\phi, \theta)$ and rare-disaster frequency $(\nu_1,\psi_f)$ by momentum matching.
For each country, we calculate the realized drift and variance of $\Delta\ln y$ and, assuming independence between $\varepsilon(t_{i+1})$ and $\eta(t_{i+1})$, estimate pollution-dependent drift and variance:
\begin{align}\label{POLL_EST}
    <\eta> &= (\mu - \frac{\sigma^2_{y}}{2}) \quad - <\Delta\ln y> \\
    Var(\eta) &= Var(\Delta\ln y) - \sigma^2_{y}.\nonumber
\end{align}

We set $\bar{f}_k$ as the time-average value of $f$ for country $k$, constructing three sets of observations $\{(\bar{f}_k, \text{Var}(\eta_k))\}, \ k=1,\dots K$— one for each macro-region under investigation — comprising $K=31$ observations for SSA, $K=27$ for LATAM, and $K=25$ for SA-APAC macro-region.
The second equation in \eqref{POLL_REAL} defines a quadratic relationship between 
$\text{Var}(\eta_k)$ and and $\nu(\bar{f}_k)$. Leveraging this relationship, we derive three sets of observations, $\{(\bar{f}_k, \nu(\bar{f}_k))\}$, that are used to estimate the parameters ($\nu_1$, $\psi_f$) in \eqref{NU_T} via a log-linear regression.
The first equation in \eqref{POLL_REAL} establishes a direct link between $\bar{f}$, the observed impact of climate risk on the log-output, $<\eta>$, the average loss in log-output caused by rare disasters, $\nu(\bar{f})/(\beta+1)$, and climate-related damage $\mu D(\overline{f})$. Using the three sets of observations $\{(\bar{f}_k, \nu(\bar{f}_k))\}$, we also estimate the parameters ($\phi$, $\theta$) of the damage function in \eqref{OUTPUT2} through log-linear regression.
The results of the two log-linear regressions are provided in Table \ref{tmean_SSA}, \ref{tmean_LATAM}, and \ref{tmean_SA_APAC}.

\begin{table}[tp]
    \centering
    \begin{tabular}{l|r|r|r|r|r|r|r}
        & Mean &   Std. Dev. &  Median & Min. & $25^{th}$ Pctile & $75^{th}$ Pctile & Max. \\\hline
         SSA & $0.048$ & $0.015$ & $0.047$ & $-0.003$ & $0.026$ & $0.059$ & $0.069$ \\
         LATAM  & $0.049$ & $0.012$ & $0.050$ & $0.017$ & $0.040$ & $0.057$ & $0.075$ \\
         SA-APAC  & $0.054$ & $0.022$ & $0.058$ & $0.022$ & $0.034$ & $0.069$ & $0.107$ \\
    \end{tabular}
    \caption{Drift of the endowment process owed to diffusion not conditional on the pollution stock ($\mu$), summary statistics (2013-2022).}
    \label{MU}
\end{table}
\begin{table}[tp]
    \centering
    \begin{tabular}{l|r|r|r|r|r|r|r}
        & Mean &   Std. Dev. &  Median & Min. & $25^{th}$ Pctile & $75^{th}$ Pctile & Max. \\\hline
         SSA & $0.054$ & $0.019$ & $0.051$ & $0.024$ & $0.043$ & $0.060$ & $0.100$\\
         LATAM & $0.050$ & $0.015$ & $0.054$ & $0.021$ & $0.040$ & $0.073$ & $0.085$\\
         SA-APAC  & $0.043$ & $0.019$ & $0.039$ & $0.019$ & $0.032$ & $0.049$ & $0.109$\\
    \end{tabular}
    \caption{Volatility of the endowment process owed to diffusion not conditional on the pollution stock $\sigma_y$, summary statistics (2013-2022).}
    \label{SIGMA_Y}
\end{table}
\begin{table}[tp]
    \centering
    \begin{tabular}{l|r|r|r|r|r|r|r}
        & Mean &   Std. Dev. &  Median & Min. & $25^{th}$ Pctile & $75^{th}$ Pctile & Max. \\\hline
        SSA & $0.72\%$ & $1.19\%$ & $0.27\%$ & $0.04\%$ & $0.15\%$ & $0.50\%$ & $5.00\%$ \\
         LATAM  & $0.50\%$ & $0.47\%$ & $0.31\%$ & $0.03\%$ & $0.12\%$ & $0.74\%$ & $1.61\%$\\
         SA-APAC  & $0.64\%$ & $0.74\%$ & $0.27\%$ & $0.02\%$ & $0.12\%$ & $1.15\%$ & $2.27\%$ \\
    \end{tabular}
    \caption{Scale parameter $\lambda_f$ of the distribution of climate-related events of unspecified severity, summary statistics (2013-2022).}
    \label{LAMBDA_F}
\end{table}
%EB: i dati sono riportati in % è corretto?SI

\clearpage\newpage
\section{The numerical scheme}
\label{SCHEME}
The numerical scheme is made up of several steps.

As a first step, we solve the asymptotic problem, i.e., the HJB equation when $t\rightarrow +\infty$. For example, the asymptotic version of the HJB (\ref{HJB_A}) is
\begin{align*}
0=  &\underset{u(t-)}{\max} \big{[} h( y -u,v^{def}) + \mathcal{L}^{def,\infty}_{u} \mkern9mu v^{def}  \big{] }+(\nu_0 + \nu_1 (f_{pre}+\widehat{f})^{\psi_f})
					            \mathbb{E} \big{[} v^{def}(Z y) - v^{def}( y) \big{]} + \nonumber \\
					            &\chi \big{[} v( \theta_D B) - v^{def}(B)\big{]}
\end{align*}
where
\begin{align*}
\mathcal{L}^{def,\infty}_{u} \mkern9mu v = &\mu (1 - \phi (f_{pre}+\widehat{f}))^{1+\theta}) y v_y +(- k_f u - \varphi \widehat{f}) v_{\widehat{f}} +\frac{1}{2} \sigma_{y}^{2} y^2 v_{yy} +
					           \frac{1}{2} \sigma_{f}^{2}  v_{\widehat{f}\widehat{f}}.
\end{align*}
The numerical solution procedure relies on a fixed point algorithm. Given guess solutions $v^{def},\,v,\,  Q^{def},\,Q$, we compute the default barriers and the controls, and then: 
\begin{enumerate}
	\item we solve the asymptotic version of the HJB \eqref{HJB_A} during the autarky regime. This allows to determine the value function during the exclusion phase $v^{def}$, given the guess function $v$ and the control $u$. We also update the control $u$ in the default regime exploiting the new solution. 
\item We solve the asymptotic version of the HJB variational inequality \eqref{HJB_VAR}, obtaining the value function in the normal regime 
	$v$, given the controls, and the guess function $Q$ . This allows to update the default barrier in the state variables space $(y, f, B)$, and the controls $c$ and $u$.
	\item We obtain the debt prices $Q$ and $Q^{def}$ solving the pricing problem for a risk-neutral investor, i.e,. the asymptotic version of the two HJBs \eqref{Q_FK} and \eqref{Qdef_FK}.
\end{enumerate}
The algorithm is iterated until the previous version of the 
controls and of the default barrier and the updated ones 
are close enough, i.e., $||c_{old}-c_{updated}||_2$ is smaller than a given tolerance threshold. 
Our solution strategy resembles the one presented in  \cite[Appendix B]{NUNO}, exploiting a finite difference scheme with upwind to solve the four asymptotic HJBs, and therefore obtaining  $v^{def,\infty},\,v^{\infty},\,Q^{def,\infty},\,Q^{\infty}$ as well as the controls. The computations are performed on the domain $[y_{\min},y_{\max}]\times [\widehat{f}_{\min},\widehat{f}_{\max}]\times [B_{\min},B_{\max}]$, where 
$y_{\min}=0.05, y_{\max}=8, \widehat{f}_{\min}=-f_{pre}, \widehat{f}_{\max}=1, B_{\min}=0, B_{\max}=6$. We set
a positive $y_{\min}$ as we deal with the logarithmic of the output in the implementation. 
We consider 70 points to discretize $z=\log(y)$, 25 for $\widehat{f}$, and 45 for $B$, for a total of 
78,750 points. To solve the four HJB equations, we set the following boundary conditions:
 \begin{itemize}
  \item on $z_{\min}$: $v^{def,\infty}_z=v^{\infty}_z=w_z$, where $w=w(z)$ is the solution
   of a related problem where the country is in default without the possibility to get back 
   to issue new debt, without pollution and disasters. The motivation of this assumption is that, 
   if the output is close to zero, the economic conditions are so bad that the country immediately 
   defaults when the exclusion period ends, and pollution and climate disasters have no 
   effect on such a bad economic condition. We also set $Q^{def,\infty}_z=Q^{\infty}_z=0$, as in \cite{NUNO};
  \item on $z_{\max}$: $v^{def,\infty}_z=v^{\infty}_z=0$, as in \cite{NUNO}, and 
  $Q^{def,\infty}=\frac{\theta_D}{1+\delta/\chi},\,Q^{\infty}=1$, i.e., the output is so high 
  that the country, in case of default will not default anymore at the end of the exclusion period, and, 
  in case of default it opts not to default, i.e., the country is default-free;
   \item on $\widehat{f}_{\min}$ and $\widehat{f}_{\max}$: zero second order derivative with respect to $\widehat{f}$ for all the four functions;
    \item on $B_{\max}$: $v^{\infty}=v^{def,\infty}(\eta y) ,\,Q^{\infty}=Q^{def,\infty}(\eta y)$, 
    the debt is so high that we assume that the country defaults.
    \end{itemize}
    No boundary conditions are needed on $B_{\min},\,B_{\max}$ for the $def$ functions, since there is no derivative with respect to $B$, and on $B_{\min}$ for the other two functions due to the use of the upwind scheme and the absence of the second order derivative with respect to $B$.

 Once the asymptotic solutions are computed, we set these solutions as terminal condition at a time $T$, with $T$ chosen such that $k_e e^{-gT}<10^{-5}$, that is, the role of the time in the coefficient of the HJB equations is negligible, and we solve backward in time with 500 time steps, to compute the solutions at time 0 and all the optimal controls. We exploit the same boundary conditions above described.
 
\clearpage\newpage
\section{Sensitivity analysis}\label{Sensitivity}
In this section we report the sensitivity plots for the SA-APAC region. Plots for the others regions are available upon request.
\begin{figure}[h!]
\centering
 \includegraphics[width=1\textwidth]{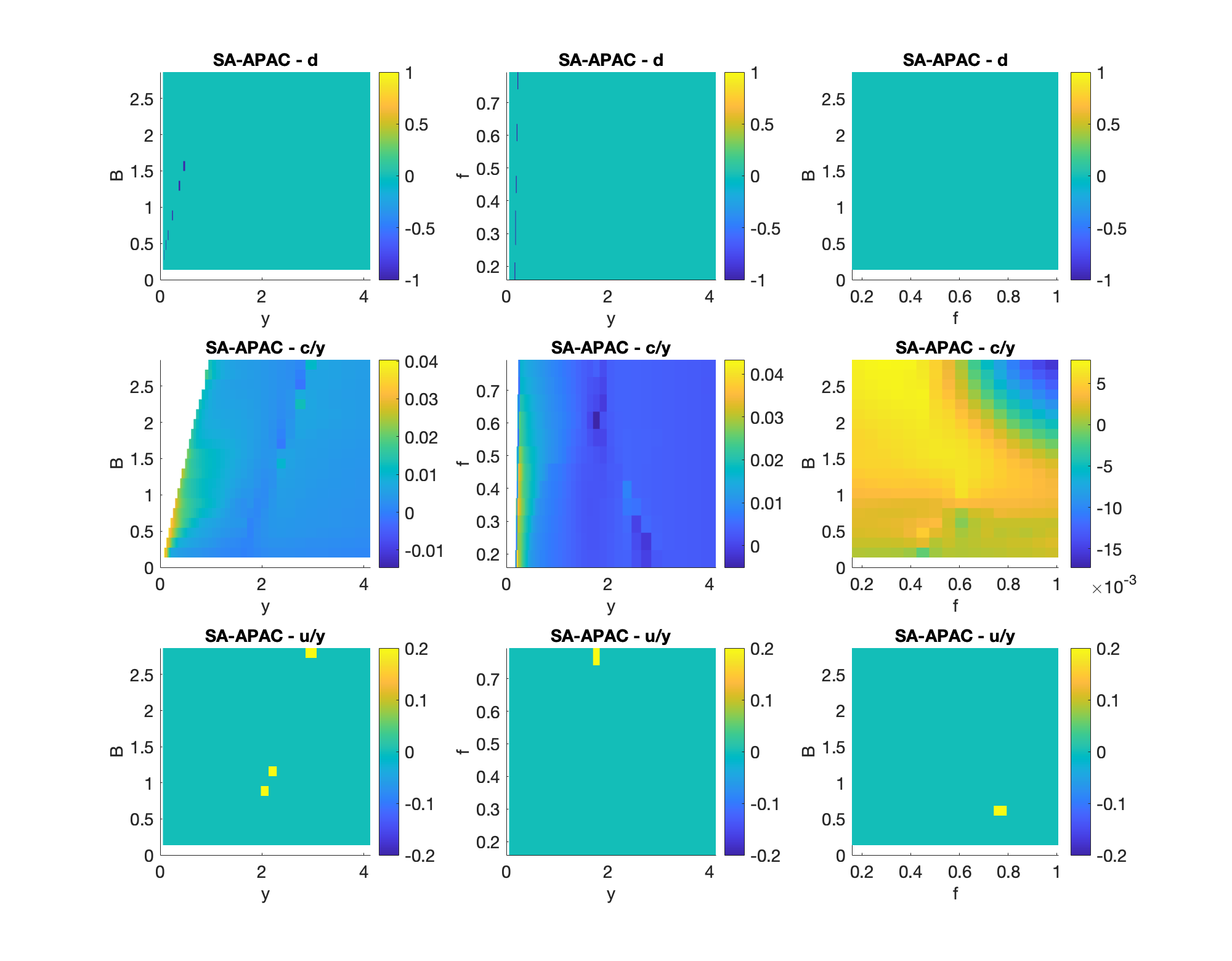}
\caption{Sensitivity analysis - parameter $\chi$, SA-APAC region. We report the differences of the optimal controls value obtained with $\chi=1.2\chi_b$ and $\chi=0.8\chi_b$, $\chi_b$ being the baseline value of the parameter, that is, if the decision to default ($d$) is considered, we plot $d_{1.2\chi_b}-d_{0.8\chi_b}$. See Figure \ref{M2-SAAPAC} for details and for the plots of the baseline value.}\label{S_chi}
 \end{figure}

\begin{figure}[ht!]
\centering
 \includegraphics[width=1\textwidth]{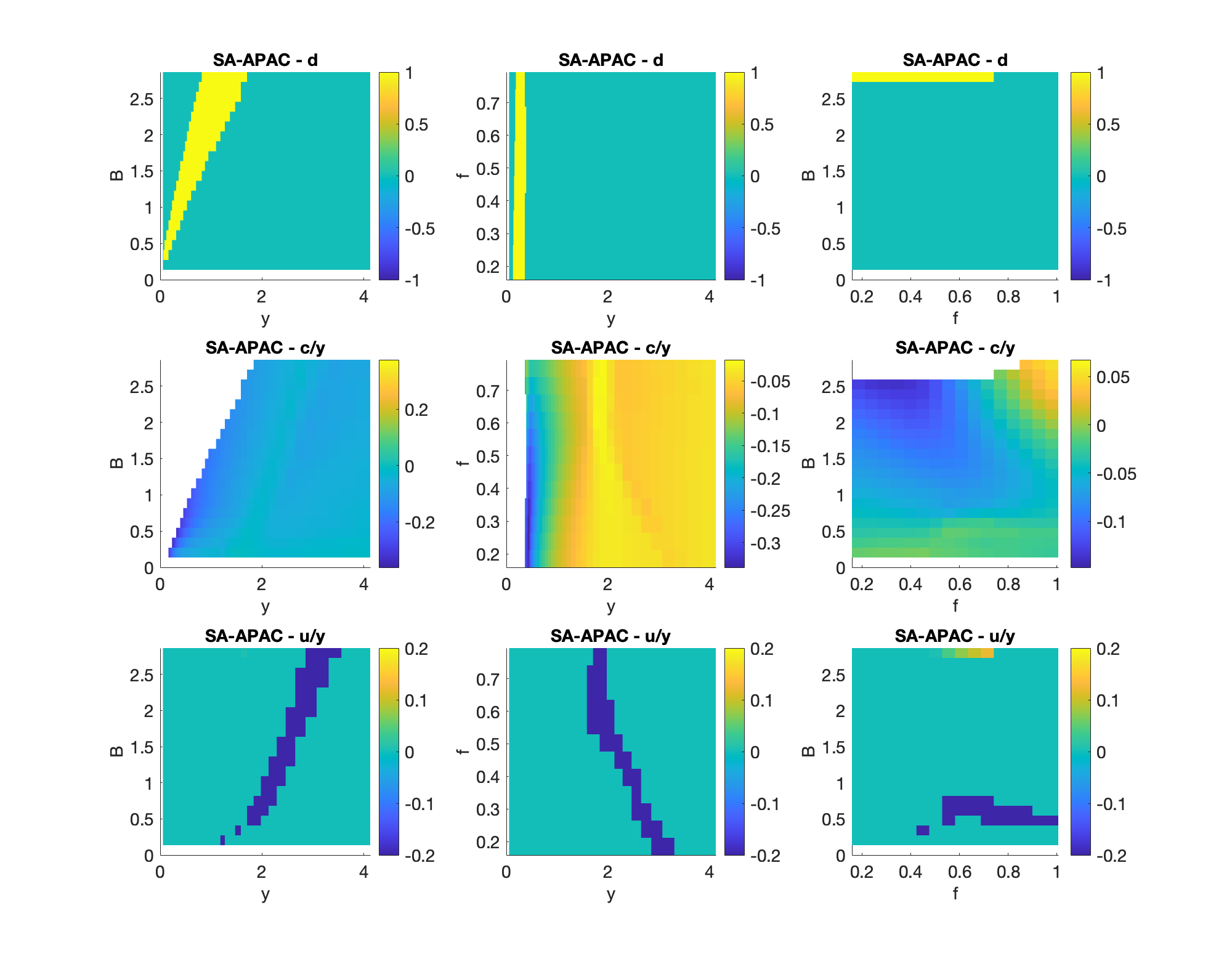}
\caption{Sensitivity analysis - parameter $\eta$, SA-APAC region. We report the differences in $d, \ c/y, \ u/y$ obtained for $\eta=1.2\eta_b$ and $\eta=0.8\eta_b$, $\eta_b$ being the baseline value of the parameter, in case of $d$ we plot $d_{1.2\eta_b}-d_{0.8\eta_b}$, see Figure \ref{M2-SAAPAC} for details and plots of the baseline parameter.}\label{S_eta}
 \end{figure}

 \begin{figure}[ht!]
\centering
 \includegraphics[width=1\textwidth]{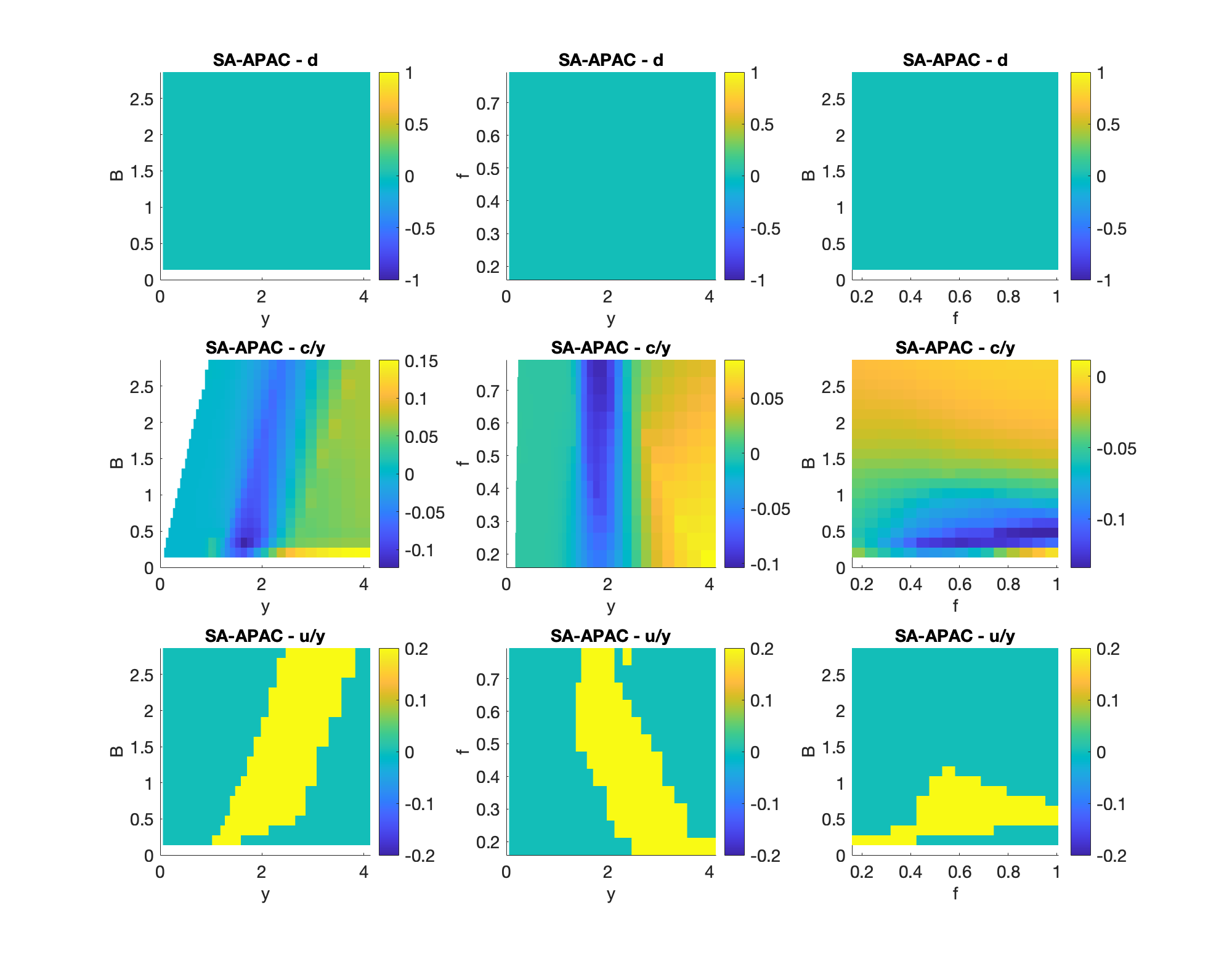}
\caption{
Sensitivity analysis - parameter $k_f$, SA-APAC region. We report the differences in $d, \ c/y, \ u/y$ obtained for $k_f=1.2k_{f,b}$ and $\eta=0.8 k_{f,b}$, $k_{f,b}$ being the baseline value of the parameter, in case of $d$ we plot $d_{1.2 k_{f,b}}-d_{0.8 k_{f,b}}$, see Figure \ref{M2-SAAPAC} for details and plots of the baseline parameter.}\label{S_kf}
 \end{figure}

  \begin{figure}[ht!]
\centering
 \includegraphics[width=1\textwidth]{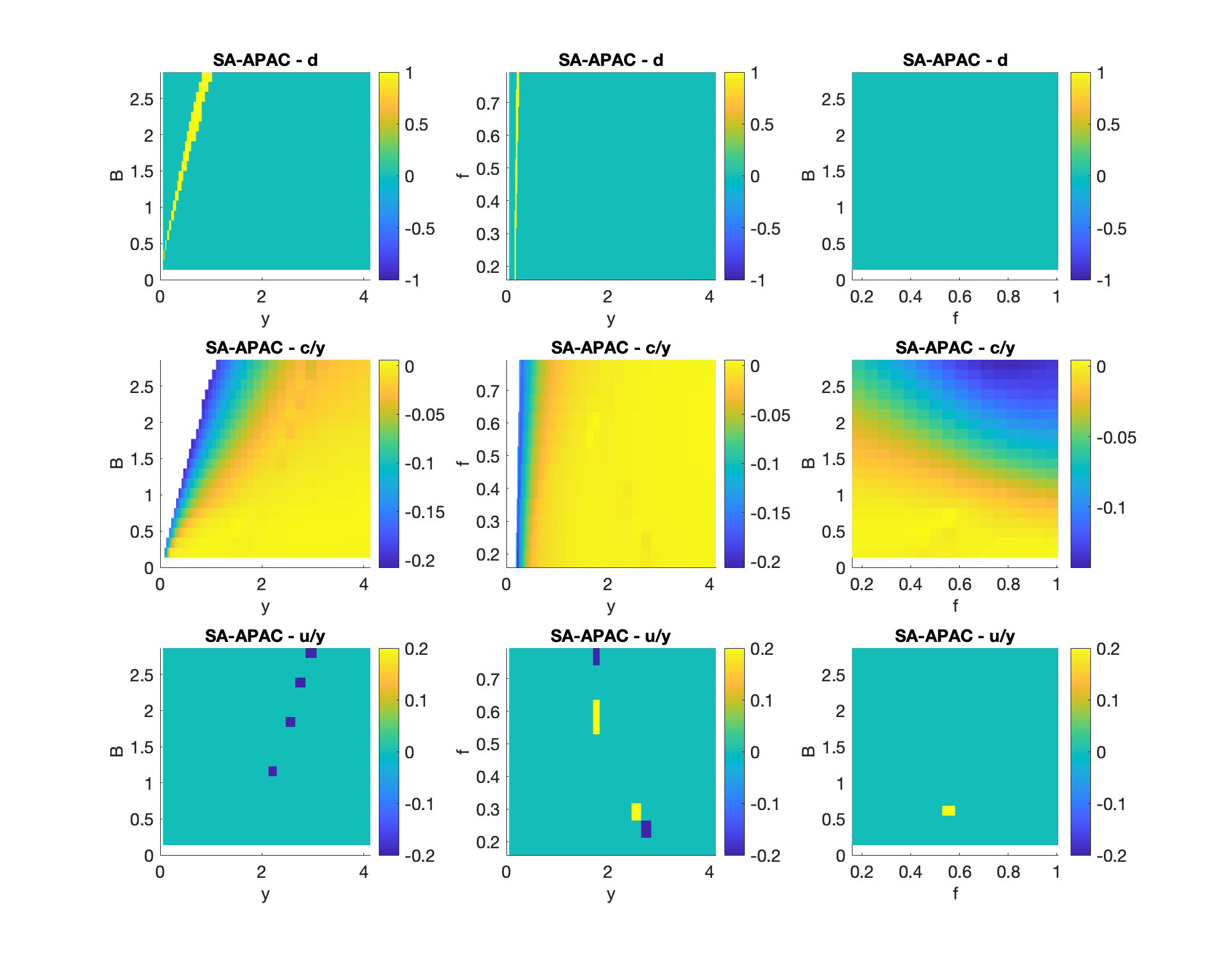}
\caption{Sensitivity analysis - parameter $\lambda$, SA-APAC region. We report the differences in $d, \ c/y, \ u/y$ obtained for $\lambda=1.2\lambda_b$ and $\lambda=0.8\lambda_b$, $\lambda_b$ being the baseline value of the parameter, in case of $d$ we plot $d_{1.2\lambda_b}-d_{0.8\lambda_b}$, see Figure \ref{M2-SAAPAC} for details and plots of the baseline parameter.}\label{S_lambda}
 \end{figure}

  \begin{figure}[ht!]
\centering
 \includegraphics[width=1\textwidth]{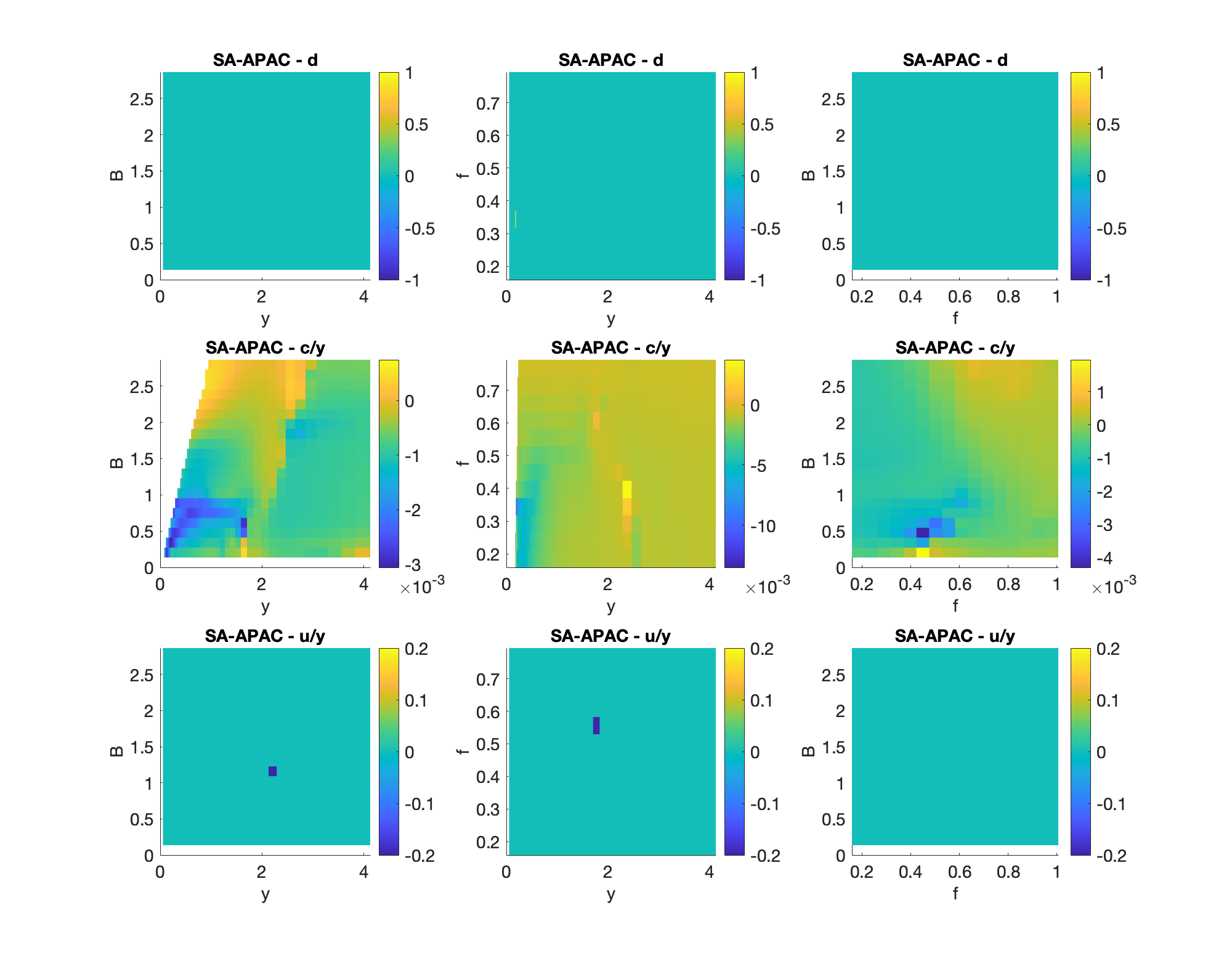}
\caption{Sensitivity analysis - parameter $\sigma_f$, SA-APAC region. We report the differences in $d, \ c/y, \ u/y$ obtained for $\sigma_f=1.2 \sigma_{f,b}$ and $\sigma=0.8 \sigma_{f,b}$, $\sigma_{f,b}$ being the baseline value of the parameter, in case of $d$ we plot $d_{1.2 \sigma_{f,b}}-d_{0.8 \sigma_{f,b}}$, see Figure \ref{M2-SAAPAC} for details and plots of the baseline parameter.}\label{S_sigmaf}
 \end{figure}

   \begin{figure}[ht!]
\centering
 \includegraphics[width=1\textwidth]{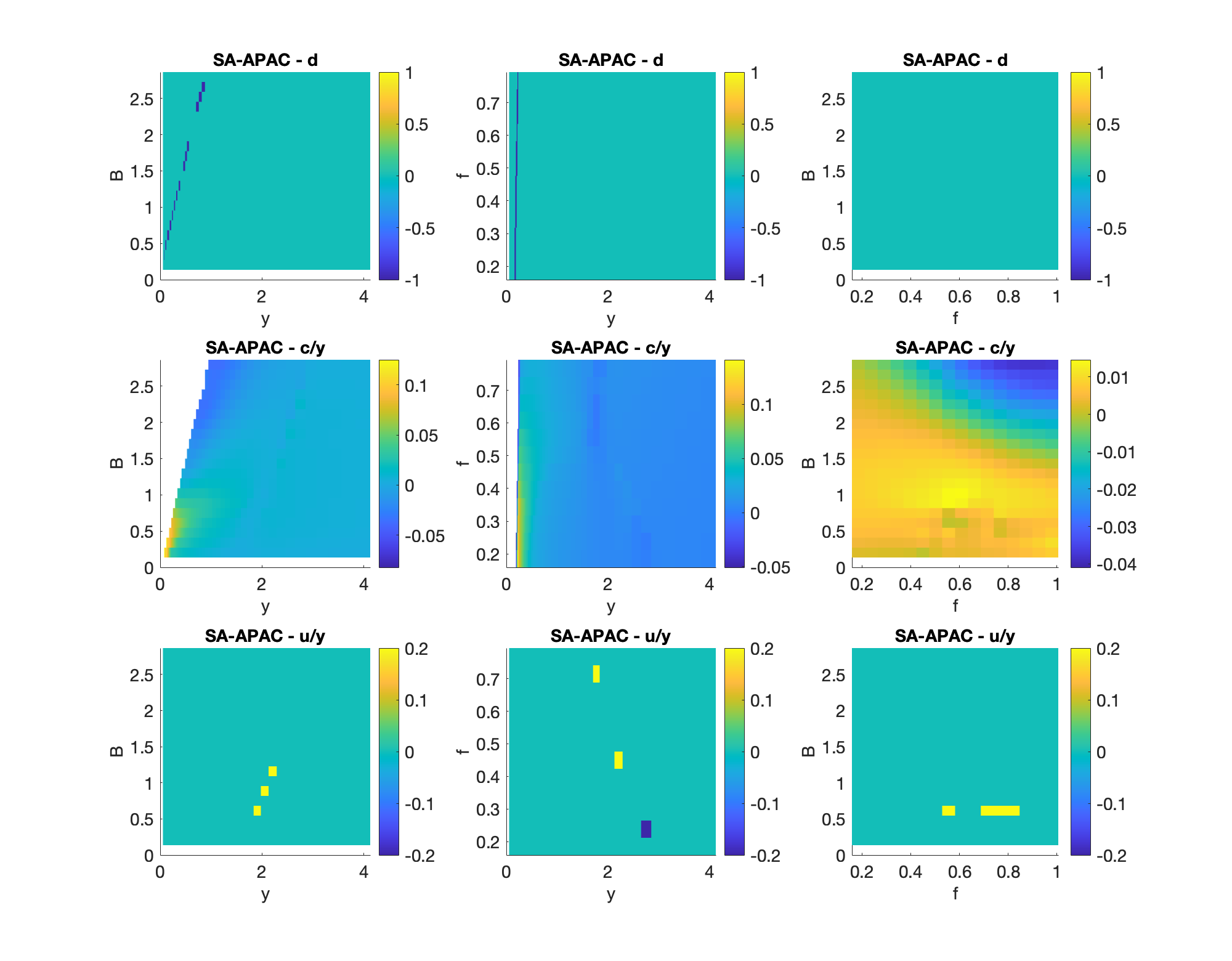}
\caption{
Sensitivity analysis - parameter $\theta_D$, SA-APAC region. We report the differences in $d, \ c/y, \ u/y$ obtained for $\theta_D=1.2\theta_{D,b}$ and $\theta_d=0.8 \theta_{D,b}$, $\theta_{D,b}$ being the baseline value of the parameter, in case of $d$ we plot $d_{1.2 \theta_{D,b}}-d_{0.8 \theta_{D,b}}$, see Figure \ref{M2-SAAPAC} for details and plots of the baseline parameter.}\label{S_thetaD}
 \end{figure}
 
   \begin{figure}[ht!]
\centering
 \includegraphics[width=1\textwidth]{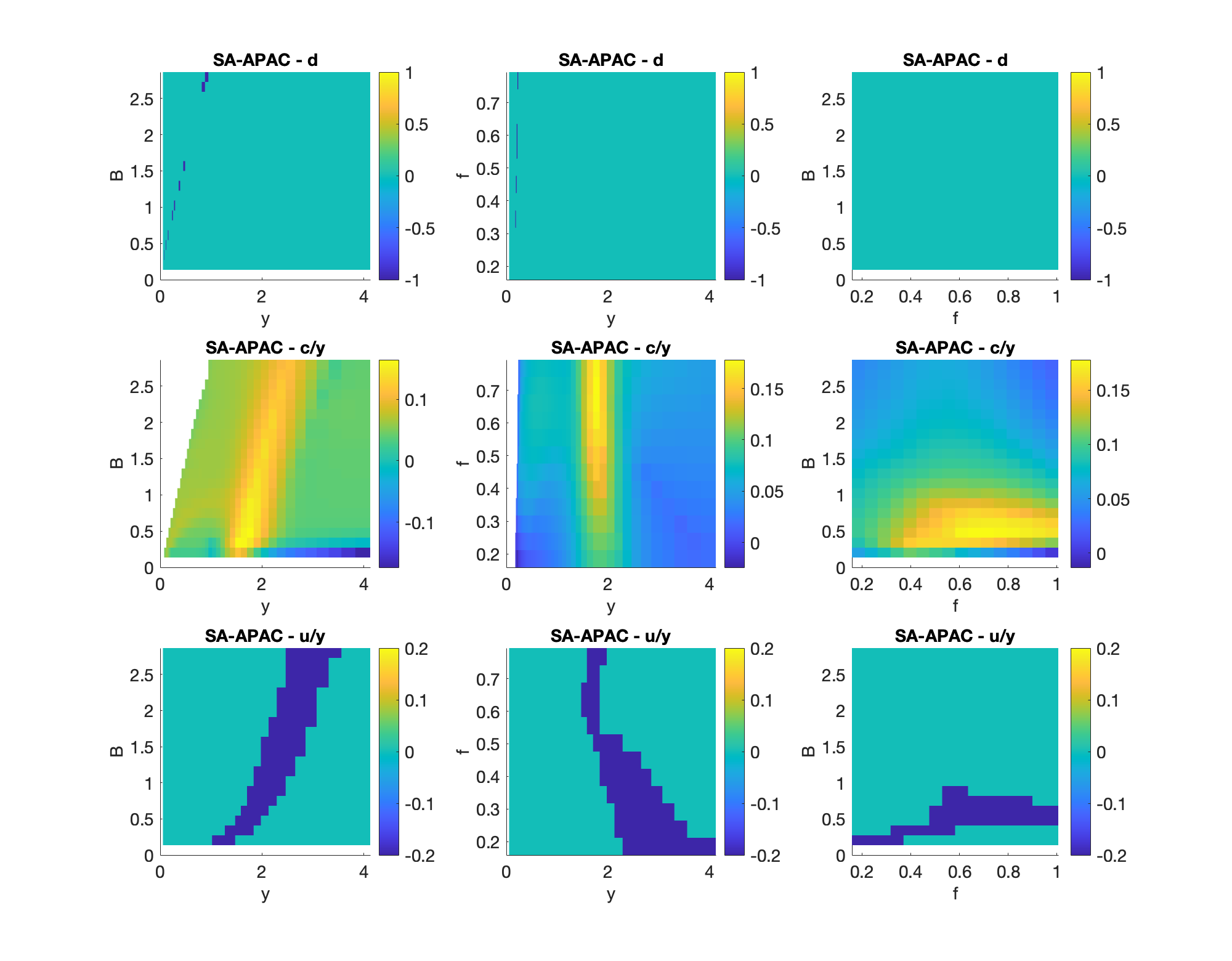}
\caption{
Sensitivity analysis - parameter $\varphi$, SA-APAC region. We report the differences in $d, \ c/y, \ u/y$ obtained for $\varphi_f=1.2\varphi_{b}$ and $\varphi=0.8 \varphi_{b}$, $\varphi_{b}$ being the baseline value of the parameter, in case of $d$ we plot $d_{1.2 \varphi_{b}}-d_{0.8 \varphi_{b}}$, see Figure \ref{M2-SAAPAC} for details and plots of the baseline parameter.}\label{S_varphi}
 \end{figure}

   \begin{figure}[ht!]
\centering
 \includegraphics[width=1\textwidth]{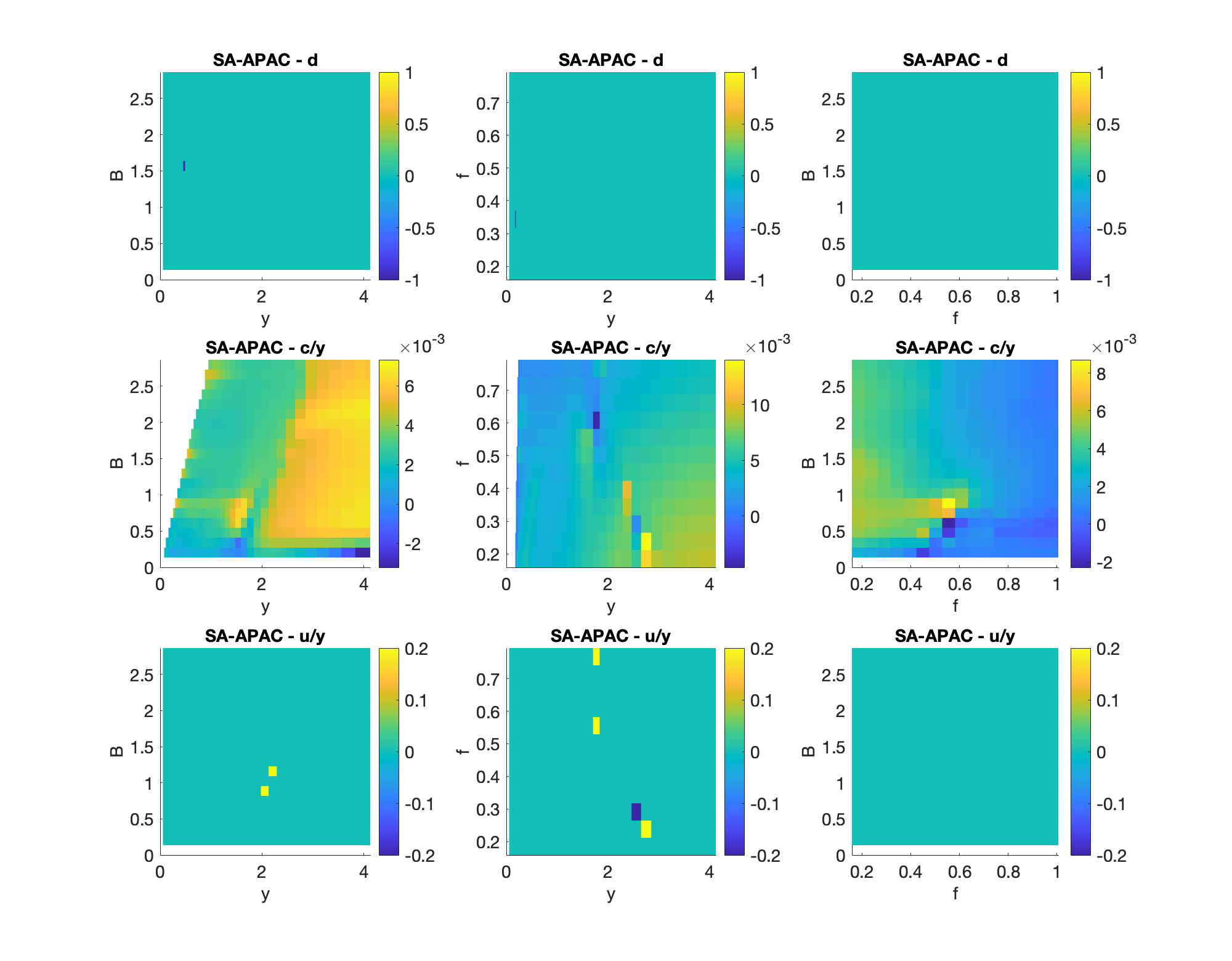}
\caption{
Sensitivity analysis - parameter $g$, SA-APAC region. We report the differences in $d, \ c/y, \ u/y$ obtained for $g=1.2g_{b}$ and $g=0.8 g_{b}$, $g_{b}$ being the baseline value of the parameter, in case of $d$ we plot $d_{1.2 g_{b}}-d_{0.8 g_{b}}$, see Figure \ref{M2-SAAPAC} for details and plots of the baseline parameter.}
\label{S_g}
 \end{figure}
 
    \begin{figure}[ht!]
\centering
 \includegraphics[width=1\textwidth]{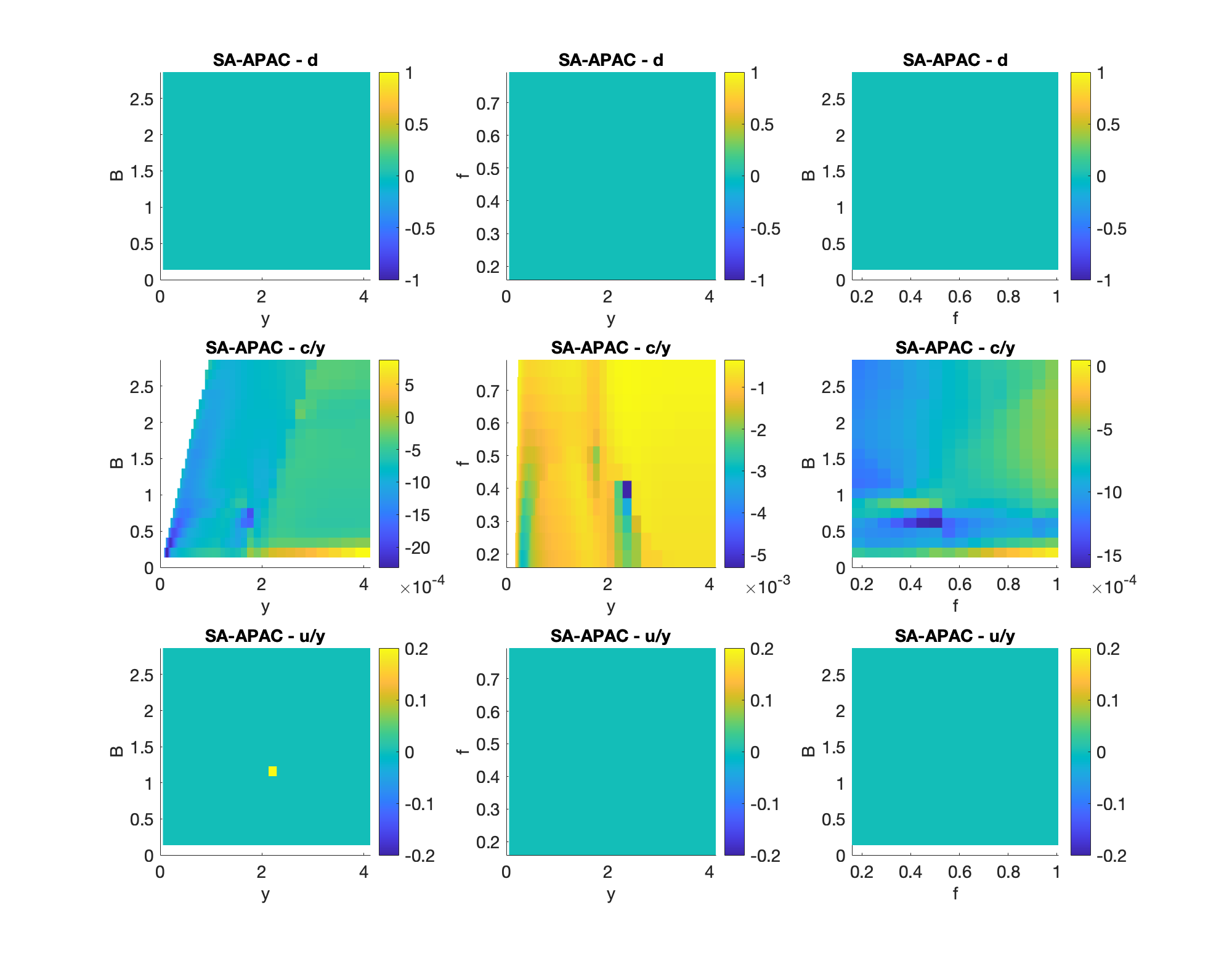}
\caption{
Sensitivity analysis - parameter $\nu_0$, SA-APAC region. We report the differences in $d, \ c/y, \ u/y$ obtained for $\nu_0=1.2\nu_{0,b}$ and $\nu_0=0.8 \nu_{0,b}$, $\nu_{0,b}$ being the baseline value of the parameter, in case of $d$ we plot $d_{1.2 \nu_{0,b}}-d_{0.8 \nu_{0,b}}$, see Figure \ref{M2-SAAPAC} for details and plots of the baseline parameter.}\label{S_nu0}
 \end{figure}
\end{document}